\documentclass[12pt,preprint]{aastex} 

\usepackage [nolists,noheads,nomarkers]{endfloat}
\usepackage{graphicx,epsfig}
\usepackage{verbatim}

\newcommand{\etal}{{et al.~}}

\newcommand{\bq}{\begin{equation}}
\newcommand{\eq}{\end{equation}}

\def\gtsim{\lower.5ex\hbox{$\buSildrel > \over\sim$}}
\def\ltsim{\lower.5ex\hbox{$\buildrel < \over\sim$}}
\def\arcsec{^{\prime\prime}}

\def\apjl{ApJL}
\def\apj{ApJ}
\def\apjs{ApJS}
\def\mnras{MNRAS}
\def\araa{ARAA}
\def\aj{AJ}
\def\aap{A\&A}
\def\aaps{A\&A Suppl.}

\begin{document}

\title{Bulge $n$ and $B/T$ in High Mass Galaxies: Constraints on the Origin of Bulges in Hierarchical Models} 

\author{Tim Weinzirl\altaffilmark{1}, Shardha Jogee\altaffilmark{1}, Sadegh Khochfar\altaffilmark{2,3}, Andreas Burkert\altaffilmark{4}, John Kormendy\altaffilmark{1}}
\altaffiltext{1}{Department of Astronomy, University of Texas at
Austin, Austin, TX }
\altaffiltext{2}{Sub-Department of Astrophysics, University of Oxford, Denys Wilkinson Bldg., Keble Road, OX1 3RH, Oxford, UK}
\altaffiltext{3}{Max Planck Institut f\"ur extraterrestrische Physik, P.O Box 1312, D-85478 Garching, 
Germany}
\altaffiltext{4}{ Universit{\"a}ts-Sternwarte M{\"u}nchen, Scheinerstr. 1, 81679 M{\"u}nchen, Germany}

\begin{abstract}
We use the bulge S\'ersic index $n$ and bulge-to-total ratio ($B/T$) to explore
the fundamental question of how bulges form. We perform  $2D$ bulge-disk-bar decomposition
on $H$-band images of 143 bright, high mass ($M_\star \geq 1.0 \times 10^{10} M_\odot$)
low-to-moderately inclined ($i<70^\circ$) spirals. Our results are:
(1)~Our $H$-band bar  fraction ($\sim58\%$) is consistent with that from ellipse fits.
(2)~70\% of the stellar mass is in disks, 10\% in bars, and 20\% in bulges.
(3)~{\it A  large fraction ($\sim$~69\%) of bright  spirals have
$B/T \leq$~0.2, and $\sim$~76\% have low $n \leq 2$ bulges. These bulges exist 
in barred and unbarred galaxies across a wide range of Hubble types.}
(4)~About 65\% (68\%) of bright spirals with $n\leq2$ ($B/T \le 0.2$)
bulges host bars, suggesting a possible link between bars and bulges.
(5)~We compare the results with  predictions from a set of $\Lambda$CDM  models.
In the models,  a high mass spiral can have a bulge with a 
present-day low  $B/T \leq$~0.2  only if it did not undergo 
a major merger since $z\le 2$.
The predicted fraction ($\sim$~1.6\%) of high mass spirals, which have 
undergone a major merger since $z\le 4$  and host a bulge with
a present-day low  $B/T \le 0.2$, is  {\it a factor of over thirty smaller}  
than the observed fraction  ($\sim66\%$) of high mass spirals  
with   $B/T \le 0.2$. 
Thus, {\it contrary to common perception, bulges built  via major mergers  since $z\le 4$ 
seriously fail to account for the bulges present in $\sim66\%$
of  high  mass spirals}. 
Most of these present-day low $B/T \leq 0.2$ bulges are likely to 
have been built by a combination of minor mergers and/or secular 
processes since  $z\le4$.
\end{abstract}



\section{Introduction}\label{sintro}

The formation of galaxies is a classic problem in astrophysics. Contemporary
galaxy formation models combine the well-established
$\Lambda$ Cold Dark Matter ($\Lambda$CDM) cosmology,
which describes behavior of dark matter on very large scales,
with baryonic physics to model galaxy formation. In the early Universe,
pockets of dark matter decoupled from the Hubble flow, collapsed into
virialized halos, and then clustered hierarchically into larger structures.
Meanwhile, gas aggregated in the interiors of the halos to form rotating
disks, which are the building blocks of galaxies (Steinmetz \& Navarro 2002;
Cole \etal 2000).
Such disks are typically destroyed during major mergers of galaxies with
mass ratio $M_1/M_2 > 1/4$ (e.g. Steinmetz \& Navarro 2002;
Naab \& Burkert 2003; Burkert \& Naab 2004; Mihos \& Hernquist 1996).
When the mass ratio is close to unity, the
remnant is a spheroid with properties close to that of a classical bulge,
namely a steep de Vaucouleurs $r^{1/4}$ surface brightness profile and a high
ratio of ordered-to-random motion ($v/\sigma$).
We shall return to this point in $\S$~\ref{sresult}.
Within this hierarchical framework, the disk of  spiral  galaxies forms
when gas of higher specific angular momentum subsequently accretes around the bulge
(Steinmetz \& Navarro 2002; Burkert \& Naab 2004).

$\Lambda$CDM-based simulations of galaxy formation face several challenges.
One issue is the angular momentum problem; simulated galaxy disks
have smaller scalelengths and, therefore, less specific angular momentum than
their counterparts in nature
(Navarro \& Steinmetz 2000; Burkert \& D'Onghia 2004; D'Onghia \etal  2006).
A second issue is the problem of bulgeless or low bulge-to-total mass
ratio ($B/T$)  spirals.  Within the $\Lambda$CDM paradigm, galaxies
that had a past major merger at a time when its mass was
a fairly large fraction of its present-day mass
are expected to have a significant bulge  with large $B/T$ and high
S\'ersic index.  Depending on the merger history and hence the fraction
of spiral galaxies that  fulfill this criterion
(see $\S$~\ref{smodel1})
we can end up with a small or  large fraction of present-day  galaxies
with low $B/T$.

There is rising evidence that low $B/T$ and bulgeless galaxies are
quite common in the local Universe, especially in low mass or late-type
galaxies.  Late-type Sd
galaxies often harbor no bulge (B{\"o}ker \etal  2002).
Kautsch \etal (2006) and
Barazza, Jogee \& Marinova (2007, 2008) also find
from the analysis of several thousand late-type SDSS galaxies that 15-20\% of such
disk galaxies out to $z \sim 0.03$ appear bulgeless.
Of the 19 local galaxies ($D<8$~Mpc) with circular velocity $V_c>150$~km~s$^{-1}$, 11 (58\%) have
pseudobulges instead of merger-built classical bulges (Kormendy \& Fisher 2008).

Theoretical work by Koda \etal (2007) conclude the survival of disk-dominated
systems in a $\Lambda$CDM universe is compatible with observational constraints
provided classical bulges form only in mergers where $M_1/M_2>0.3$ and the
primary halo has virial velocity $V_{vir}>55$~km~s$^{-1}$.

Evidence also suggests that bulges with  low $B/T$ and low S\'ersic index $n$
may be common even in high mass and/or early-type spirals.
Balcells \etal (2003) report that early-type disk galaxies  
tend to have $n <$~3 and often from 1 to 2.
Laurikainen \etal (2007) find barred and unbarred early-type disk galaxies
to have mean $B/T$ between 0.25 and 0.35, while later Hubble types have $B/T<0.2$;
they also find mean bulge S\'ersic indices to be $\sim2.5$ or less across the Hubble
sequence.
Graham \& Worley (2008) report low $B/D$ ratios across the Hubble sequence
based on bulge-disk decomposition of $K$-band images of local spiral galaxies.
They suggest that these low
values  are problematic for $\Lambda$CDM simulations, but no quantitative
assessment of the extent of the problems is presented.

These emerging statistics on the  fraction of bulgeless ($B/T\sim0$)
galaxies, and  galaxies with low $B/T$ and low $n$ bulges provide
important first constraints. More work is needed to fully explore the
the distribution of bulge properties in both high and low mass galaxies.
In particular, we need to explore how the observed distributions of
bulge $B/T$  and $n$ compare with the predictions from $\Lambda$CDM-based
simulations of galaxy evolution. To the best of our knowledge, few such
quantitative comparisons have  been attempted, so that  it remains unclear
how serious the problem of low $B/T$ galaxies is. This study is an attempt
to derive robust observational constraints on bulge properties in high mass
spirals and to  attempt such a comparison with models.

Completely resolving the issue of low $B/T$ systems will require understanding
the different types of bulges and their formation pathways.  Bulges are
commonly divided in  several groups:  classical bulges, boxy/peanut bulges,
and `pseudobulges' or  disky bulges.  Classical bulges are believed
to be built by major mergers ($M_1/M_2 \ge 1/4$)  and the associated
violent relaxation of stars.
They are associated with modest-to-high bulge  S\'ersic indices, in the range
$2<n<6$ (Hopkins \etal 2008; Springel \etal 2005; Robertson \etal
2006; $\S$~\ref{smodel1}).
Boxy/peanut bulges are believed to be the result of vertical resonances and
buckling instabilities in bars, which are viewed at high inclination
(Combes \& Sanders 1981; Combes \etal 1990; Pfenniger  \& Norman 1990;  Bureau \& Athanassoula 2005;
Athanassoula 2005; Martinez-Valpuesta \etal 2006).
Pseudobulges or  disky bulges are believed to form as a result of
gas inflow into the central kiloparsec and  subsequent star formation building
a  compact disky, high $v/\sigma$  stellar component
(Kormendy 1993; Jogee 1999; Kormendy \& Kennicutt 2004, hereafter KK04;  Jogee,
Scoville, \& Kenney 2005;  Athanassoula 2005; Kormendy  \& Fisher 2005).
Pseudobulges  tend to have a bulge $n<2.5$ (Kormendy  \& Fisher 2005; Fisher \& Drory 2008).

One possibility for the formation of  disky bulges or pseudobulges  is the
idea of secular evolution (Kormendy 1993; KK04; Jogee,
Scoville, \& Kenney 2005),  where a stellar bar or globally oval structure
in a  {\it non-interacting}
galaxy drives the gas inflow into the inner kpc via  shocks and gravitational
torque. Another idea for building disky bulges is that the gas inflow into the
inner kiloparsec is driven by  {\it external non-secular processes}, such as tidal
interaction and minor mergers. The gas inflow in such cases can be caused
by a  tidally induced non-axisymmetric feature,
such as a bar (e.g.,  Quinn et al. 1993; Hernquist \& Mihos 1995),
and by tidal torques from  the companion.  The  subsequent central star formation can
still form a compact  high $v/\sigma$  stellar component, $aka$ a pseudobulge.

Throughout this paper,  we avoid making any
{\it a priori} assumptions about the origin of different types of bulges by
simply referring to them according to their bulge S\'ersic index
$n$ or  bulge-to-total mass ratio ($B/T$).  We consider bulges of
high  ($n\ge$~4), intermediate ($2<n<4$) and low ($n \leq 2$)
index, as well as those of low or high $B/T$.

The structural properties of galaxy components, such as  bulges, disks, and bars
can be derived through the decomposition of the $2D$ light distribution, taking into
account the PSF.
Many early studies have performed  only   two
component $2D$  bulge-disk decomposition (e.g., Allen \etal 2006; Byun
\& Freeman 1995; de Jong 1996; Simard 1998; Wadadekar \etal 1999),
ignoring the contribution of the bar, even in strongly barred galaxies.
However, recent work has shown that it is important to include the bar in $2D$
decomposition of barred galaxies, else the $B/T$ ratio can be artificially
inflated, and bulge properties skewed (e.g., Laurikainen \etal  2005, 2007).
Furthermore, since most ($\ge60\%$) bright spiral galaxies are
barred in the NIR (Eskridge  \etal  2000; Laurikainen \etal 2004; Marinova \&
Jogee 2007, hereafter MJ07;  Menendez-Delmestre \etal 2007), the inclusion of the bar is quite
important.  This has led to several recent studies, where $2D$ bulge-disk-bar
decomposition are being performed  (e.g.   Laurikainen \etal  2007;
Reese \etal 2007; Gadotti \& Kauffmann 2007).

Another advantage of bulge-disk-bar decomposition over bulge-disk decomposition
is that the former allows us  to constrain the properties of the bar itself.
Bars provide the most important internal mechanism for redistributing
angular momentum in baryonic and dark matter components
(e.g.  Weinberg 1985; Debattista \& Sellwood 1998, 2000;
Athanassoula 2002; Berentzen, Shlosman, \& Jogee 2006).
They efficiently drive gas inflows into the central
kpc, feed central starbursts  (Elmegreen 1994; Knapen
\etal  1995; Hunt \& Malakan 1999; Jogee \etal 1999;  Jogee,
Scoville, \& Kenney 2005; Jogee 2006) and lead to the formation of
disky or pseudobulges (see above). Furthermore, the prominence of strong bars out
to $z\sim 1$  over the last 8 Gyr  (Jogee \etal 2004; Sheth \etal 2008)
suggest that bars have been present over cosmological times and can
shape the  dynamical  and secular evolution of disks.
Thus, quantifying bar properties, such  as the fractional light and mass ratio
(Bar/$T$), can yield insight into these processes.

In this paper, we constrain the properties of bulges and bars along the Hubble
sequence, and compare our results with $\Lambda$CDM-based simulations of galaxy
evolution. In $\S$~\ref{sdata}, we define our complete sample of 143
bright ($M_B\leq-19.3$) low-to-moderately inclined ($i \le 70^\circ$) spirals from the Ohio State
University Bright Spiral Galaxy Survey (OSUBSGS; Eskridge \etal 2002), which is
widely used as the local reference sample for  bright spirals by numerous
studies  (e.g.,  Eskridge \etal 2000; Block \etal 2002;  Buta \etal 2005;
MJ07 ; Laurikainen \etal 2004, 2007).
In $\S$~\ref{sanaly}, we  perform $2D$ bulge-disk and bulge-disk-bar decompositions of $H$-band
images using GALFIT (Peng \etal 2002), and derive  fractional light ratios ($B/T$, Bar/$T$,
Disk/$T$), as well as  S\'ersic indices and half light radii or scale lengths.
Tests to verify the  robustness  of our decompositions  are presented in $\S$~\ref{stests}.
In $\S$~\ref{sresult}, we present our results.
Specifically, the total stellar mass present in bulges, disks, and bars is calculated  ($\S$~\ref{smass}).
In  $\S$~\ref{sbulgenbt}, the distribution of bulge S\'ersic  index $n$ and $B/T$
as a function of galaxy Hubble type and stellar mass is presented, and the
surprising prevalence of bulges with low S\'ersic index $n$ and low $B/T$ is established.
A comparison with other works is presented in $\S$~\ref{scomp1}.
We  examine how Bar/$T$ and bar fraction (the fraction barred disks) change as a function of host galaxy properties
in $\S$~\ref{sbar2}.
In  $\S$~\ref{smodel1}, we compare our  observed  distribution of
bulge $B/T$ and $n$ in  high mass ($M_\star \geq  1.0\times 10^{10} M_\odot$)
spirals with predictions from $\Lambda$CDM  cosmological semi-analytical models.
$\S$~\ref{ssumm} summarizes our results.

\section{Sample Properties}\label{sdata}
\subsection{OSUBSGS}\label{sdataosu}
Our dataset is derived from the 182 $H$-band images from the public 
data release of the Ohio State University Bright Spiral Galaxy Survey 
(OSUBSGS; Eskridge \etal 2002). These galaxies are a subset of the RC3 catalog
that have $m_B \leq 12$, Hubble types $0 \leq T \leq 9$ (S0/a to Sm), 
$D_{25} \leq 6'.5$,
and $ -80^\circ < \delta < +50^\circ$. Imaging of OSUBSGS galaxies
spans optical and near infrared (NIR) wavelengths with $BVRJHK$ images 
available for most galaxies.  OSUBSGS images were acquired on a wide
range of telescopes with apertures ranging from 0.9-2.4m.  The $JHK$
data were acquired with a variety of telescopes and detectors, but mainly 
with the 1.8 m Perkins reflector at Lowell Observatory and the CTIO 
1.5 m telescope with the OSIRIS detector, 
having 18.5 micron pixels (Depoy \etal 1993).  Pixel scale is dependent on the telescope
and for these observations ranged between $1-1.50\arcsec$/pix. Exposure
times were heterogeneous, but the total observing time per object 
was typically between 10-15 minutes in $H$.  The resulting limiting 
$H$-band surface brightness is $\sim20$ mag arcsec$^{-2}$.
The typical limiting surface brightnesses of the images
$\sim 26$ mag arcsec$^{-2}$ in $B$-band and 
$\sim 20$ mag arcsec$^{-2}$ in $H$-band (Eskridge \etal 2002).
The seeing depends on observing time and location.  We find the 
$H$-band images have seeing of $\sim3\arcsec$.

We choose to use the  NIR images rather than optical
ones for several reasons. Firstly, NIR images are better tracers of the
stellar mass than optical images, and the mass-to-light ratio is less
affected by age or dust gradients.
Secondly, obscuration by dust and SF are minimized in the NIR, compared
to the optical. As the $K$-band images are of poor quality, we settle on 
using the $H$-band images.


The OSUBSGS is widely used as the local reference sample for  bright spirals by numerous studies  
(e.g. Eskridge \etal 2000; Block \etal 2002;  Buta \etal 2005;
MJ07; Laurikainen \etal 2004, 2007).  Thus, there are numerous complementary results
that we can use or compare with. In particular, MJ07 have identified bars in this sample
using quantitative criteria based on ellipse fitting, and characterized their
sizes, position angles, and ellipticities.

\begin{figure}[]
\centering
\scalebox{1.00}{\includegraphics*[0in,2.6in][5in,10in]{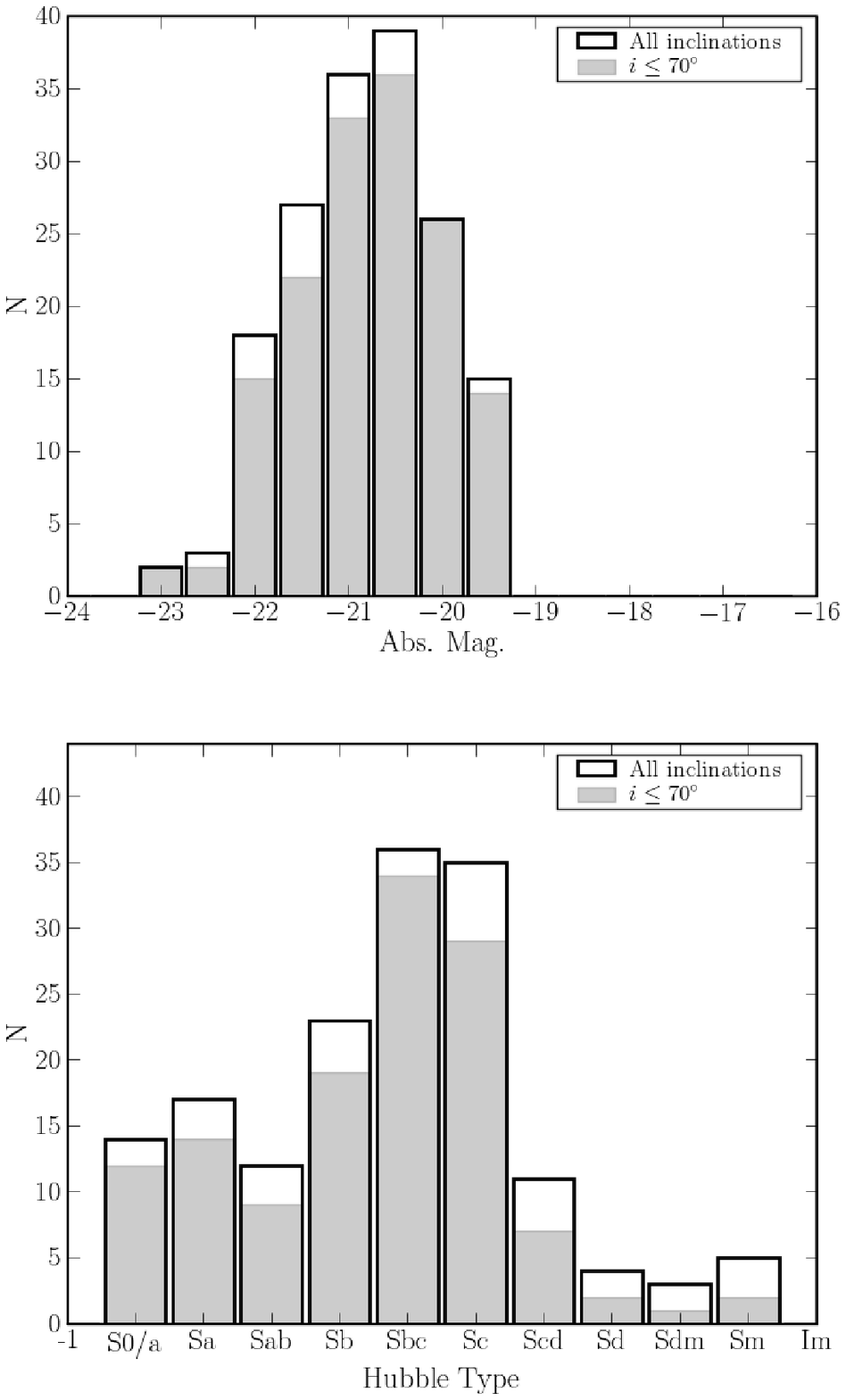}}
\caption{
Our final sample S1 consists of 143  bright ($M_B\leq-19.3$)
low-to-moderately inclined ($i \le 70^\circ$)  spirals  in the OSUBSGS  survey.
The distribution of absolute $B$-band magnitude for the
sample of bright spirals in the  OSUBSGS  survey
is shown in the top panel before  (solid line) and after (shaded greyscale)
the cut to remove highly inclined ($i > 70^\circ$) spirals.
The distribution of Hubble types for the
sample is shown in the bottom panel before (solid line) and
after (shaded greyscale) the cut to remove highly
inclined ($i > 70^\circ$) spirals. \label{2posu-hist-hubb}}
\end{figure}

\begin{figure}[]
\centering
\plotone{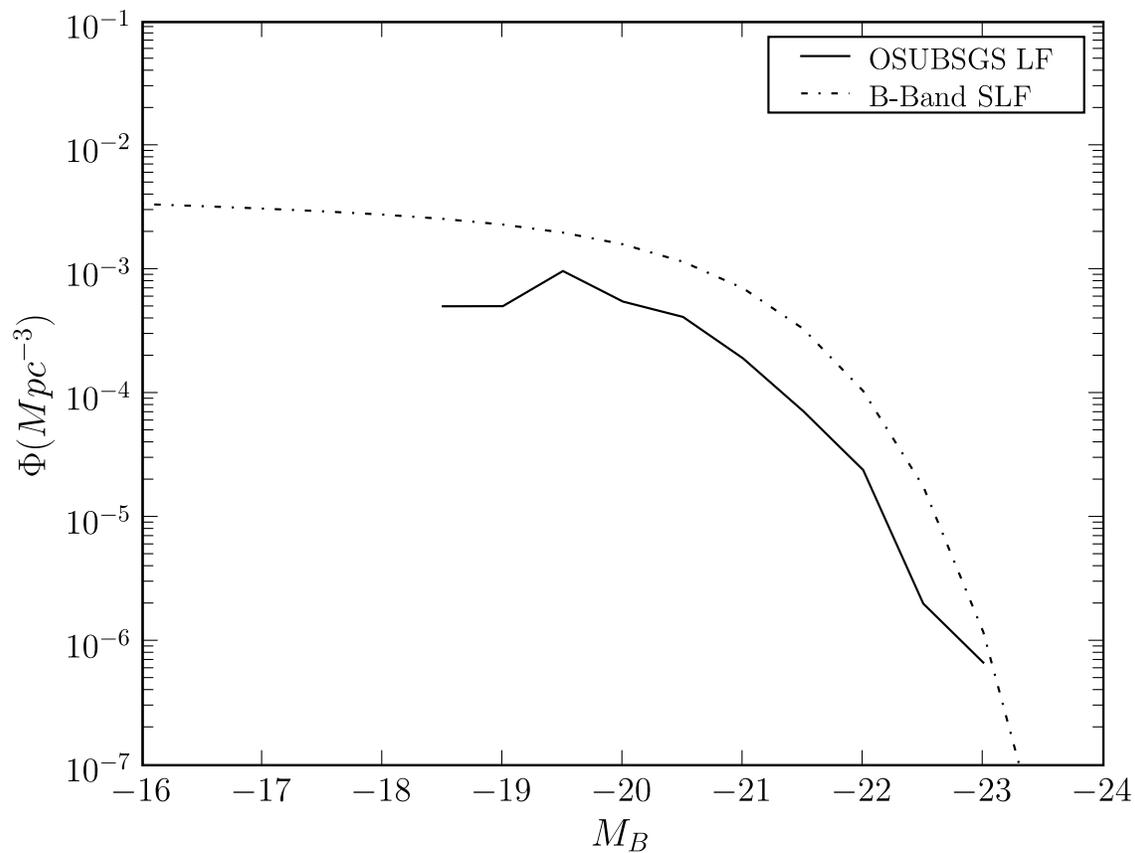}
\caption{The luminosity function of the full OSUBSGS sample is compared with
the $B$-band Schechter luminosity function (SLF).  The former is calculated as described in
\S~\ref{sdataosu} using equation~(\ref{vmax}).
The parameters for the SLF are $\Phi^*=5.488\times10^{-3}$
Mpc$^{-3}$, $\alpha=-1.07$, and $M^*_B=-20.5$ (Efstathiou, Ellis \& Peterson  1988), 
corresponding to $H_0$=70 km/s Mpc$^{-1}$. \label{osu_uncorr}}
\end{figure}

OSUBSGS is a magnitude-limited survey with objects whose distances range up to
$\sim60$ Mpc.  Faint galaxies are inevitably missed at larger distances,
resulting in the absolute magnitude distribution in Figure~\ref{2posu-hist-hubb}.
We compare the  $B$-band LF of this sample  with  a Schechter (Schechter 1976) 
LF (SLF) with $\Phi^*=5.488\times10^{-3}$ Mpc$^{-3}$, $\alpha=-1.07$,
and $M^*_B=-20.5$ (Efstathiou, Ellis \& Peterson  1988)
in Figure~\ref{osu_uncorr}.  The volume used to determine the number density
in each magnitude bin is
\begin{equation} \label{vmax}
V_{max}=\frac{4\pi}{3}d^3_{max}(M),
\end{equation}
where
\begin{equation} \label{dmax}
d_{max}(M)=10^{ 1+0.2(m_c-M)  }
\end{equation}
is the maximum distance out to which a galaxy of absolute
magnitude $M$ can be observed given the cutoff magnitude $m_c$.
If the SLF is representative of the true LF, then Figure~\ref{osu_uncorr}
suggests that the OSUBSGS sample is seriously
incomplete at $M_B>-19.3$, while at the brighter end (-19.3 to -23)
the shape of its LF matches fairly well the SLF. We thus conclude that
the sample is reasonably complete for bright ($M_B\leq-19.3$ or
$L_B >$ 0.33 $L^*$) galaxies.

\begin{figure}[]
\centering
\plotone{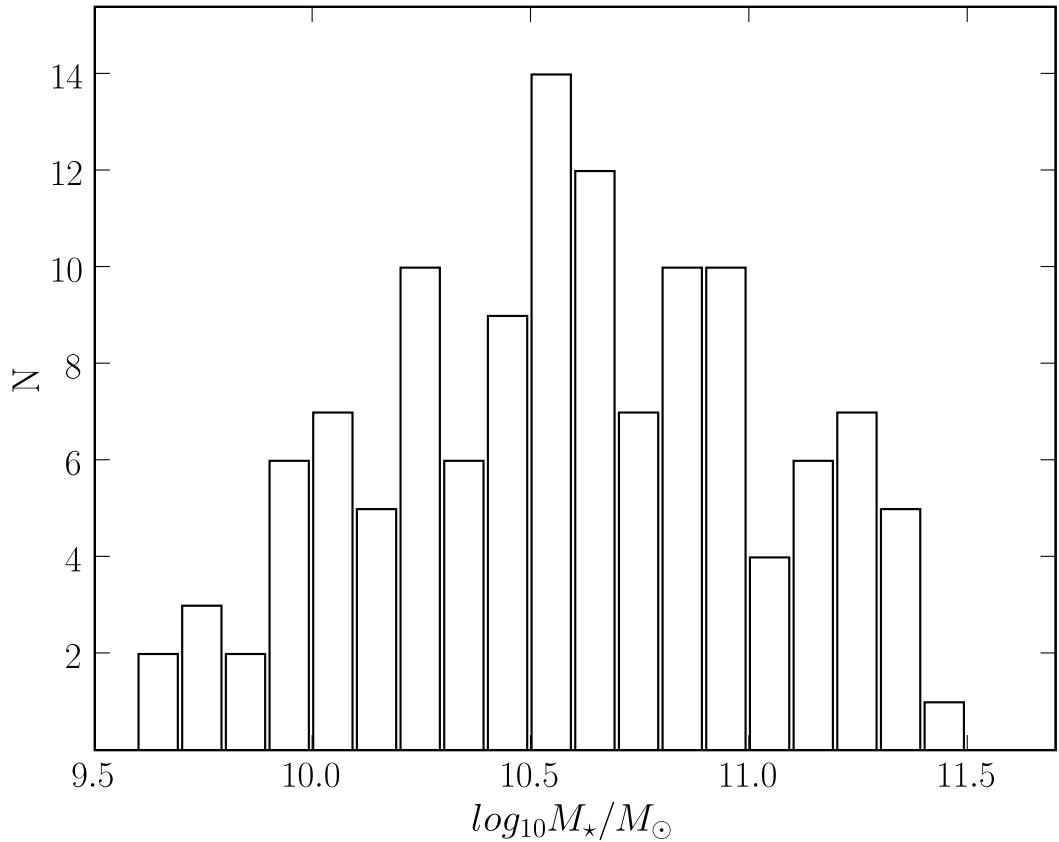}
\caption{
Out of our final sample S1  of 143  bright ($M_B\leq-19.3$) low-to-moderately
inclined ($i \le 70^\circ$)  OSUBSGS spirals, stellar
masses could be estimated for   126 galaxies. Their stellar mass 
distribution is shown, as determined in
$\S$~\ref{stellmass}. Most have stellar masses $M_{\star} \geq 1.0 \times 10^{10} M_\odot$.
This sample  of 126 galaxies is referenced henceforth as the sample S2.
}
\label{stellmasshist}
\end{figure}

We exclude highly inclined ($i > 70^\circ$) galaxies for which
structural decomposition does not yield accurate results.
Thus, our final sample S1 consists of 143 bright ($M_B\leq-19.3$)
low-to-moderately inclined ($i \le 70^\circ$) spirals
with Hubble types mainly in the range S0/a to Sc (Figure~\ref{2posu-hist-hubb}).
Of the 126 for which we could derive stellar masses (see $\S$~\ref{stellmass}, most
have stellar masses $M_{\star} \geq 1.0 \times 10^{10} M_\odot$
(Figure~\ref{stellmasshist}).
Table~\ref{tosu} summarizes the morphologies, luminosities, and stellar masses of the
sample. Note that there are few galaxies of late  Hubble types (Scd or later) and
we do not draw any conclusions on  such systems from our study. In a future paper,
we will tackle galaxies  of lower mass and later Hubble types.

\subsection{Stellar Masses}\label{stellmass}
We derive global stellar masses for most of the OSUBSGS sample galaxies
using  the relation between stellar mass and rest-frame $B-V$ color from 
Bell \etal (2003). 
Using population synthesis models, the latter study  
calculates stellar $M/L$ ratio as a function of color using 
functions of the form $log_{10}(M/L)=a_\lambda+b_\lambda\times Color+C$, 
where $a_\lambda$ and $b_\lambda$ are bandpass dependent constants and 
C is a constant 
that depends on the stellar initial mass function (IMF). 
For the $V$ band Bell \etal (2003) find $a_\lambda=-0.628$ and
$b_\lambda=1.305$; assuming a Kroupa (1993) IMF, they find $C=-0.10$.
This yields an expression for the stellar mass in  $M_{\odot}$ for a 
given $B-V$ color:

\begin{equation}\label{masseq}
M_{\star}=v_{lum}10^{-0.628+1.305(B-V)-0.10},
\end{equation}
where
\begin{equation}\label{vlumeq}
v_{lum}=10^{-0.4(V-4.82)}.
\end{equation}
Here, $v_{lum}$ is the luminosity parametrized in terms of absolute $V$
magnitude. 

How reliable are stellar masses determined from this procedure?  Clearly, the
above relationship between $M_\star$ and $B-V$  cannot apply to all galaxies, 
and must depend on the assumed stellar IMF, and range of  ages, dust, and metallicity. 
However, it is encouraging  to note that several studies
(Bell \etal 2003; Drory \etal 2004; Salucci, Yegorova, \& Drory 2008) 
find  generally good agreement between
masses based on   broad-band colors  and those from  spectroscopic
(e.g.  Kauffmann \etal 2003)  and dynamical (Drory \etal 2004) techniques.
Typical errors are within a factor of two to three.
Salucci \etal (2008) derive disk masses with both photometric
and kinematic methods and find the two methods are equivalent on average.
For a sample of 18, they find an rms scatter of 0.23 dex, 
while on an individual basis the deviation can be as high as 0.5 dex.

We used this relation to compute stellar masses for 126 of 143 (88\%) objects.
The remainder did not have $B-V$ colors available in the Hyperleda
database or RC3.  The mass distribution is summarized in Figure~\ref{stellmasshist}.
Individual masses are listed in Table~\ref{tosu}. 
This sample of 126 galaxies is referenced henceforth as sample S2.

\section{Method and Analysis}\label{sanaly}
The structural properties of galaxy components, such as  bulges,
disks, and bars can be derived through the decomposition of the
2D light distribution, taking into account the PSF.
There are several algorithms for  $2D$ luminosity
decomposition, including GIM2D (Simard \etal 2002),
GALFIT (Peng \etal 2002), and BUDDA (de Souza \etal 2004).
The latter two allow bulge-disk-bar decomposition, 
while the former only allows bulge-disk decomposition.

Most previous work has addressed $2D$ bulge-disk decomposition only.
Allen \etal (2006), for example, performed bulge-disk decomposition of $B$-band 
images with GIM2D on 10,095  galaxies from the Millennium Galaxy Catalog
(Liske \etal 2003; Driver \etal 2005).
However, recent work (e.g., Laurikainen \etal  2005; 
 Graham \& Balcells, in preparation)  has shown that the $B/T$ ratio can
be artificially  inflated in a barred galaxy unless the bar
component is included in the $2D$ decomposition. The fact that
most ($\ge60\%$) bright spiral galaxies are barred in the NIR
(Eskridge  \etal  2000; Laurikainen \etal 2004; MJ07;
Menendez-Delmestre \etal 2007), further warrants the inclusion
of the bar.
Another advantage of bulge-disk-bar decomposition is that
it allows us  to constrain the properties of the bar itself,
and to constrain scenarios of bar-driven evolution
(see $\S$~\ref{sintro}).

Motivated by these considerations, several studies have tackled the problem of
$2D$ bulge-disk-bar decomposition.
Laurikainen \etal (2005, 2007) have developed a $2D$ multicomponent
decomposition code designed to model bulges, disks, primary and
secondary bars, and lenses; they apply S\'ersic functions to bulges
and use either S\'ersic or Ferrers functions to describe bars and lenses.
Reese \etal (2007) have written a non-parametric algorithm to model
bars in $\sim70$ $I$-band images.
Gadotti \& Kauffmann (2007) are performing $2D$ bulge-disk-bar and
bulge-disk decomposition of 1000 barred and unbarred galaxies from SDSS
with the BUDDA software.

In this study, we  perform $2D$  two-component bulge-disk decomposition and three-component
bulge-disk-bar decomposition of the OSUBSGS sample with GALFIT.
We note that  Laurikainen \etal (2007) have also performed bulge-disk-bar
decomposition on the OSUBSGS sample. However, there are also important
complementary differences between our study and theirs.
The decomposition algorithm and tests on the robustness performed in
our study are different (see $\S$~\ref{sanaly} and 
$\S$~\ref{stests}). Furthermore, unlike
Laurikainen \etal (2007), we also compare the bulge-to-total ratio ($B/T$) with predictions
from hierarchical models of galaxy evolution\ ($\S$~\ref{sresult}),
and also present the distribution of  bar-to-total  ratio (Bar/$T$).

\subsection{Image Preparation}\label{sdecom1}
Running GALFIT on an image requires initial preparation.  
The desired fitting region and sky background must be known, and
the PSF image, bad pixel mask (if needed), and pixel noise map must 
be generated.  We addressed these issues as follows:
(1)  The GALFIT fitting region must be large enough to include the outer galaxy disk, 
as well as some sky region.  Since cutting out
empty regions of sky can drastically reduce GALFIT run-time, a balance was 
sought between including the entire galaxy and some decent sky region, while 
excluding large extraneous blank sky areas.
(2)  It is possible for GALFIT to fit the sky background, but this is not
recommended.  When the sky is a free parameter, the wings of the bulge S\'ersic 
profile can become inappropriately extended, resulting in a S\'ersic index
that is too high.  
Sky backgrounds were measured separately and designated as fixed parameters; 
(3) GALFIT requires a PSF image to correct for seeing effects.  Statistics of
many stars in each frame can be used to determine an average PSF.
However, many of our images contain merely a few stars. Instead, 
a high S/N star from each frame was used as a PSF.
(4) We used  ordered lists of pixel coordinates to make bad pixel masks,
which  are useful for blocking out bright stars and other image
artifacts.
(5) We had GALFIT internally calculate pixel noise maps for an image from the noise 
associated with each pixel.  Noise values are determined from image
header information concerning gain, read noise, exposure time, and the 
number of combined exposures.

\subsection{Decomposition Steps}\label{sdecom2}
\begin{figure}[]
\centering
\plotone{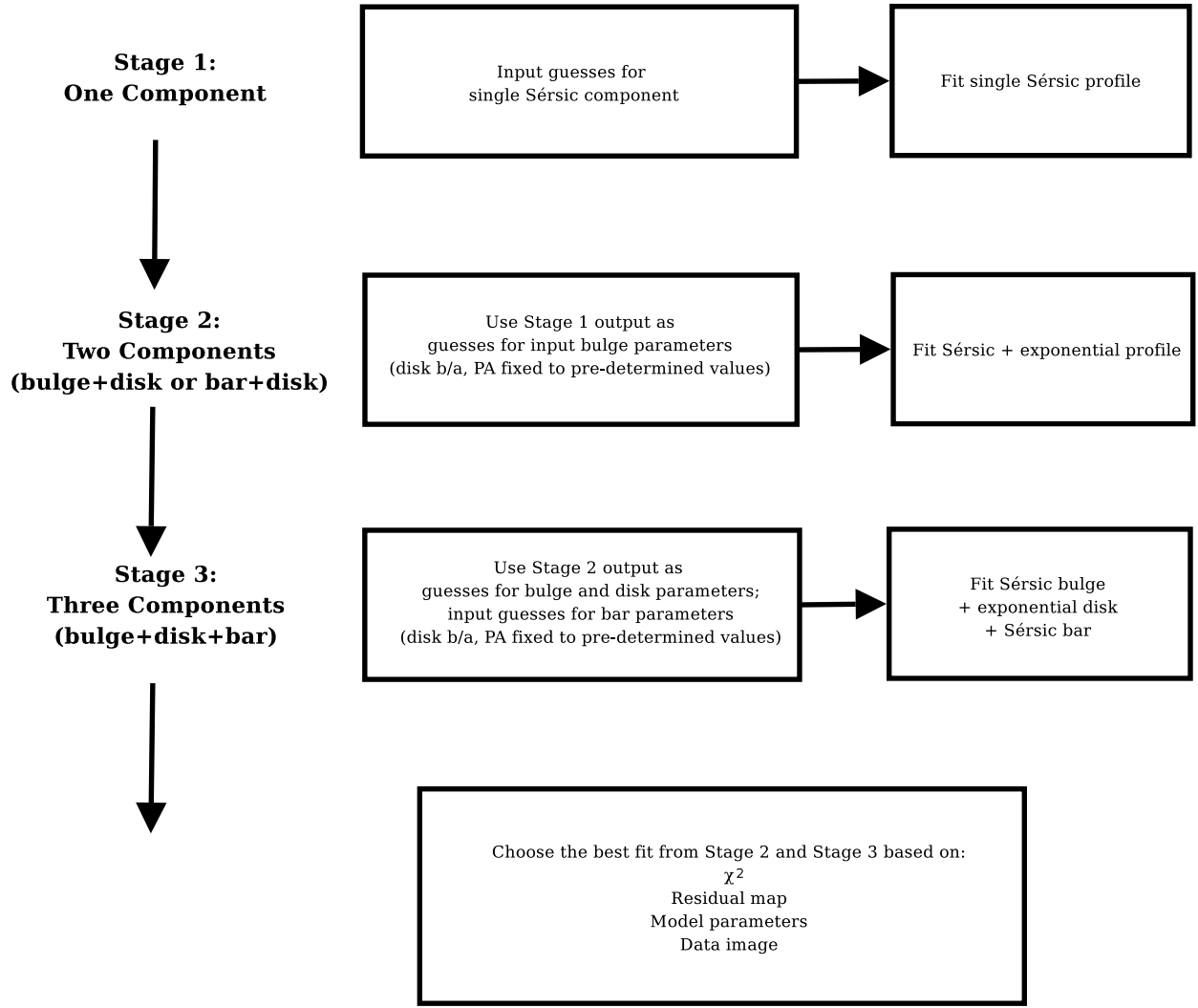}
\caption{An overview of the method of decomposition. All images are
subjected to Stages 1, 2, and 3.  Either the best fit of Stage 2 or Stage 3 is chosen
as the best model.   \label{flowchart}}
\end{figure}

Figure~\ref{flowchart} summarizes our method of decomposition, which we now detail.  
GALFIT requires initial guesses for each component it fits.  It uses 
a Levenberg-Marquardt downhill-gradient algorithm  to determine the
minimum $\chi^2$ based on the input guesses.  GALFIT continues iterating
until the $\chi^2$ changes by less than 5e-04 for five iterations 
(Peng \etal 2002).  
We recognize that a drawback to any least-squares method is that a local 
minimum, rather than a global minimum, in $\chi^2$ space may be converged upon.  
We explore this possibility with multiple tests described in \S~\ref{stests}. 
We adopted an iterative process,  involving three 
separate invocations of GALFIT, to perform 1-component, 2-component,
and 3-component decomposition:

\begin{enumerate}
\item 
Stage 1 (single S\'ersic):  In Stage 1, a single S\'ersic component is fitted 
to the galaxy. This serves the purpose of measuring the total luminosity, which is 
conserved in later Stages, and the centroid of the galaxy, which is invariant 
in later fits. 
\item 
Stage 2 (exponential plus S\'ersic):  In Stage 2, the image is fit with the sum 
of an exponential disk and a S\'ersic component. 
During the Stage 2 fit,  the  disk $b/a$ and $PA$ are held constant at  values, 
which we take from the  published ellipse fits of MJ07,  as well as ellipse fits of our 
own.  
This procedure reduces the number of free parameters in the fit by fixing the 
disk  $b/a$ and $PA$, which are easily measurable parameters. 
It also prevents GALFIT from  confusing the disk and bar, and artificially 
stretching the disk along the bar PA in an attempt to  mimic  the bar.
As initial guesses for the S\'ersic component in Stage 2, the output of Stage 
1 is used.
The  S\'ersic component in Stage 2 usually represents the bulge, in which case
Stage 2  corresponds to a standard bulge-disk decomposition.

However, in a few rare cases, where the galaxy has just a 
bar and a disk, the   S\'ersic component in Stage 2  represents a bar. 
The latter  is recognizable by a low  S\'ersic index and large half-light radius.
\item 
Stage 3 (exponential plus two S\'ersic components):  In Stage 3, a three-component 
model consisting of  an exponential disk, a S\'ersic bulge, and a  S\'ersic bar is 
fit. As suggested by Peng \etal (2002), the bar can be well described by 
an elongated, low-index S\'ersic ($n < 1$) profile.  
As in Stage 2, the  disk $b/a$ and $PA$ are held constant at values 
predetermined from ellipse fits.
We provide initial guesses for the bar $b/a$ and $PA$, based on
ellipse fits of the images from MJ07 or analysis of the images in DS9.
We provide GALFIT with input guesses for the bulge parameters, 
based on the output from Stage 2. 
In principle, it is also  possible to generate reasonable guess parameters 
for the bulge and disk  from a bulge-disk decomposition 
on a $1D$ profile taken along a select PA.
As described in    $\S$~\ref{s1dcut},  we also experiment with initial guesses 
derived in this way, and find that the final convergence solution is the same.
We also note that GALFIT fixes the bulge $b/a$ and does not
allow it to vary with radius, while real bulges may have a varying  $b/a$. 
We tested the impact of fixed and varying bulge $b/a$ on the derived $B/T$ 
($\S$~\ref{stest1}) and find that there is no significant change in $B/T$.
\end{enumerate}

For objects with central point sources, the bulge S\'ersic index
in the Stage 2 and Stage 3 models can be inadvertently overestimated unless
an extra nuclear component is added to the model.
Balcells \etal (2003) note that for galaxies imaged both from the ground and
$HST$, the combination of unresolved central sources and seeing effects mimic
high-index bulge S\'ersic profiles in the ground images. 
Depending on sample and resolution, the frequency of central sources 
can range from 50\% to 90\% 
(Ravindranath \etal 2001; B{\"o}ker \etal 2002; Balcells \etal 2003).
Ravindranath \etal (2001) find a frequency of 50\% in early type (E, S0, S0/a) 
galaxies, while B{\"o}ker \etal (2002) measure a frequency of 75\% for
spirals with Hubble types Scd to Sm. Balcells \etal (2003) determine a
frequency of 84\% for S0-Sbc galaxies imaged with $HST$.
Our dataset most closely resembles the latter sample, so we might expect that, 
as an upper limit, a similar fraction of our galaxies will need to be 
corrected with an extra compact component.

We added point sources as third or fourth components to the initial
models.  
For those cases where the model successfully converged with the extra component,
the images were visually inspected to verify the presence of a central bright
source. Sometimes, the model converged to significantly different and unreasonable 
parameters for all components.  Other times, the model would converge to a
very dim point source without changing any of the other model parameters.
Where new model parameters were not unreasonable and not identical to the case
without the point source, the new model was adopted.  This was the case for 111 of 143 (78\%) of 
our sample. The point sources contribute less than 1\%, 3\%, and 5\% 
of the total luminosity 55\%, 86\%, and 95\% of the time, respectively.
As the point sources take up such a small fraction of
the light distribution, their contribution is folded back into $B/T$ in all cases where
a point source was modeled.  Inclusion of the point source reduced bulge index by 
$\sim0.8$ for both barred and unbarred galaxies.  Such a change is expected based on 
the above discussion. The decline in bulge index caused a minor decrease
in $B/T$; on the mean, this change was 1.04\% for barred galaxies and 
0.32\% for unbarred galaxies.
For barred galaxies, this light most primarily added to $D/T$ rather than Bar/$T$.

It is important to recognize the physical significance of the added nuclear
components.
We began by determining which objects show evidence for AGN activity.  
The sample was checked against the catalog 
of Ho \etal (1997), the V{\'e}ron Catalog of Quasars \& AGN, 12th Edition
(V{\'e}ron-Cetty \& V{\'e}ron 2006), and NED.  Of the 111 objects fit with 
point sources, 43 (39\%) contain AGN.  An additional 20 (18\%) possess HII nuclei 
according to Ho \etal (1997) and visibly show bright compact nuclei.  
The remaining 48 (43\%) probably contain neither AGN nor HII nuclei, 
but could house nuclear star clusters.  
For these objects, we visually inspected the images to ensure
there was a bright compact source at the center.
We are confident that the fitted point source components have physical 
counterparts in the data images.

GALFIT also allows a diskiness/boxiness parameter to be added to any S\'ersic or
exponential profile. We did not use this parameter for any bulge or disk 
profiles. Bars in general have boxy isophotes, and we could have included the
diskiness/boxiness parameter in the bar profiles.
However, it was found that adding boxiness to the bar
did not change any model parameters, including fractional luminosities
$B/T$, $D/T$, and Bar/$T$, by more than a small percentage, even though
the appearance of the residual images improved in some cases due to the
change in bar shape. Consequently, we chose to neglect bar boxiness altogether.

\subsection{Choosing the Best Fit Between Stage 2 and Stage 3}\label{schoice23}
All objects in our sample were subjected to Stages 1, 2, and 3.  
Depending on whether a galaxy with a bulge is unbarred or barred, 
its best fit should be taken from the Stage 2 bulge-disk decomposition or the 
Stage 3 bulge-disk-bar decomposition, respectively.
For objects with prominent bars, it is obvious that  the  Stage 3 
model provides the best fit. However, it is more difficult to  decide
between  Stage 2 versus Stage 3 fits in galaxies which host weak bars 
with no strong  visual signature.
In practice, we therefore applied  the set of  criteria  below  to each
galaxy  in order to select between  the Stage 2 bulge-disk decomposition 
and  Stage 3 bulge-disk-bar decomposition.
Table~\ref{tosu}  lists the model chosen for each galaxy. 
Table~\ref{tmodparam} summarizes the model parameters from the best fits.

For completeness, we note that for the few rare galaxies  
(see $\S$~\ref{sdecom2}) that have just a bar and a disk, the choice of 
a final solution is between the  Stage 2 bar-disk decomposition and 
Stage 3 bulge-disk-bar decomposition. The same guidelines below can be 
used to identify the best model.
 
\begin{figure}[]
\centering
\plotone{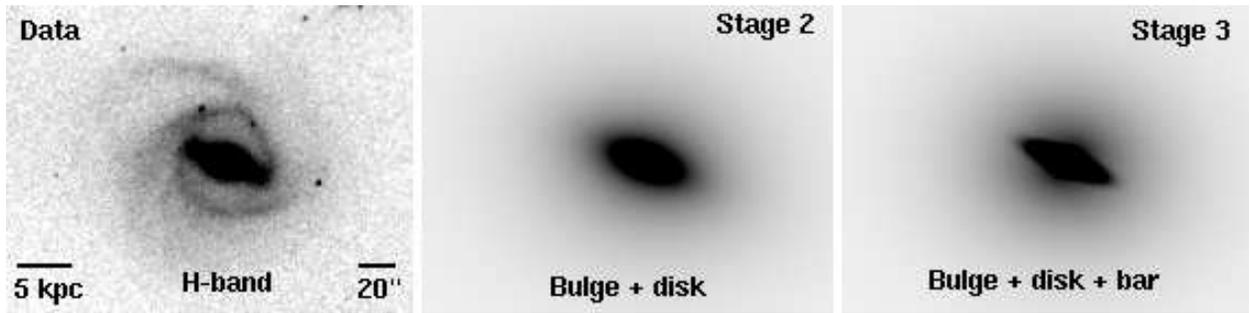}
\caption{This plot shows the data image, Stage 2 model, and Stage 3 model
for NGC 4902.  
The Stage 2 bulge is too bright and is extended along the major axis of the bar
($B/T$=31.2\% and $b/a$=0.45).  In Stage 3, the bulge and bar are fit with
distinct components ($B/T$=5.59\%, bulge $b/a$=0.68, Bar/$T$=9.97\%, bar $b/a$=0.22).
All other fit parameters appear in Table~\ref{tngc4902}.
\label{bulgedistort}}
\end{figure}

\begin{figure}[]
\centering
\plotone{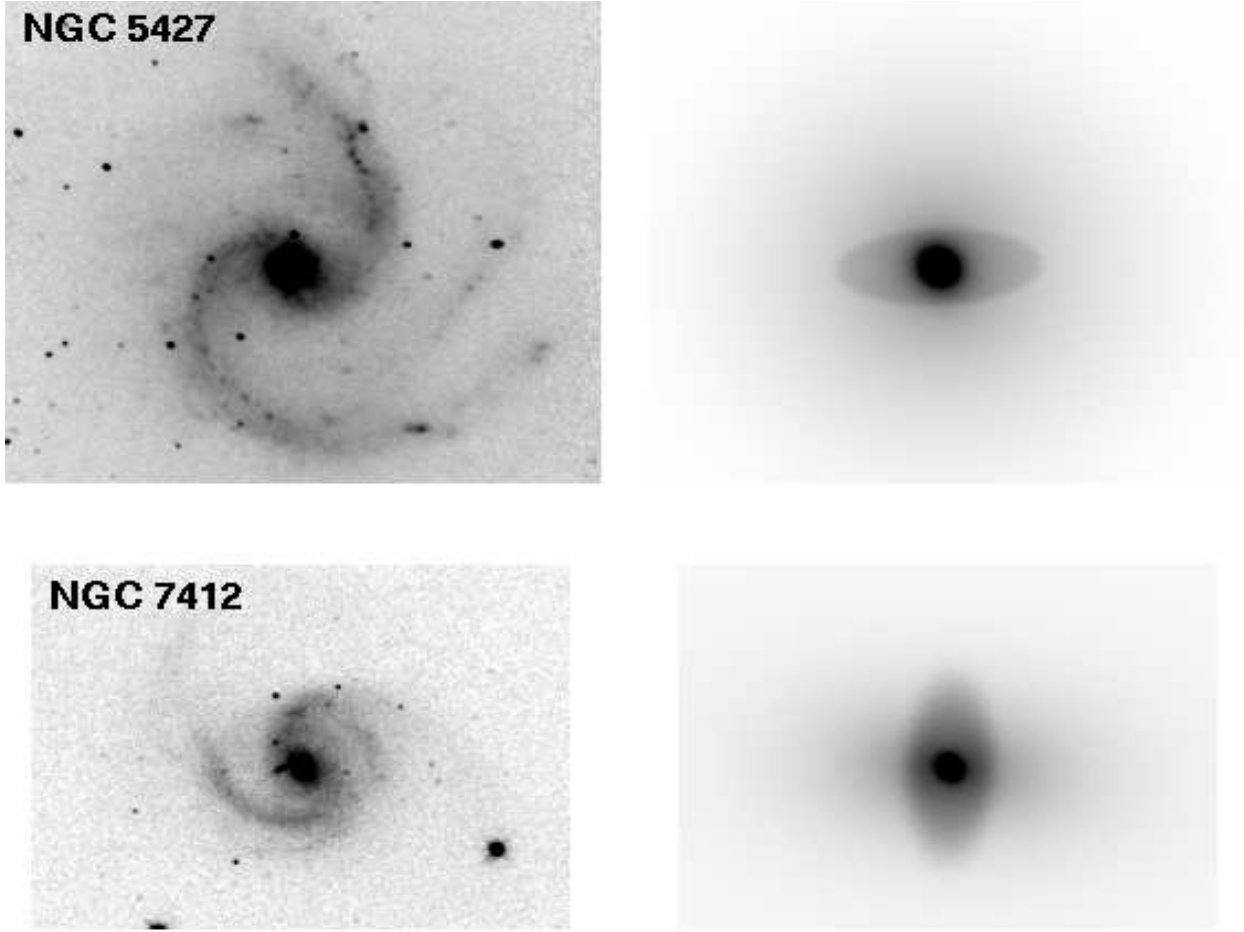}
\caption{The data images and Stage 3 bulge-disk-bar decomposition models of NGC 5427 
and NGC 7412 are shown.
The Stage 3 models each distinctly show a false bar component, which is not present in the
data images.  The false components can be inspired by prominent spiral arms, such as those
present in these galaxies. Such cases are flagged during the visual inspection of fits and
the Stage 3  bulge-disk-bar decomposition  is discarded in favor of the Stage 2 bulge-disk
decomposition.
\label{falsebars}}
\end{figure}

\begin{enumerate}
\item GALFIT calculates a $\chi^2$ and $\chi^2_\nu$ for each model.  It was 
found that $\chi^2$ almost universally declines between the Stage 2 and Stage 3
fits for a given object.  This is because in the Stage 3 fit, five extra free 
parameters (bar luminosity, $r_e$, S\'ersic index, $b/a$, and $PA$)
are added with the S\'ersic bar component, allowing GALFIT to almost always make
a lower $\chi^2$ model during Stage 3.  However, this does not necessarily mean
that the solution in Stage 3 is more correct physically.  Thus, an increasing 
$\chi^2$ was interpreted as a sign that the Stage 3 fit should not be adopted, 
but a decreasing $\chi^2$ was not considered as a sufficient condition to adopt
Stage 3.

\item In cases with prominent bars, a symmetric light distribution due to 
unsubtracted  bar light was often found in the Stage 1 and 
Stage 2 bulge-disk residuals.  This was strong evidence that the Stage 3 
bulge-disk-bar fit be selected. NGC 4643 is shown
in Figure~\ref{ngc4643} because it has a particularly striking bar residual;
the corresponding fit parameters appear in Table~\ref{tngc4643}.
Note that in all figures and tables, we adopt the convention  that PA values are
positive/negative if they are measured from North counterclockwise/clockwise.

\item 
The Stage 2 and Stage 3 models were selected only so long as the model
parameters were all well behaved.  
In unbarred galaxies, the Stage 3 model parameters might be 
unphysically large or small, in which case the Stage 2 fit 
was favored. 
Conversely, in galaxies with prominent bars, the bulge component 
of the Stage 2 bulge-disk fit tends to grow too extended in 
size.  Addition of a bar in the Stage 3 bulge-disk-bar fit removes this artifact,
giving a more physical solution.  
An extreme example of this situation is the barred galaxy NGC 4548, which has
a prominent bar and a faint disk.
The Stage 2 fit, based on a S\'ersic bulge and exponential  disk, is
highly inadequate to describe the bulge, disk, {\it and} the bar because 
it leads to an extremely extended bulge.    
The Stage 3 bulge-disk-bar fit, however, yields a believable fit with a prominent bar.  
The results of Stage 1, Stage 2, and Stage 3 are displayed in Figure~\ref{ngc4548} and 
Table~\ref{tngc4548}.

\item Not all barred galaxies had unphysical Stage 2 models.  
Instead, the bulge could be stretched along the $PA$ of the bar, giving the bulge
a lower S\'ersic index and larger effective radius.  A Stage 3 model that 
returned the bulge to a size and shape more representative of the input 
image was favored over the Stage 2 fit.  Figure~\ref{bulgedistort}  and Table~\ref{tngc4902} 
demonstrate this behavior in NGC 4902. 
We distinguish this effect from cases like NGC 4548 (Figure~\ref{ngc4548} and  
Table~\ref{tngc4548})  where the Stage 2 fit is completely wrong.

\item In cases where there was no bar, GALFIT can sometimes be enticed into 
fitting a bar to any existing spiral arms, rings, or the clumpy disks of 
late-type  spirals.  Stage 3 fits in these cases could
be discarded by noting the resulting discrepancies in appearance 
between the galaxy images  and the Stage 3 model images.  
Examples of false bars are shown in Figure~\ref{falsebars}.
\end{enumerate}

After fitting the whole sample and picking the best fit from 
either the Stage 2 bulge-disk decomposition or  the Stage 3 bulge-disk-bar 
decomposition, we also performed the following extra tests.  
For  our sample S1 of  143  bright ($M_B\leq-19.3$) 
low-to-moderately inclined  ($i \le 70^\circ$) spirals  in the
OSUBSGS survey, we determine the  fraction  (75 of 143  or $\sim52\%$) of spiral galaxies 
where a bulge-disk-bar  decomposition was picked as the best fit for the $H$-band 
image. There are also eight galaxies with  pure bar-disk fits.
The $H$-band  bar fraction, which is defined as the  fraction of disk 
galaxies that are barred, is therefore $58.0 \pm 4.13\%$ (83 of 143).
We then compared our results ($58.0\% \pm 4.13\%$) with the  $H$-band bar 
fraction  (60\%) determined from  ellipse fits of the OSUBSGS sample 
by MJ07, with  a slightly more conservative inclination cut 
($i\leq 60^\circ$). The two numbers are in excellent agreement.
As a further check to our fits, we compare the bar and unbarred
classification for individual galaxies  from our fits with those from
MJ07, which were based on ellipse fits. Of the 73 galaxies that we classify as barred, 
and that are mutually fitted by MJ07, 54 (74\%) are also classified as barred 
by MJ07.  The remaining 19 (26\%) galaxies are mainly weakly barred (with
Bar/$T$ below 0.08).  Their RC3 optical types are  weakly barred 
AB (10), barred B (7), and unbarred A (2). 

In most previous bulge-disk and bulge-disk-bar $2D$ decomposition,
the issue of parameter coupling  and the systematic exploration of local versus global 
minima in  $\chi^2$ have been ignored.  Quantifying how the
parameters are coupled is important in measuring error bars for the model
parameters.  With $2D$ models containing several free parameters, this is not an easy
task.  Although we also do not address this problem in rigorous detail, we describe 
in \S~\ref{coupling} simple test that explores parameter coupling in our models.

\section{Extra Tests to Verify Correctness of Fits} \label{stests}

\subsection{Varying $b/a$ as a Function of Radius}\label{stest1}
Models generated with GALFIT do not allow the $b/a$ of the bulge, disk, 
or bar to vary with radius.  
Since real bulges may have a varying  $b/a$, it is legitimate to investigate 
the impact of fixing the bulge  $b/a$  on the estimated $B/T$.
We therefore performed the following test on  NGC 4548.
To mimic a model bulge of varying $b/a$, we fitted  the bulge light of 
NGC 4548 with ten concentric  S\'ersic profiles of increasing   $r_e$  and 
varying  $b/a$.   The $r_e$ of the outermost profile comes from the 
original bulge model (see Table~\ref{tngc4548}) where $b/a$ was kept 
constant with radius.  
The separation in $r_e$ between adjacent profiles is 0.5 pixels (0.75'').
The luminosity, S\'ersic index, $b/a$, and $PA$ of each profile were free 
parameters. 
The disk and bar components were fixed to the values in Table~\ref{tngc4548},
as the emphasis was on the change in the bulge.

\begin{figure}[]
\centering
\epsscale{0.8}
\plotone{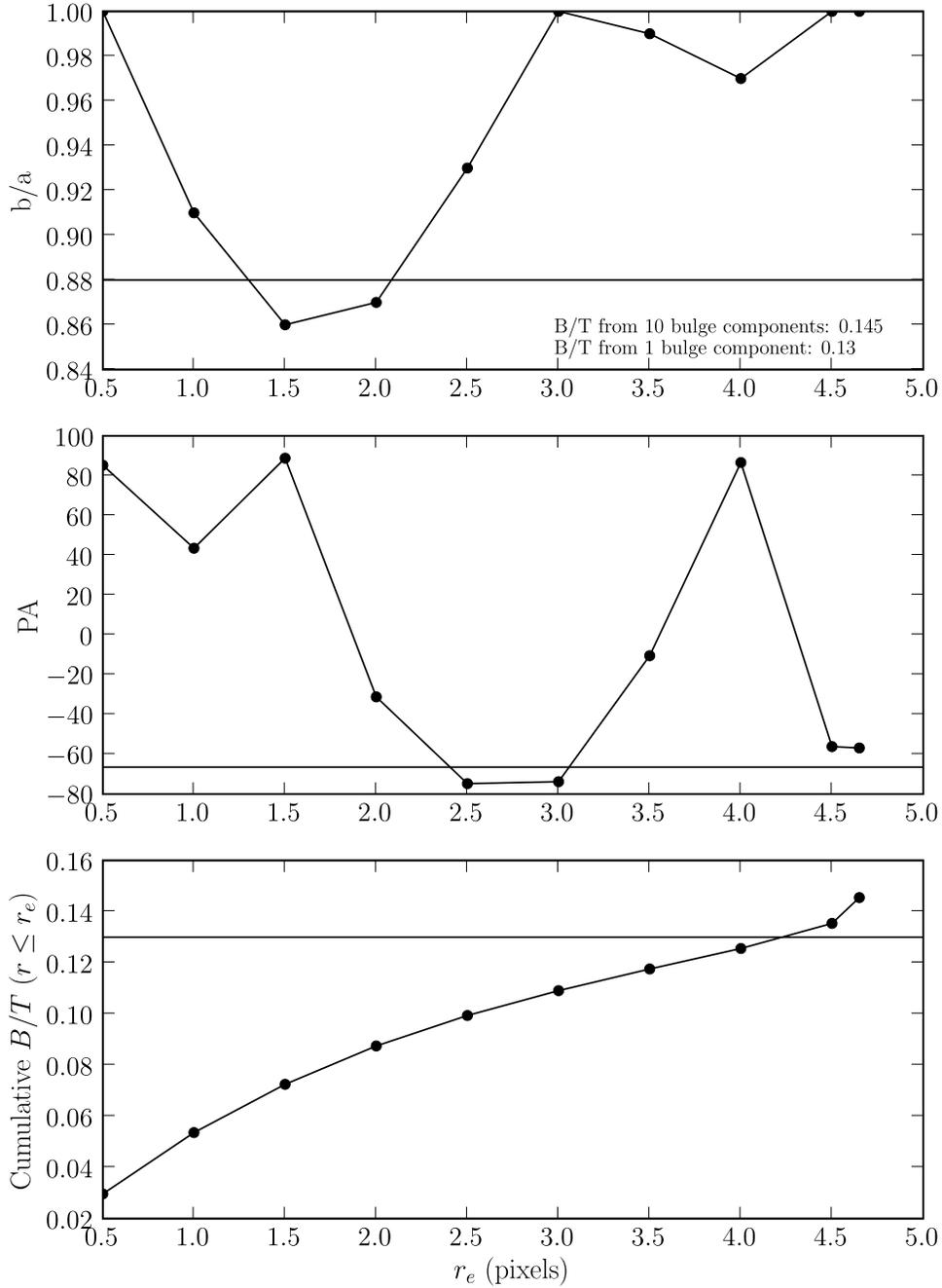}
\caption{
This plot compares the $B/T$ obtained by fitting the bulge of NGC~4548 with a 
S\'ersic model of constant $b/a$  as opposed to a S\'ersic model varying 
$b/a$.  To mimic a S\'ersic model with varying $b/a$ in GALFIT,  the bulge 
was fitted with  ten concentric S\'ersic profiles with fixed $r_e$, each 
separated by 0.75''.  The top two panels show the run of $b/a$ and $PA$ of 
the ten concentric S\'ersic profiles. The bottom panel shows the cumulative 
$B/T$ calculated by summing all models with $r \le r_e$. 
The bulge $b/a$ (0.88), $PA$ (-66.5), and $B/T$ (13\%) from the original 
S\'ersic fit of constant $b/a$ (Table~\ref{tngc4548}) are indicated with 
horizontal lines on the 3 panels.
\label{qtest}}
\end{figure}

Figure~\ref{qtest}   compares the $B/T$ obtained by fitting the bulge of NGC~4548 
with a S\'ersic model of constant $b/a$  as opposed to a S\'ersic model varying $b/a$. 
The bulge $b/a$ (0.88), $PA$ (-66.5), and $B/T$ (13\%) from the original S\'ersic fit 
of constant $b/a$ (Table~\ref{tngc4548}) are indicated with horizontal lines 
on the 3 panels.
The top two panels show the run of $b/a$ (0.85 to 1.0) and $PA$ 
($-90^\circ$ to $+90^\circ$)  of the ten concentric S\'ersic profiles. 
It can be seen that the S\'ersic indices of the ten bulge models  were 
generally higher toward the center and declined at larger $r_e$, indicating 
that the `fitted bulge is more concentrated  at the center. 
The bottom panel shows the cumulative $B/T$ calculated by summing all models with 
$r \le r_e$. The last point representing the summed $B/T$ from all ten components 
is 14.5\%, in good agreement with the 13.0\% value from the  S\'ersic fit  of 
constant $b/a$. 
Thus,  using a S\'ersic model of constant $b/a$, does not have any significant 
adverse impact on our derived $B/T$ in NGC 4548.

\subsection{Fitting Artificially Simulated Images}\label{sartif}
An elementary test is to determine if GALFIT can recover the known parameters
of artificially simulated noisy images.
The images were simulated by taking parametric model images produced by GALFIT,
and adding noise to the images with the PyFITS module for Python
(Barrett \& Bridgman  1999).
Noise in each pixel was calculated by adding in quadrature the noise due to
the source, sky, and read noise. The standard deviation of pixel noise in
electrons was computed as
\begin{equation} \label{nse}
\sigma=\sqrt{ T_{source}+T_{sky}+T_{read}^2},
\end{equation}
where $T_{source}$ is the number of electrons due to the source,
$T_{sky}$ is the number of electrons due to the sky,
and $T_{read}$ is the detector read noise.
The contribution due to detector dark current was very small and
therefore neglected.
The offset added to each pixel was drawn
from a normal distribution centered at zero with standard deviation $\sigma$.

\begin{figure}[]
\centering
\plotone{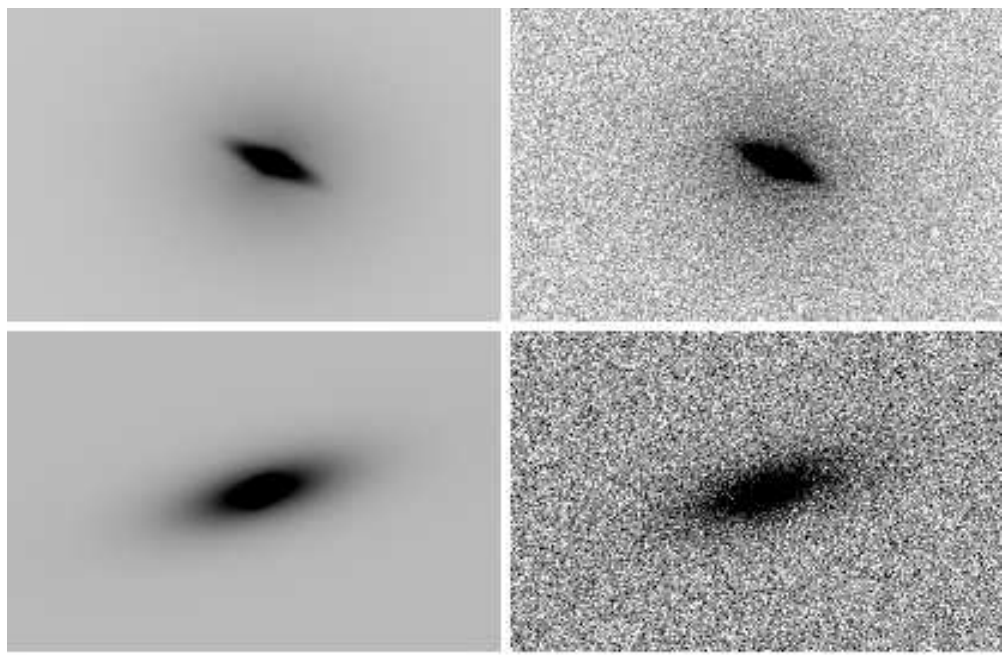}
\caption{
An elementary test is to determine if GALFIT can recover the known parameters 
of artificial noisy images. 
Noisy images were simulated by taking parametric model images 
(left panels) produced by GALFIT, and adding noise and sky background
(right panels).  The noisy images were then fitted to see if the original known
parameters can be recovered. See   \S~\ref{sartif} for details. }
\label{nsemodel}
\end{figure}

Our test sample consisted of 40 models (20 bulge-disk and 
20 bulge-disk-bar) with noise added as described.  
Thirty of the images included point sources as extra components.
The range explored for each parameter in the model images is a fair representation 
of the parameter space covered by our full sample 
(e.g., $B/T$ ranges from $0.02-0.70$, the bulge index ranges from $\sim0 - 5$, and the full range of possible bulge 
and bar PA was also tested ($-90^\circ$ to $+90^\circ$)). 
In terms of surface brightness, the models span five magnitudes in mean surface 
brightness inside the disk scalelength.
Examples of the noise-added models are shown in Figure~\ref{nsemodel}.

\begin{figure}[]
\centering
\plotone{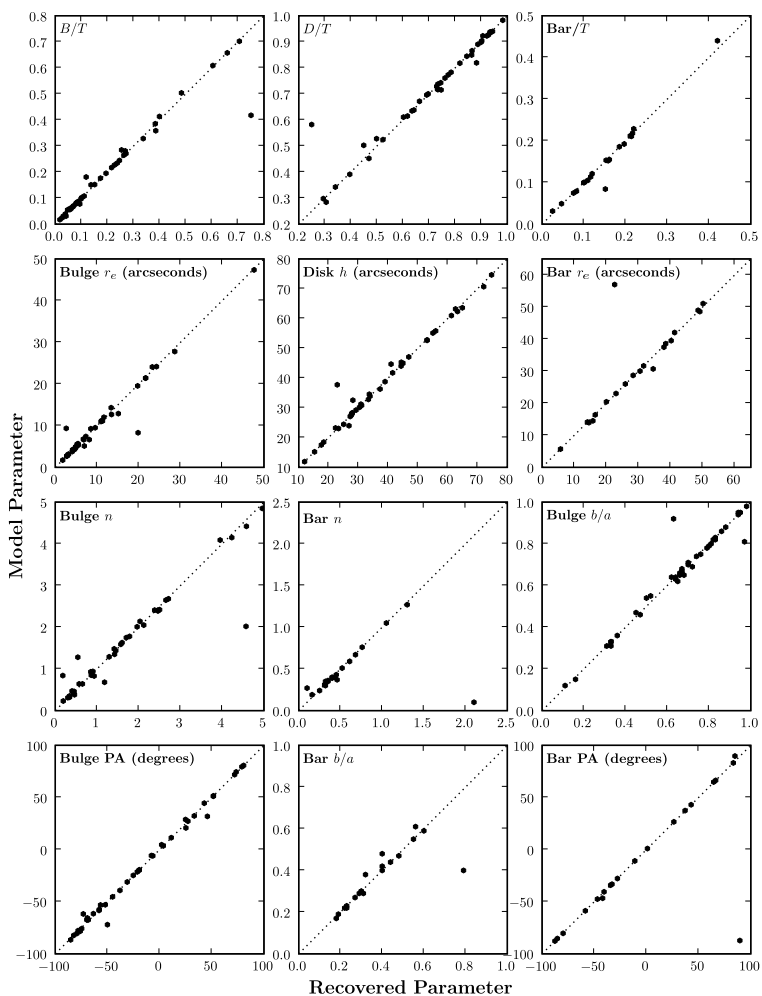}
\caption{
The plots compare recovered versus original model parameters for the simulated
images discussd in \S~\ref{sartif}. The vertical axis limits demonstrate the
range explored for each parameter. The dotted line shows $y=x$ for comparison.
Except for some extreme cases where the images were highly distorted by noise,
all parameters were recovered to within a few percent.
}
\label{simsum}
\end{figure}

\begin{figure}[]
\centering
\plotone{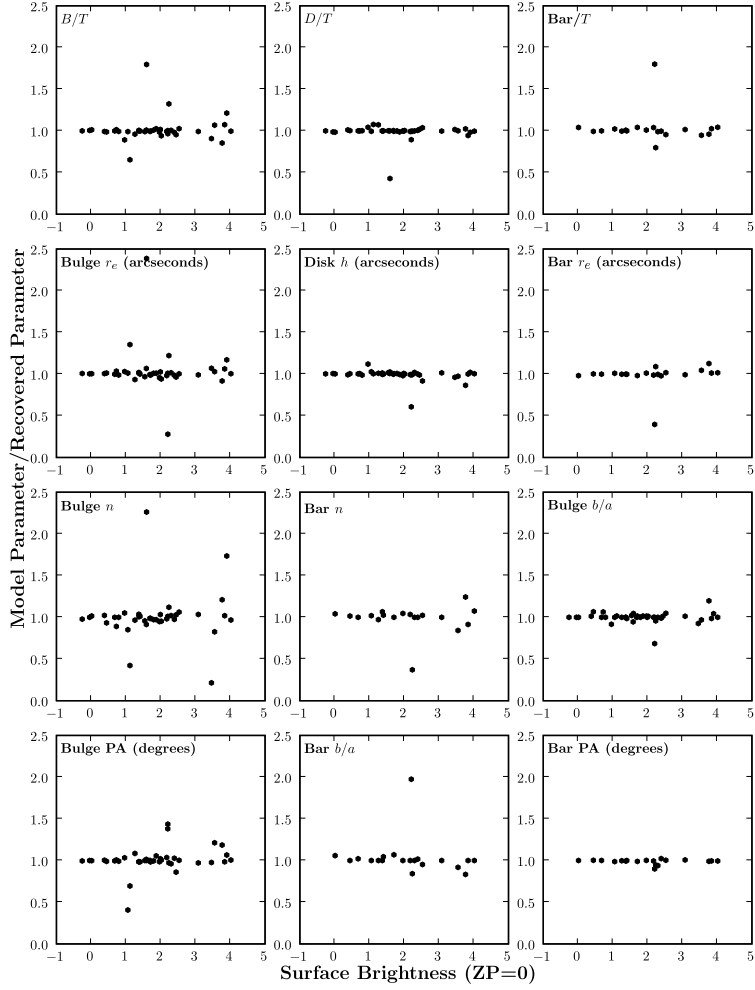}
\caption{
The ratio of model-to-recovered parameter is plotted against mean surface brightness 
inside the disk scalelength, $\mu=mag+2.5\times log_{10}(2\times \pi \times b/a \times h^2)$, 
for the simulated images discussed in \S~\ref{sartif}.  Surface brightness is not photometrically
calibrated and is shown for a zeropoint of 0.
}
\label{simsum_sb}
\end{figure}

The noisy images were subjected to the $2D$ decomposition procedure outlined in 
Figure~\ref{flowchart}.  GALFIT reproduced the model 
(bulge, disk, bar, and point source) parameters quite closely for the
majority of the test cases.
Figure \ref{simsum} compares the recovered versus original model parameters.
Except for some extreme cases where the images were highly distorted by noise,
all parameters were recovered to within a few percent.
Figure \ref{simsum_sb} plots the ratio of model-to-recovered parameter against
mean surface brightness inside the disk scalelength; there is no strong trend
in error with dimming surface brightness.
This suggests our decompositions are effective across
the parameter space spanned by our sample.
The overall success of this test is evidence that GALFIT is able to
converge to the absolute minimum in $\chi^2$ space for our bulge-disk and
bulge-disk-bar decompositions when the input is the sum of parametric functions.

\subsection{Using  $1D$ Decomposition To Generate Guesses for Bulge Parameters}\label{s1dcut}
It is important to verify that GALFIT converges to the same solution even
if the initial guesses for the bulge parameters in Stage 2 and 3  are different.
Bulge-disk decomposition from $1D$ profiles provides an alternative means of 
generating initial guesses.
While $1D$ bulge-disk decompositions of radial profiles along the bar major axis
can be influenced by the bar, decomposition of cuts along the bar minor axis 
will not be influenced as heavily.  The resulting bulge and disk parameters should be 
adequate guesses for Stage 3 of our $2D$ decomposition method. 

We tested the robustness of our Stage 3 fits by extracting initial guesses for
the bulge and disk using $1D$ decomposition along the bar minor axis. 
The nonlinear least-squares algorithm designed to perform the $1D$ 
decomposition simultaneously fits the sky-subtracted profiles 
with the sum of a S\'ersic bulge and an exponential disk, while ignoring
the PSF.
The results from the $1D$ decomposition include a bulge magnitude, $r_e$, 
S\'ersic index, disk magnitude, and disk scalelength.  

The robustness of several bulge-disk-bar fits were tested by using the results 
of the $1D$ decomposition as input to Stage 3.  The $1D$ 
decompositions do not provide information about the axis ratio 
($b/a$) or $PA$, so these
parameters for the bulge were estimated by eye; for the disk, the $b/a$ and
$PA$ were fixed to the values determined by ellipse fitting, as described in
$\S$~\ref{sdecom2}.  The initial bar parameters were unchanged from the earlier 
Stage 3 fits.  In all cases, the new models were identical to the Stage 3 
models.  As an example, Table~\ref{t1dfit} compares Stage 3 input derived 
from 1D decomposition and GALFIT for NGC 4548 and NGC 4643.  In each case, 
both sets of input reproduced the same results. 

\subsection{Parameter Coupling}\label{coupling}
Assessing the coupling between model parameters is complicated when
models have a large number of free parameters.
A standard approach 
is to calculate confidence regions using multi-dimensional ellipsoids 
for a given $\Delta\chi^2$ contour.  As the errors in the GALFIT 
models are not normally distributed, but are instead dominated by
the systematics of galaxy structure, this approach does not 
yield meaningful results because of the ambiguity in assessing which
$\Delta\chi^2$ contour levels are statistically significant.

We carry out a simple test for representative galaxies to determine
not only the effects of parameter coupling, but also the
the effect parameter coupling has on model parameter errors,
paying particularly close attention to $B/T$, $D/T$, and Bar/$T$, as
they are of primary interest.
We perform this test on four representative galaxies 
(NGC 3885, NGC 4151, NGC 4643, and NGC 7213).  
Two are barred (NGC 4151 and NGC 4643).  
Two have high ($n>2$) bulge indices (NGC 4643 and NGC 7213), 
and the other two have low ($n<2$) bulge indices.

We fit two and three-component bulge-disk and bulge-disk-bar models using
fixed bulge indices of $n=1$ and $n=4$.  The initial inputs to these
fits were the same as those used to generate the model in which bulge
$n$ is a free parameter.  We then had GALFIT re-fit the
models with bulge index as a free parameter using these output model 
parameters as input initial guesses.  We compare the output of these 
two fits with the the best fit as selected in \S~\ref{schoice23}, in which
bulge $n$ is a free parameter. Ideally, the re-fits should 
converge to the same final parameters as the selected best fit.  

Table~\ref{tcoupling} displays the outcome of this test for the four
representative systems.  For barred galaxies NGC 4643 and NGC 4151, the
$n=1$ and $n=4$ re-fits converged to the same $\chi^2$ as when bulge 
$n$ is initially left a free parameter.  
$B/T$, $D/T$, and Bar/$T$ are precisely equal for 
NGC 4643, while for NGC 4151 there is a small dispersion of 0.1-0.2\%.  
For unbarred galaxies NGC 3885 and NGC 7213, the $n=1$ re-fits again 
converged to the same $\chi^2$ and model parameters as when bulge $n$ is  
initially left a free parameter.  This is not true for the $n=4$ re-fits.  
When the $n=4$ condition is enforced, the bulges in these cases become too
extended and luminous while the $\chi^2$ drop below those when 
bulge $n$ is initially left a free parameter.  
During the $n=4$ re-fits, $B/T$ increases further at the
expense of $D/T$ and the $\chi^2$ remain unchanged or decrease further.
The $B/D$ ratio from the $n=4$ re-fit for NGC 3885 is 1.3, roughly 3.5 times
higher than when $n$ is free.  Given that NGC 3885 is an S0/a galaxy with a bulge 
embedded in a smooth extended disk, the latter $B/D$ is arguably too large.
For NGC 7213, the $n=4$ re-fit yields starkly unphysical values. The $B/D$ 
ratio is 11.3, and $r_e/h$, the ratio between bulge effective radius and 
disk scalelength is 9.0.
The lower $\chi^2$ in these cases cannot be taken seriously as the
bulges are too luminous and the resulting $B/D$ ratios do not match 
the data images.

As illustrated by the above discussion and Table~\ref{tcoupling}, this test shows that 
in some cases (e.g., NGC 4643 and NGC 4151) GALFIT converges to the similar model 
parameters and $B/T$, $D/T$, and Bar/$T$ while starting from highly different initial 
input guesses (e.g. bulge $n=1$, bulge $n=4$, and bulge $n$ based on the Stage 2 fits) 
in different regions of parameter space. For NGC 3885 and NGC 7213, however, during 
the $n=4$ re-fits GALFIT converged to models that were unphysical and 
different compared with the reasonable models generated with input guesses
corresponding to the bulge fixed at $n=1$ or the bulge $n$ based on output from Stage 1.
In effect, when the initial input guesses were very different from the data images,
the resulting models were found, in spite of the lower $\chi^2$, to be unphysical
through comparison with the input data images.
We emphasize that for all sources analyzed in the paper, the data, converged model
output, and residuals are always inspected before adopting the best final fit (see
\S~\ref{schoice23}).

Table \ref{tcoupling} also provides hints as to how the model parameters
are coupled to bulge index.
As suggested in the above discussion, 
fixing the bulge index to $n=4$ leads to a more extended
and luminous bulge, causing bulge $r_e$ and $B/T$ to rise without fail
for increasing bulge index.  
The disk is coupled with the bulge such that 
increasing bulge index, bulge $r_e$, and $B/T$ yields a reduction in $D/T$.  
At the same time, disk scalelength either increases
(NGC 3885 and NGC 4643) or decreases (NGC 3885 and NGC 4151);
in the latter two cases, the disk becomes very compact and the bulge
quite extended.
The behavior of the bar is coupled with \emph{both} the
bulge and disk.  In the case of NGC 4643, as bulge index is raised to $n=4$,
bar $r_e$ becomes slightly larger, but Bar/$T$ falls by a factor of 2.6
as light is redistributed from the bar and disk to the bulge.
Bar index also declines as the bar assumes a flatter profile.  
For NGC 4151, bar $r_e$ again increases slightly,
but this time Bar/$T$ rises by a factor of 1.7
as light is transferred from the disk to the bulge and bar. 

Based on the above test, we stress that GALFIT was able to overcome
this parameter coupling in the cases where the input guess parameters 
well-reflected the data images.

\section{Results and  Discussion}\label{sresult}

\subsection{Impact of Bars in $2D$ Decomposition }\label{sbar1}
From the Stage 2 bulge-disk decomposition  and Stage 3 bulge-disk-bar 
decompositions, which we performed on all objects ($\S$~\ref{sdecom2})
we saw firsthand the effects of adding a bar to the fit of a barred galaxy.
We summarize below some of these effects in order to underscore 
the importance of including a bar component in the $2D$ luminosity 
decomposition of barred galaxies 

\begin{enumerate}
 \item 
During the Stage 2 bulge-disk decomposition of a barred galaxy, the 
luminosity which comes from the galaxy's disk, bulge, {\it and bar} 
gets distributed only between two model components: the model bulge and disk. 
Since the disk $b/a$ and $PA$ are measured independently and 
held constant during the fits, the Stage 2 model tends to distort 
the bulge in order to fit the bar. Thus, the bulge in the Stage 2 
bulge-disk decomposition  of a barred galaxy can be artificially
long or too bright and extended.
When a model bar component is added in the Stage 3  bulge-disk-bar 
decomposition of a barred galaxy, it forces a reshuffling of the
luminosity between the three components. Generally, the bulge declines 
in luminosity, whereas light can be either taken from, or added back, 
to the disk.
 \item 
We find that the inclusion of a bar component in the  Stage 3 
bulge-disk-bar decomposition  of a barred galaxy   reduces the 
bulge fractional luminosity  $B/T$, compared with the Stage 2 
bulge-disk decomposition. For our 75 barred galaxies, the 
reductions correspond to factors of less than two, 
2 to 4, and above 4, in   36\%,  25\%, and 39\% of barred galaxies, 
respectively.
The larger changes in $B/T$  occur in very strongly barred galaxies, 
where a prominent bar cause the Stage 2 bulge-disk decomposition to 
overestimate the  bulge. For instance, $B/T$ declines in both of
NGC 4643 (Figure~\ref{ngc4643} and Table~\ref{tngc4643}) 
and NGC 4548 (Figure~\ref{ngc4548} and Table~\ref{tngc4548}).  
In the latter case, $B/T$ is reduced by a factor of $\sim5$ between Stage 2
and Stage 3. These examples underscore the importance of including 
bars in $2D$ luminosity decomposition of very strongly barred galaxies.
 \item 
The scalelength of the disk is generally unchanged by including
the bar. NGC 4548 (Figure~\ref{ngc4548} and Table~\ref{tngc4548}) 
is a good example.  Sometimes, however, the disk from the 
Stage 2 bulge-disk decomposition of a barred galaxy is erroneous due to a
poor fit.  The disk parameters from the Stage 3 bulge-disk-bar decomposition
are quite different in such cases. NGC 4643 (Figure~\ref{ngc4643} and 
Table~\ref{tngc4643}) illustrates this behavior. 
\end{enumerate}

We find that for  our sample S1 of 143  bright ($M_B\leq-19.3$) low-to-moderately
inclined  ($i \le 70^\circ$) spirals  (Figure~\ref{2posu-hist-hubb})  in the OSUBSGS survey,  
75 of 143  or $\sim52\%$ are better fit with a Stage 3 bulge-disk-bar 
decomposition than a Stage 2 bulge-disk decomposition.  
There are also eight galaxies with pure bar-disk fits.  As stated in  
$\S$~\ref{schoice23}, the resulting $H$-band bar fraction 
($58.0\% \pm 4.13\%$ or 83 of 143)  is in excellent agreement with the 
$H$-band bar fraction of 60\% reported by MJ07  based on ellipse fits of the 
OSUBSGS sample, with  a slightly more conservative inclination cut 
($i\leq 60^\circ$).

\subsection{Mass in Bulges, Disks, and Bars}\label{smass}
The  fractional $H$-band luminosities in the bulge, disk, and bar  
($B/T$, $D/T$, Bar/$T$) of each galaxy can be considered as a 
fractional mass if we assume that the same mass-to-light ($M/L$) 
ratio can be used to convert the $H$-band luminosities  of both 
the numerator ($B$, $D$, or Bar) and the denominator ($T$) terms 
into a stellar mass.
This is not an unreasonable assumption as the $H$-band $M/L$ ratio 
is not very sensitive to differences in dust or age that might  
exist between  the bulge, disk, and bar.  The uncertainties in  $M/L$  
can  be estimated by looking at population synthesis models.
Charlot, Worthey, \& Bressan (1996) find that for idealized galaxies with 
a single generation of stars, the  uncertainties in $M/L$ ratio  due 
to different input stellar models and spectra are roughly $\pm35\%$ for a 
fixed metallicity and IMF.  Furthermore, as the age of a stellar population 
varies from $\sim0.5$ Gyr to 10 Gyr, the $K$-band $M/L$ ratio rises by a 
factor of $\sim$ 2 to 3  (Charlot 1996).  Asymptotic giant branch (AGB) 
stars dominate 
the  NIR light for ages between 0.1 and 1 Gyr, while  red giant branch 
(RGB) and super-giant branch (SGB) stars dominate between 1 Gyr and 
10 Gyr.

In this paper, we convert the $B/T$ light ratio determined from 
$H$-band images to a $B/T$ mass ratio by assuming a constant 
mass-to-light ($M/L$) in the $H$-band for both the bulge and the
rest of the galaxy. However, Moorthy \& Holtzman (2006) present 
line strengths of bulges and inner
disks for 38 spirals with Hubble types S0 to Sc.  They show 76\% of
spirals have negative metallicity gradients. $B-K$ color gradients 
are shown to largely match metallicity gradients and are likewise 
negative outward, indicating bulge $M/L$ is higher than in the disk 
and bar. If bulges are much older than the disks, then our
prescription  would  underestimate  the true $B/T$ mass ratio. If we
assume an extreme case where bulges are $\sim12$ Gyr and the disk light
is dominated  by a young 3 Gyr population, our assumption of a constant $H$-band
$M/L$ ratio  would underestimate the true $B/T$ by a factor of $\le2$ (Charlot 1996).
In several sections of the paper (e.g., $\S$~\ref{sbulgenbt}, $\S$~\ref{smodel1}),
we illustrate how our main results  would change if the true  $B/T$ was higher
by up to a factor of two.
On the other hand, central regions of galaxies may harbor intense 
episodes of star formation. If the bulge is younger than the disk and 
happens to have star formation and a significant young population of 
massive stars, then our prescription could overestimate the 
true $B/T$ mass ratio. This would make our current results on the 
high fraction of low $B/T$ bulges even stronger.  

Using the total galaxy stellar mass from $\S$~\ref{stellmass}, 
the fractional masses can be converted into absolute masses.
(We do not convert the $H$-band luminosity directly into a mass 
as the $H$-band images do not have photometric calibration).
The results are shown in Table~\ref{masstable}.
For  our sample S1 of  143  bright ($M_B\leq-19.3$) low-to-moderately 
inclined  ($i \le 70^\circ$) spirals with a mass-weighted mean Hubble
type of Sab-Sb, we find that 
{\it $\sim70\%$ of the stellar mass is in disks,  $\sim$~10\%  is in stellar bars and 
$\sim20\%$ is in bulges (with $\sim11\%$ in $n>2$ bulges and $\sim9\%$ in 
$n \leq 2$ bulges).} 
Thus while bulges with $n\le2$  are highly ubiquitous (see 
next section), they only account for a small fraction of the total stellar mass
budget.
 
\begin{figure}[]
\centering
\epsscale{0.90}
\plotone{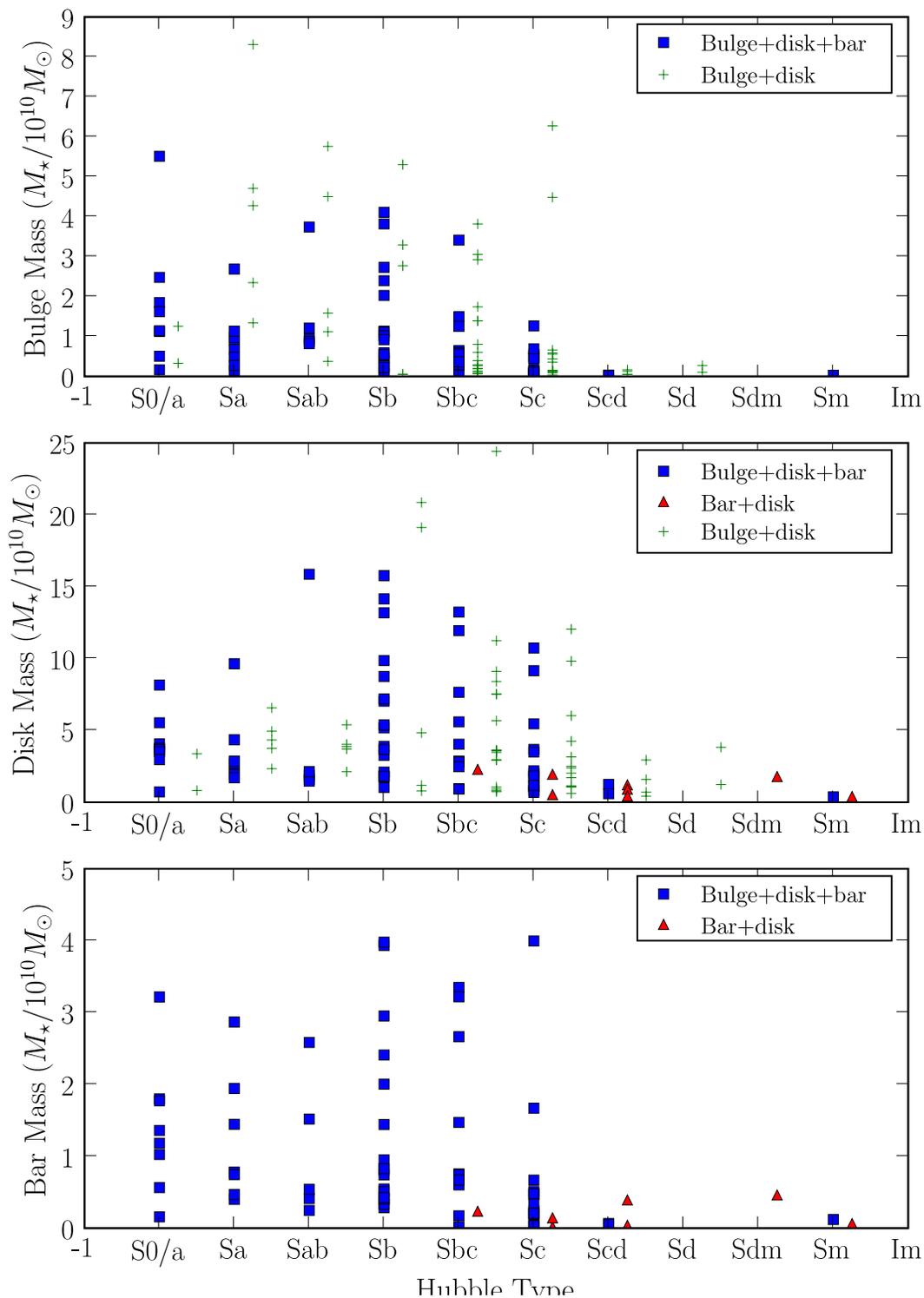}
\caption{ The top, middle, and bottom panels show stellar mass for bulges,
disks, and bars, respectively, along the Hubble sequence.   Values are
shown for sample S2 of 126 galaxies in Fig.~\ref{stellmasshist}.  
The legend in each panel indicates the type of decomposition used for
each data point.
\label{3p-plot2}}
\end{figure}

Figure~\ref{3p-plot2} shows the stellar mass for bulges, disks, and bars  along 
the Hubble sequence.  

It is useful to compare our results with those of Driver \etal (2006), who   
performed bulge-disk decomposition of $B$-band images with GIM2D on 10,095 
galaxies from the Millennium Galaxy Catalog  (Liske \etal 2003; Driver \etal 
2005).  They found  68.6\% of the stellar mass to be in disks, and 32.6\% in bulges 
(with 30.8\% in high $n$ bulges, and 1.8\% in low $n\le2$ bulges). 
 Their study thus finds a  higher stellar mass fraction  in all bulges (32.6\% vs  our 
18.9\%), and in high $n$ bulges (30.8\% vs our 10.4\%),  and a lower fraction in  
low $n\le2$ bulges  (1.8\% vs our  8.4\%),  and disks+bars (68.6\% vs our  71.6\% + 9.6\%).
This difference can be attributed to the fact that  the  Driver \etal (2006) study
did not perform  bulge-disk-bar fits and thus, their $B/T$ ratios may be skewed to 
higher values. 

\subsection{Distribution of Bulge Index and $B/T$}\label{sbulgenbt}
Figure~\ref{4p-bplot1}  shows the individual and mean  $B/T$ 
and bulge S\'ersic index, plotted, as a function  of Hubble type and 
galaxy stellar mass. Barred and unbarred galaxies are shown separately.
Figure~\ref{2p-bplot2} shows the relationship between bulge index and 
$B/T$.

\begin{figure}[]
\centering
\plotone{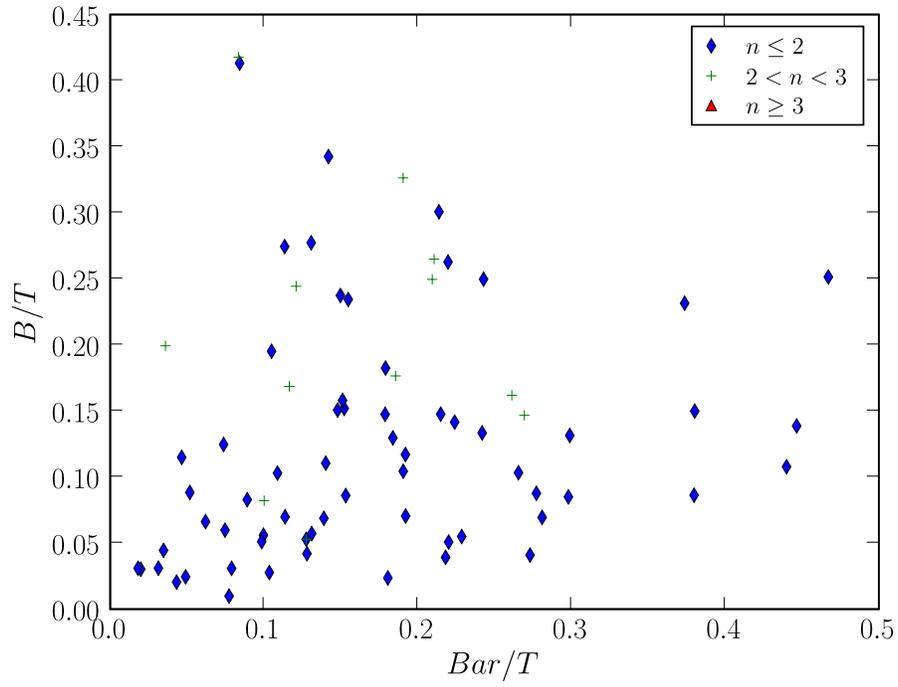}
\caption{$B/T$ is plotted against Bar/$T$ and sorted by bulge S\'ersic index.
There are six galaxies with Bar/$T$ $\geq0.3$.
\label{2p-brplot5}}
\end{figure}

We first consider the $B/T$ values in  Figure~\ref{4p-bplot1}.
The mean $B/T$  in barred galaxies is lower than in unbarred 
galaxies, but there is a large overlap in the individual 
values. The offset in the mean $B/T$ of  barred and unbarred galaxies 
reported here,  agrees with the result of Laurikainen \etal (2007; 
see $\S$~\ref{scomp1}) on the same sample. 
We also note that  $B/T$ does not correlate with Bar/$T$ 
(Fig.~\ref{2p-brplot5}):  aside from the six galaxies with large Bar/$T$ ($>0.3$), 
most galaxies have low-to-moderate Bar/$T$ and a wide range of $B/T$ is seen 
at each Bar/$T$. This is reassuring and suggests that the bar fit is not 
arbitrarily biasing the $B/T$ values.  The distribution of Bar/$T$ is further
discussed in $\S$~\ref{sbar2}.

How does  the $B/T$ vary as a function of Hubble type and galaxy stellar mass?
Bulges with very high $B/T$ ($>0.4$)  exist primarily in galaxies 
with high mass ($M_\star > 6 \times 10^{10} M_\odot$) and early types (S0/a to Sab).
Bulges with very low  $B/T$ ($<0.1$) lie primarily in lower mass galaxies 
with later morphologies (Sb to Sc).
{\it 
It is striking that  $\sim69\%$   of  bright ($M_B\leq-19.3$) 
low-to-moderately inclined  ($i \le 70^\circ$)
spirals have $B/T \leq 0.2$: these bulges are pervasive   
and exist across  the whole spectrum of S0/a to Sd.
}
The results are summarized in Table~\ref{bulge1}.
We shall return to this point in \S~\ref{smodel1}.
We note again that these $B/T$ mass ratios were calculated assuming a 
constant $M/L$ ratio in the $H$-band for the bulge and disk components.  
As noted in $\S$~\ref{smass}, if the bulge in these high mass spirals 
is much younger (older) than the disk and bar, then the $B/T$ can be 
overestimated  (underestimated) by up to a factor of two, 
and the limiting value 
of  0.2 for the $B/T$ cited in the above fraction, would have to be modified 
in the extreme case to  0.1 and 0.4, respectively. 

\begin{figure}[] 
\centering
\plotone{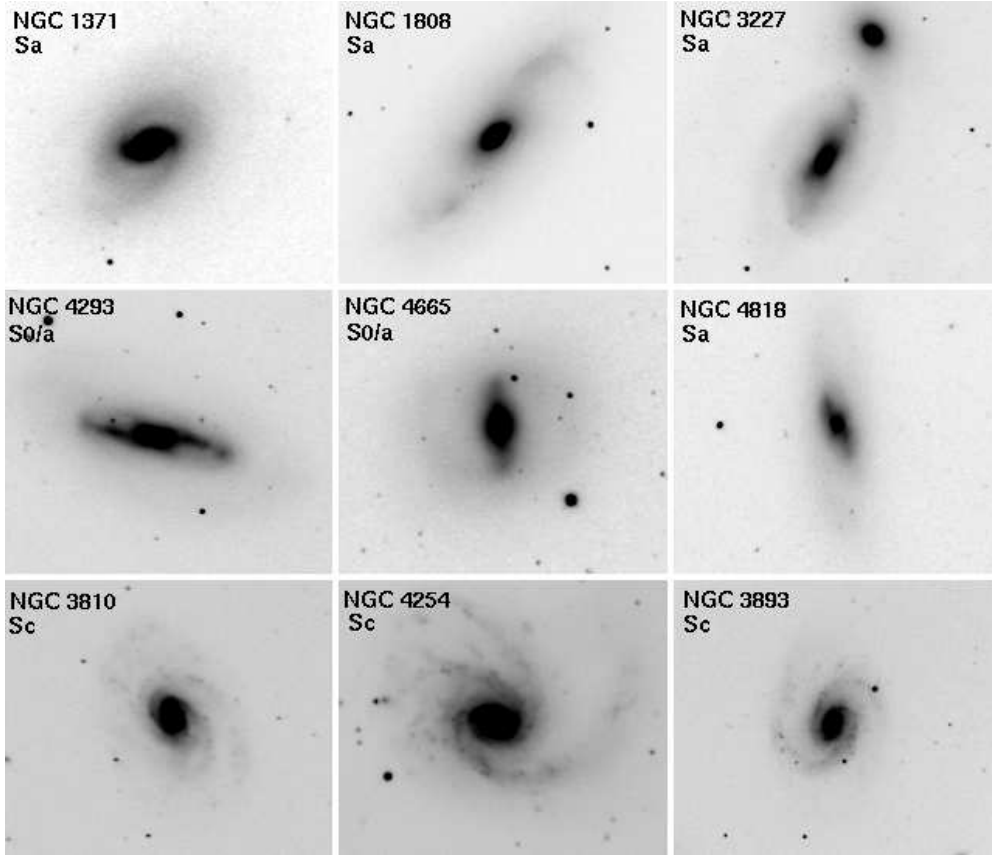}
\caption{The {\bf top two rows} show $H$-band images of barred galaxies, which 
have early RC3 Hubble types, but yet have  $B/T\leq0.2$.
The {\bf bottom row} shows $H$-band images of unbarred galaxies, which 
have late RC3 Hubble types, but yet have $B/T \sim 0.4$.  
The Hubble types assigned to these objects more reflect disk smoothness
and spiral arm topology than $B/T$.
All images are from OSUBSGS with characteristics as described  in \S~\ref{sdata}.
\label{lowBT}}
\end{figure}

Some of the low $B/T \le 0.2$  values for six barred S0/a and Sa galaxies
in Figure~\ref{4p-bplot1} may at first look suspicious. 
Balcells \etal (2007) report the mean $B/T$ for S0 galaxies to be 0.25,
so much smaller $B/T$ are potentially worrisome.
OSUBSGS $H$-band images of these objects in Figure~\ref{lowBT} show 
a smooth extended disk around the bulge.
It should be noted that Hubble types were originally assigned on a 
combination of criteria including disk smoothness and spiral arm topology 
in addition to the prominence of the bulge.
It is likely that these galaxies were assigned early Hubble
types due to their smooth extended disks, in spite of their low bulge-to-disk
ratio.  

Similarly, some of the high  $B/T \sim 0.4$ bulges in three of the 
Sc galaxies may at first seem odd. However, again, visual inspection  of their image
(Figure~\ref{lowBT}) reveals prominent spiral arms and clumpiness,
which may explain why they were assigned late Hubble types.

How does  the bulge S\'ersic index $n$ vary as a function of 
Hubble type, and galaxy stellar mass (Figure~\ref{4p-bplot1}), as well
as $B/T$ (Figure~\ref{2p-bplot2})?
The results are summarized in Table~\ref{bulge1}.
Only a small fraction ($\sim1\%$) of bright spirals have high $n\ge4$
bulges; such bulges lie primarily in S0/a to Sab, and have a large $B/T > 0.2$. 
A  moderate fraction ($\sim22\%$)   have intermediate 
$2<n<4$ bulges; these exist in barred and unbarred S0/a to Sd,  and 
their $B/T$ spans a wide range  (0.03 to 0.5) with a mean of 0.29.
{\it 
A strikingly large fraction ($\sim76\%$) of bright  spirals have 
low $n \leq 2$ bulges; such bulges exist  in barred and unbarred galaxies 
across a wide range of Hubble  types, and  their $B/T$  varies from 0.01 to 0.4, 
with most having $B/T \leq0.2$. }

\subsection{Comparison With Other Work}\label{scomp1}
As an independent check of our decomposition method, we compare 
our results with independently published decompositions.

\begin{figure}[]
\centering
\plotone{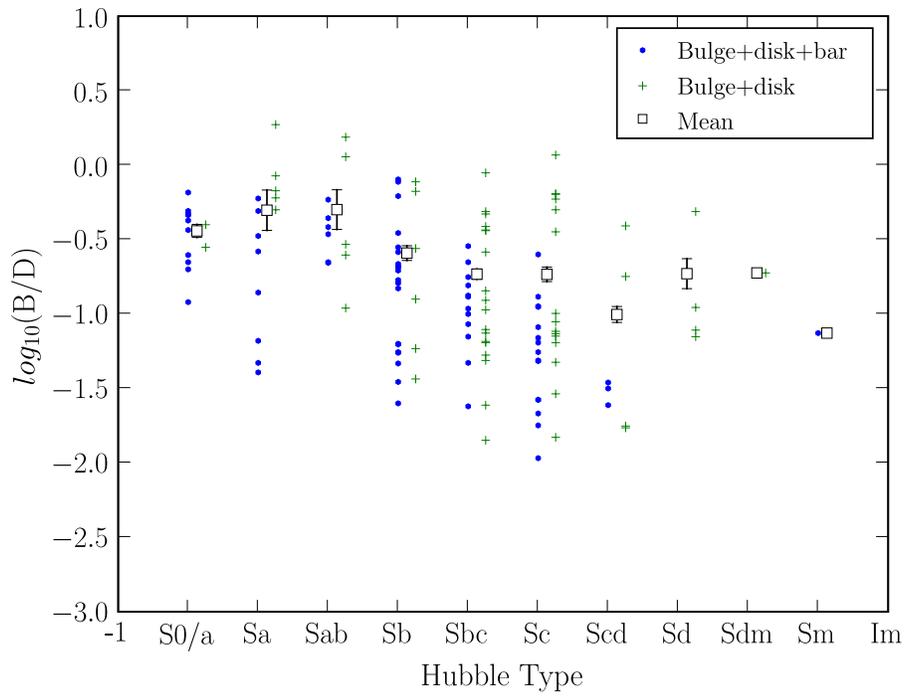}
\caption{ 
$B/D$  is plotted  against Hubble type. 
The legend indicates the type of decomposition used for each data point.
The mean values for barred and unbarred together in each bin are shown. 
}
\label{log10BD}
\end{figure}

We find our mean $H$-band $B/D$ (Figure~\ref{log10BD}) ratios are comparable
to the $K$-band $B/D$ derived with the $1D$ bulge-disk decompositions
of Graham (2001) and Trujillo \etal (2002).
Like Graham (2001) and Trujillo \etal (2002), we find $B/D$ is 
widely variable with Hubble type and that mean $B/D$ steadily declines from Sa
through Scd galaxies.  Graham (2001) and Trujillo \etal (2002) find bulge 
indices are widely variable within a Hubble type, but they are in 
general $>1$ for early types and $<1$ for late types.  
We likewise find wide scatter in bulge index with
$n<1$ bulges existing in both early and late types.
Figure~\ref{log10BD} is also in good agreement with the more recent
results of Graham \& Worley (2008), who have calculated, as a function
of morphology, inclination and dust-corrected $B/D$ and bulge
S\'ersic indices. They find $B/D$ values are typically $<1/3$.

Another meaningful comparison can be made with Laurikainen \etal (2007)  who,
using their own $2D$ decomposition code, fit a hybrid sample containing some
OSUBSGS galaxies combined with additional S0 galaxies.  One difference between their 
work and ours is that
they typically model bars with a Ferrers function, but may sometimes use a
S\'ersic profile, while we use only the latter.  
Also, they include additional components to model secondary bars or inner disks. 
They report a distinct offset in the mean $B/T$ between barred and unbarred 
galaxies,  which we confirm in Figure~\ref{4p-bplot1}.  Their
mean $B/T$ are similar to ours, and they conclude that pseudobulges 
exist throughout the Hubble sequence.  
The S\'ersic indices derived by Laurikainen 
\etal (2007) are likewise similar, on the mean, to ours for both barred
and unbarred systems.  They likewise find $n\le2$
bulges across early and late Hubble type galaxies.

Balcells \etal (2003) emphasized that bulges typically have
indices $\sim3$ or lower. Our results in Figure~\ref{4p-bplot1} are 
consistent.  We find a low frequency ($\sim1\%$) of high $n\geq4$ bulges,
with most bulges having $n\leq3$.

\subsection{Bar Strength}\label{sbar2}
Stellar bars exert gravitational torques on the gaseous component and are 
particularly efficient in driving  gas from the outer disk into the inner kiloparsec 
(see $\S$~\ref{sintro}).  Thus, it would be natural to have  a measure of bar strength 
that is  sensitive to the strength of the gravitational torque and  hence measures 
both the shape and mass of the bar.

Many measures of bar strength have been formulated. The $Q_b$ method of (Block \etal 2002; 
Buta \etal 2003; Buta \etal 2005) measures directly the gravitational
torque at a single point along the bar.  This method requires a value of 
scaleheight
for the disk and a model of the potential to be made from the image.  
In the bar/inter-bar contrast method of Elmegreen \& Elmegreen (1985) and
Elmegreen \etal (1996), bar strength is parameterized as the ratio between
peak surface brightness in the bar region and the minimum surface brightness
in the inter-bar region.  Elmegreen \& Elmegreen (1985) and
Elmegreen \etal (1996) also characterize bar strength with the maximum
amplitude of the $m=2$ mode from Fourier decomposition.
When ellipse fitting is applied,
the maximum ellipticity of the bar, $e_{bar}$, can be used to characterize
bar strength (e.g. MJ07).  This constitutes a partial measure of bar 
strength only, however, as it offers no information about mass of 
the bar.

Bulge-disk-bar decomposition  in the $H$-band  provides another 
\emph{partial} measure of bar strength through the $H$-band Bar/$T$ 
light ratio, which is a measure of the Bar/$T$ mass ratio under the 
assumption that the $H$-band $M/L$ ratio is the same for the bar and the rest
of the galaxy, as discussed in $\S$~\ref{smass}. 
Figures~\ref{fbar1} and \ref{fbar2} explore the derived bar properties.

\begin{figure}[]
\centering
\epsscale{1}
\plotone{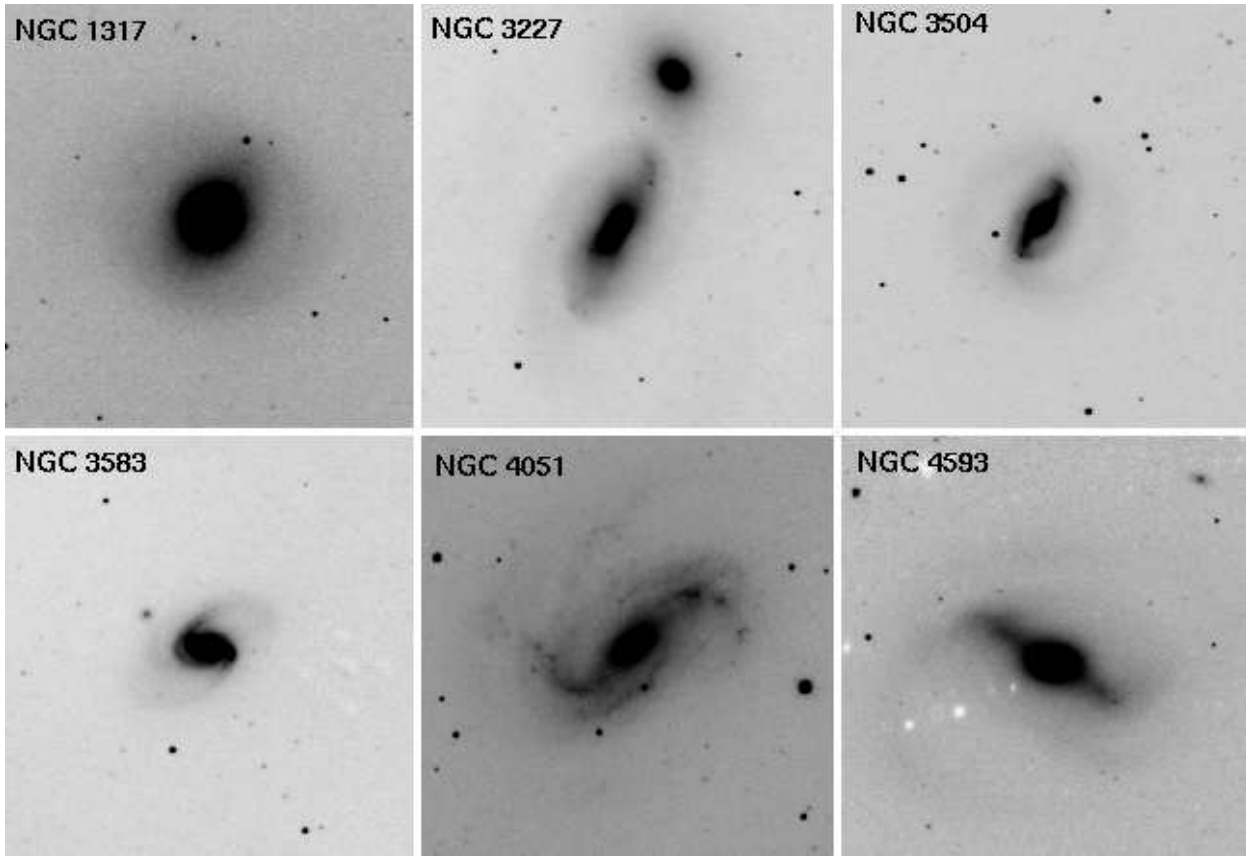}
\caption{$H$-band OSUBSGS images of spirals with prominent bars of
Bar/$T$ ($>0.3$) are shown. 
An interesting example is the oval or lens  galaxy NGC 1317:  
the  bar has a low ellipticity, but its $B/T$ is large because it is 
extended and massive. Such bars/lenses may exert significant gravitational 
torques although they are not very elongated. 
\label{highBrT}}
\end{figure}

The upper left panel of Figure~\ref{fbar1}
plots the individual and mean Bar/$T$ against Hubble type.
There is a wide range ($\sim0.03$ to  $\sim0.47$) in the  individual  
Bar/$T$ at a given Hubble type.  The mean Bar/$T$ remains fairly 
constant with Hubble type from Sa to Sb, but shows a possible weak decline by about 
0.1 from Sb to Sc.    Their number statistics are too small to make any robust statement
for later Hubble types. We also note that six systems have high  Bar/$T$ above $0.3$; 
these are displayed  in Figure~\ref{highBrT}.  

Bar S\'ersic indices are mostly below unity.  
Neither the individual, nor the mean bar S\'ersic index, show any trend 
with Hubble type or with stellar mass, for Sa to Sc galaxies (Fig.~\ref{fbar1}).
Thus, the steepness of the bar profile does not seem to depend on the Hubble type.
Is the bar mass ratio and its mass profile related?  There is a wide range in the  
individual   Bar/$T$  at a given bar S\'ersic index (Fig.~\ref{fbar2}). 
The mean Bar/$T$  rises with bar index  out to a bar index of $\sim0.6$, and then 
flattens out. This suggests that on the mean, bars of lower Bar/$T$ have flatter 
profiles.

Is there a relation between the bar strength and the bulge present in a galaxy?
There is a wide range in the  individual   Bar/$T$  at a given $B/T$, 
and at a given bulge S\'ersic index (Fig.~\ref{fbar2}). 
The  mean  Bar/$T$  shows a weak decline for bulge S\'ersic indices above two.
Similarly  the mean  Bar/$T$  shows a weak rise from 0.1 to 0.25 as  $B/T$ rises 
out to 0.15, after which the trend flattens or reverses.

Both Bar/$T$ and maximum bar ellipticity $e_{bar}$ are partial measures of 
bar strength.  Figure~\ref{fbar1} shows mean Bar/$T$ may scale weakly with
Hubble type.  The bars with highest $e_{bar}$  (i.e, thin bars) are often 
termed strong bars, and  $e_{bar}$ has been shown to correlate with $Q_b$.
Total bar strength should scale with both bar mass and bar 
ellipticity. Does bar strength have a dependence on Hubble type? 
The upper left panel of Figure~\ref{fbar2} plots the product of 
Bar/$T$ and $e_{bar}$, as determined by MJ07 for galaxies mutually 
classified as barred, against Hubble type.
There is a wide range in Bar/$T\times e_{bar}$ in each bin, and mean
bar strength shows no definite trend with Hubble type.
We note that bars with high Bar/$T$ and high $e_{bar}$ 
should exert the largest  gravitational torque and be most effective at driving 
gas inflows.
A nice example is the oval or lens  galaxy NGC 1317 (Figure~\ref{highBrT}); 
the bar has a low ellipticity, but its $B/T$ is large as it is extended and 
massive. Such bars or lenses may exert significant gravitational torques 
although they are not very elongated. 

\subsection{Bar Fraction as a Function of $B/T$ and Bulge Index} \label{sbar3}
As outlined in $\S$~\ref{sbulgenbt}, we found that as many as  $\sim76\%$ 
of bright spirals have bulges with $n \leq 2$; such bulges exist  
in barred and unbarred galaxies 
across a wide range of Hubble types, and  their spread in $B/T$ is from 0.01 to 0.4, with most 
having $B/T \leq$ 0.2.
The variation of the bar fraction as a function of $B/T$ and bulge $n$ can 
provide important constraints  on bulge formation scenarios ($\S$~\ref{smodel1}).
Table~\ref{barFraction} shows our results.
The bar fraction declines with bulge index; $\sim65\%$ of the spirals 
with low $n\le2$ bulges host bars while intermediate $2<n<4$ bulges 
have a lower bar fraction ($\sim38\%$). The high $n\ge4$ bulges 
in the sample are unbarred, so the bar fraction is $0\%$.
Systems with low $B/T$ are more likely to be barred.  
For $B/T \le 0.2$, the bar fraction is high ($\sim68\%$).  Systems with 
$0.2<B/T<0.4$ and $B/T \ge 0.4$ have lower bar fraction 
($\sim42\%$ and $\sim17\%$). 

Overall, Table~\ref{barFraction} shows bulges with low $n\leq2$ and 
low $B/T \le 0.2$ preferentially exist in barred galaxies.  
This is consistent with earlier work 
(Odewahn 1996; Barazza \etal 2008; Marinova \etal 2008; Aguerri \etal 2008, in prep.)
where an enhanced optical bar fraction is seen is galaxies with late Hubble 
types or low $B/D$.
It may be tempting to infer this result to mean the formation pathway 
of two-thirds of low-$B/T$ bulges is related to bars in that 
spontaneous or/and tidally induced bars play a role in bulge formation 
(with the remaining one-third of such bulges may have been formed either by 
mechanisms like retrograde minor mergers or short-lived bars).
We caution that this type of cause-effect relationship is not the
only scenario consistent with this result.  It may also be possible that
bar instabilities are favored in galaxies with low $B/T$ and no inner Lindblad
resonances (ILR).  Under these conditions, the swing amplifier model with
a feedback loop
(Julian \& Toomre 1966; Toomre 1981; Binney \& Tremaine 1987) 
may be responsible for bar formation and partly account for the high 
bar fraction in  galaxies of low $B/T$.

\subsection{Formation of Bulges}\label{bulgeFormation}
Our observational results provide some interesting challenges for models of 
galaxy evolution that try to  address the origin of present-day bulges.
Any successful model must be able to account for the observed distribution 
of bulge $B/T$ and $n$ in  high mass 
($M_\star \geq  1.0\times 10^{10} M_\odot$) spirals,
as shown in  Table~\ref{bulge1} and Table~\ref{barFraction}.
In particular, the following  results must be reproduced:
 
\begin{enumerate}
\item 
In terms of the overall distribution of bulge $n$, as much as  
($\sim74\%$) of high mass spirals have bulge $n\leq2$:  
such bulges exist  in barred and unbarred galaxies and  their $B/T$  
ranges from 0.01 to 0.4,  with most having $B/T \leq0.2$ (Table~\ref{bulge1}).
A moderate  fraction   ($\sim24\%$) of high mass spirals 
have $2<n<4$, and just ($\sim2\%$) have $n\geq4$. 

\item  
Theoretical models often make more robust
predictions on the bulge-to-total mass ratio $B/T$  than on the bulge 
index $n$, so we consider the  empirical $B/T$ distribution in detail.
We note  that as much as $\sim66\%$ of high mass spirals have bulges 
with $B/T \leq 0.2$ (Table~\ref{bulge1}). 
In terms of bar fraction, $\sim68\%$ are barred  (Table~\ref{barFraction}).

\item 
The fraction of bars rises among spirals with low bulge index $n$. 
About 63\% of spirals with low $n\leq2$ bulges host bars, 
while the bar fraction in spirals with $2<n<4$ bulges 
(44\%) is two-thirds as large (Table~\ref{barFraction}).
\end{enumerate}

In a hierarchical Universe, there are several physical processes that 
contribute to the assembly of bulges: major mergers, minor mergers, and 
secular evolution. We briefly describe these, expanding on our introduction 
in $\S$~\ref{sintro}.

\begin{figure}[]
\centering
\plotone{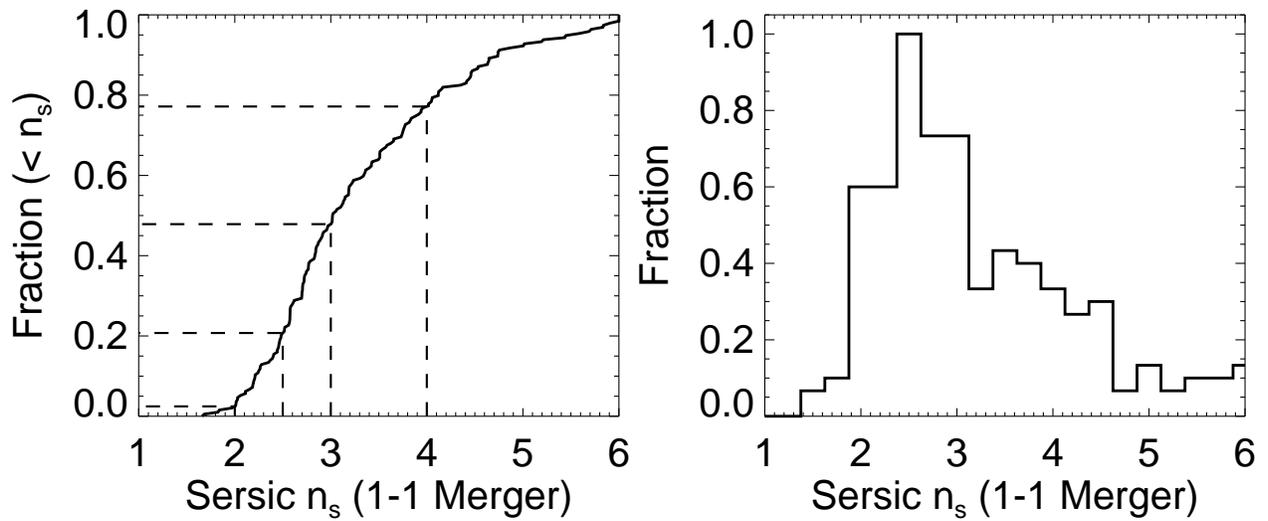}
\caption{
Shown is the distribution of S\'ersic indices $n$ for remnants of
1:1 gas-rich major mergers in the simulations of 
Hopkins \etal 2008: they lie in the range of $2<n<6$. Specifically,  
$\sim$~22\% of the remnants have classical $n\geq4$, as much as  20\%  
have low  $n\leq 2.5$, while 50\% have $n \leq 3$.
Almost none have $n \leq 2$. [Figure: courtesy of Phil Hopkins]
}
\label{fhopki1}
\end{figure}

Major mergers,  defined as those with mass ratio   $M_1/M_2 \ge 1/4$,  
typically destroy the extended outer stellar disks during violent relaxation, 
leaving behind  a classical bulge.  Such bulges are associated with 
modest-to-high bulge  S\'ersic indices, in the range $2<n<6$ 
(Hopkins \etal 2008; Springel \etal 2005; Robertson \etal 2006; 
$\S$~\ref{smodel1}) in simulations. This trend is also consistent with the 
fact that among ellipticals, high luminosity ones tend to have  a  
S\'ersic index $n>4$, while  low luminosity ones tend to have  
$2 \le n \le 3.5$  (Caon \etal 1993; Kormendy et al. 2008).  
The final  S\'ersic index depends on the amount of residual gas the
settles into a somewhat disky component. Simulations by Hopkins \etal 
(2008) find that the S\'ersic indices  of remnants from 1:1 
gas-rich major mergers lie in the range of $2<n<4$, with most above 
2.5 (see Fig.~\ref{fhopki1}). This body of evidence strongly suggests that 
many bulges with $n>2$ might have a major merger origin.

Minor mergers, typically defined as those with mass ratio 
$1/10 < M_1/M_2 < 1/4$, do not destroy the stellar disk of the 
primary system,  but can contribute to building bulges via three pathways.
Firstly,  a fraction  $F_{sat}$  of the satellite's stellar mass can end up 
in the central region of the primary galaxy. The value of  $F_{sat}$ 
depends on how centrally concentrated the in-falling satellite is. Typically, 
the more diffuse outer stellar body is tidally stripped, while the  central 
core sinks by dynamical friction to the central region (e.g., Quinn et al. 1993; Walker et al. 
1996). 
Secondly,  a non-axisymmetric feature 
(e.g., a stellar bar or bar-like feature) can be induced in the main disk, 
and gravitational torques exerted by the feature can drive gas into the inner 
kpc (e.g., Hernquist \& Mihos 1995; Jogee 2006 and references therein), where subsequent SF 
forms a compact  high $v/\sigma$  stellar component, or disky pseudobulge.
Most of the gas inflow happens during the merger phase and large gas inflow
rates (e.g., $\gg1$  $M_\odot$ per year) may  be generated. 
Thirdly,  gas inflow can also be caused by direct tidal torques from  the 
companion (e.g., Hernquist \& Mihos 1995; Eliche-Moral \etal 2006). 
It is to be noted that in the simulations by  Hernquist \& Mihos (1995),  the 
gas inflow  generated by non-axisymmetric features 
(e.g., bar-like features) in the inner part of the disk is much larger than 
that caused by direct tidal torques from the satellite.
In the recent work of Eliche-Moral \etal (2006), 
N-body simulations of minor mergers followed by fits of 1D S\'ersic+exponential 
models to the remnants, suggest that
the bulge S\'ersic index and $B/D$ ratio can grow as a result of
the central re-concentration of stellar disk material in the primary
system by tidal forces.
Minor mergers are frequent under $\Lambda$CDM, and the likelihood of multiple 
successive minor mergers occurring during the formation of a galaxy is high.
Bournaud, Jog, \& Combes (2007) study the effects of repeated minor mergers on
galaxy structure.  They show that a disk galaxy undergoing successive minor mergers
will eventually transform into an elliptical galaxy with an $r^{1/4}$-law profile
and high $V/\sigma$.  However, galaxy growth is not completely merger-driven and
the efficiency of minor mergers at creating ellipticals must be regulated
by other mechanisms (e.g., cold gas accretion).

In addition, the process of secular evolution can build  a disky bulge (pseudobulge)
between merger events.  Here  a stellar bar or globally oval structure 
in a  {\it non-interacting} galaxy drives  gas inflow into the inner kpc, 
where subsequent star formation forms a compact  high $v/\sigma$  stellar component 
(e.g., Kormendy 1993; Jogee 1999; KK04;  Jogee, 
Scoville, \& Kenney 2005;  Athanassoula 2005; Kormendy \& Fisher 2005; Kormendy 2008).  
This process is different from that of minor mergers in the sense that it 
happens in the quiescent phase of the galaxy, between minor or major 
merger events. 
The prevalence of pseudobulges in galaxies of different Hubble types is 
discussed in KK04, and select examples of S0 galaxies with pseudobulges are also 
shown in Kormendy \& Cornell (2004) and KK04.

The present-day bulge mass can be written as the sum of mass 
contributed from each process: 
\begin{equation} \label{bulgemass}
M_{bulge}=M_{bulge}\times(f_{\rm maj} +  f_{\rm min1} + f_{\rm min2}  
+ f_{\rm min3} + f_{\rm sec}),
\end{equation}

where 

\begin{itemize}
\item 
$f_{\rm maj}$ is the percentage of the  bulge stellar mass, which is built by  major mergers,  
\item 
$f_{\rm min1}$ is the percentage of the bulge stellar mass, which is built during minor mergers 
from stars of the satellite. This depends on the  fraction  $F_{sat}$  of the satellite's 
stellar mass, which ends up in the central region of the primary galaxy during each minor merger.
\item  
$f_{\rm min2}$ is the percentage of the bulge stellar mass, which is built during minor mergers  
from gas inflow  caused by a tidally induced bar.
\item  
$f_{\rm min3}$ is the percentage of the stellar mass, which is built during minor mergers 
from gas inflow caused by tidal torques from the companion.
\item  
$f_{\rm sec}$ is the percentage of the stellar mass, which is  built secularly  
between  merger events from gas inflow  caused by  bars or ovals
\end{itemize}

In \S~\ref{smodel1}, we compare our derived distribution of bulge $n$ and $B/T$ 
with hierarchical models that model major and minor mergers, but not secular
evolution.  The main goal of the model is to see 
whether bulges built via major mergers can account for the large fraction of high
mass spirals  with bulges of low $B/T$  or/and low $n$.
A secondary goal is to see if a first order simplified prescription for minor 
mergers can broadly account for the observations. We stress here that bulge-building 
during minor mergers is  modeled in a very simple way: all the stars in the satellite
are assumed to contribute to the bulge of the larger galaxy (i.e., $F_{sat}=100\%$), 
and bulge-building via gas inflow driven through tidal torques and via 
gravitational torques from induced bars are ignored (i.e., $f_{\rm min2}$~=~0, 
and $f_{\rm min3}$~=~0).  Furthermore, the models entirely ignore secular evolution 
between mergers.  In a future paper, these extra terms will be addressed and 
a comprehensive picture built of the relative importance of  minor mergers and 
secular processes in making present-day bulges.

\subsection{Comparison of $B/T$ With Hierarchical Models of Galaxy Evolution}\label{smodel1}
We compare our data with the  predictions from  cosmological semi-analytical models 
based on Khochfar \& Burkert (2005) and Khochfar \& Silk (2006).  We briefly 
describe the models first.
The merger trees of dark matter (DM)  halos are derived by using
the  extended Press-Schechter formalism (Press \& Schechter 1974), as in 
Somerville \& Kolatt (1999).  
When two DM halos merge, the merger time scale of the galaxies is calculated 
by considering the timescale it would take the satellite galaxies to reach the 
central galaxy at the center of the halo via dynamical friction (e.g., 
Kauffmann \etal 1999;  Springel \etal 2001).
The baryonic physics, which includes radiative cooling, star formation, and
feedback from supernovae, is treated via semi-analytic prescriptions (see 
Khochfar \& Silk (2006) and references therein).
Baryonic mass inside the dark matter halos is divided between hot gas, cold
gas, and stars.  The hot gas is initially shock-heated to the halo virial
temperature.  As the gas radiatively cools, it settles down into a 
rotationally supported disk at the halo center.  Cold disk gas is allowed
to fragment and subsequently form stars according to the Schmidt-Kennicutt 
law (Kennicutt 1998). Star formation is regulated by feedback from 
supernovae using the prescription in Kauffmann \etal (1999).

Major mergers are typically considered as those with stellar mass ratio 
$M_1/M_2 \ge 1/4$. In the simulations, one assumes that during a major
merger any  existing stellar disk is destroyed,  gas is converted to stars 
with some star formation efficiency (SFE), and all stars present undergo 
violent relaxation to form a bulge.  Therefore, the  bulge-to-total stellar 
mass ratio  ($B/T$)  of a bulge immediately after a major merger is always one. 
Note  that the SFE during a major merger is not assumed to be 100\%  as there is 
mounting  evidence from  SPH simulations (Springel \& Hernquist 2005; Cox \etal 2008) 
that not all cold gas is converted to stars. Instead, the burst efficiency 
defined by Cox \etal (2008) is applied to control the fraction of stars 
formed due to the interaction.
This efficiency is dependent on the relative masses of merging galaxies
and is expressed as
\begin{equation} \label{cox}
e=e_{1:1} {\left(\frac {M_{Satellite}} {M_{Primary}}\right)}^\gamma,
\end{equation}
where $e_{1:1}$ is the burst efficiency for a 1:1 merger and $\gamma$ fixes the
dependence on mass ratio; Cox \etal (2008) find $e_{1:1}=0.55$ and $\gamma=0.69$.
The remaining fraction (1-$e$) of gas is added to the gaseous disk and can start 
making stars.

As stated above, immediately after a major merger, the remnant is a bulge  with  
a $B/T$ equal to one. As time proceeds, $B/T$ falls because a stellar disk 
grows around the bulge as hot gas in the halo cools, settles into a disk, and 
forms stars. The formation of stars by any residual cold gas left at the end of 
the major merger also helps to grow the disk.  Thus $B/T$ falls until the next 
major merger happens, at which point $B/T$ is reset to one in the models. 

The bulge may also grow in stellar mass due to minor mergers. 
Minor mergers are defined as mergers with mass ratio $ 1/10 < M_1/M_2 < 1/4$, 
and the stellar disk of the large companion is not destroyed during such mergers. 
The models assume that during minor mergers, all  the stars in the  satellite 
are added to the bulge of the host, while the gas settles in the disk. 
When DM halos grow by accretion or minor mergers, the  hot gas that comes in with 
a satellite is immediately stripped and added to the hot gas component of the 
host. The cold gas in the disk of the satellite is only added to the cold gas 
of the host if they merge.  Until they merge the satellite is using up its own 
cold gas to make stars.

Fig.~\ref{mergehis} shows the relationship between the present-day $B/T$ of a 
a high mass ($M_\star \geq  1.0\times 10^{10} M_\odot$) spiral and the redshift 
$z_{\rm last}$  of its last major merger.
As expected,  systems where the last major merger occurred at earlier times, 
have had more time to grow a disk and have a lower $B/T$.  
The dispersion in the present-day $B/T$ at a given $z_{\rm last}$ 
is  due to the different times spent by a galaxy in terms of being a satellite 
versus a central galaxy in a DM halo, since the cooling of gas and the 
growth of a  disk is stopped when a galaxy becomes a satellite. 
Thus, galaxies that became a satellite galaxy shortly after their last major 
merger  stayed at high $B/T$. Conversely,  galaxies that continued to be 
a central galaxy for a long time after their last major merger will have 
low $B/T$.

The present-day  $B/T$ of a high mass ($M_\star \geq  1.0\times 10^{10} M_\odot$) 
spiral depends on its major merger history. 
In particular, we note from Fig.~\ref{mergehis} that a  high mass 
($M_\star \geq  1.0\times 10^{10} M_\odot$) galaxy, which has undergone a 
past major merger since $z \le 2$ will end up hosting a 
present-day  $B/T > 0.2$. In effect, {\it 
a   high mass spiral can have  a  present-day $B/T \le0.2$ only if its last major 
merger occurred at $z > 2$  (lookback times $>10$ Gyr).}

The predicted distribution of present-day $B/T$  depends
on the galaxy merger history in the models and it is relevant 
to ask how well the latter is constrained  observationally.   
Over the redshift range $z\sim0.24$ to 0.80 (lookback times of  3 to 7 Gyr), 
recent studies by Jogee \etal (2008, 2009) find that   among  high 
mass ($M_\star \geq  2.5\times 10^{10} M_\odot$) galaxies, $\sim$~10\% 
of galaxies are undergoing mergers of  mass ratio $>1/10$, and 
 $\sim$~3\% are undergoing major mergers of  mass ratio $>1/4$.
These findings agree within a factor of  less than 
$\sim2$  with the merger rates  from the models of Khochfar \& Burkert 
(2001) over $z\sim$~0.24 to 0.80.
At higher redshifts, the empirical merger rate/fraction is uncertain 
due to relatively modest volumes and bandpass shifting effects, 
but there is  a general trend towards  higher merger fractions at higher 
redshifts (e.g., Conselice \etal 2003). 
The models used here (Khochfar \& Burkert 2001) agree with this 
trend and  predict  that $\sim$~13.5\% 
and $\sim$~20\%  of high mass ($M_\star \geq  1.0\times 10^{10} M_\odot$) 
spirals have undergone major mergers since $z\le 2$ and  $z\le 4$, 
respectively (see  Table~\ref{tmodel1}).

The contribution of 
galaxies with different merger histories to the present-day 
$B/T$ distribution 
are  shown in  Table~\ref{tmodel1}. The top and middle parts of 
the table describe systems with and without major mergers since 
 $z\le 2$ and  $z\le 4$, respectively.
In the model, $\sim$~13.5\% of present-day high mass 
($M_\star \geq  1.0\times 10^{10} M_\odot$) spirals, experienced 
a major merger since $z\le 2$, causing most of them  ($\sim$~11.2\%) 
to have a present-day  high $B/T > 0.4$ and a negligible fraction 
($\sim$~0.1\%) to have a low present-day $B/T \leq 0.2$.
In contrast, the remaining $\sim$~86.5\% spirals experienced no 
major merger at $z\le$~2, and most (67.2\%) of them  have a present-day 
low $B/T \leq 0.2$.
If the comparisons are extended to systems without a major merger 
since $z\le 4$, the numbers are very similar (see middle part of
Table~\ref{tmodel1}).

Table~\ref{tmodel1}  shows that there is good agreement 
between the model and data for the fraction of high mass spirals with 
present-day low $B/T \leq 0.2$ ($\sim$~67\% in the model versus $\sim$~66\% 
in the data).
The model contribution to low $B/T \leq 0.2$ comes almost entirely from
galaxies, which have not had a major merger since  $z\le 2$ (see
column 4 in Table~\ref{tmodel1}). 
In fact, most of these galaxies have  not even had a major merger 
since  $z\le4$, as illustrated by the bottom part of Table~\ref{tmodel1}.
{\it 
In the model, the fraction ($\sim1.6\%$; column 3 of Table~\ref{tmodel1}) 
of high mass spirals, which have undergone a  major merger since $z\le 4$  
and host a bulge with a present-day $B/T \le 0.2$, is  
{\it a factor of over thirty smaller}  than the observed 
fraction  ($\sim66\%$) of high mass spirals  with  $B/T \le 0.2$.  
}
Thus, bulges built  via major mergers  since $z\le 4$ 
seriously fail to account for the bulges present in $\sim66\%$
of  high  mass spirals. 
These results are also illustrated in Fig.~\ref{pbulgeless1}, which 
shows the comparison between data and models for the cumulative
fraction of high mass spirals as a function of  present-day $B/T$.

It is also interesting to note from Table~\ref{tmodel1} that although
the models reproduce well the frequency of bulges with 
present-day low $B/T \leq 0.2$, they tend to over-produce the
frequency of present-day  high $B/T > 0.4$ systems by nearly a 
factor of  two ($\sim$~14\% in the model versus $\sim$~8\% in the data; see 
columns 6 and 2 in middle part of Table 11).
Most of this overproduction stems from major mergers at $z\le4$ 
($\sim$~13\%; column 4 in middle part  of  Table~\ref{tmodel1}).
This suggests that major mergers, as currently modeled here, are 
{\it 
building  bulges too efficiently.}

One possible solution to this problem might relate to the suggestion  by Hopkins 
\etal (2009) that the efficiency of bulge-building during the major merger of two 
spirals depends not only  on the mass ratio $M_{\rm 1}/M_{\rm 2}$ of the merger, 
but also depends on the cold gas mass fraction $f_{\rm gas}$ in the disk.  In 
their semi-analytic  models,  the entire stellar mass of the satellite violently 
relaxes, but the fraction of stellar mass in the primary disk that violently  
relaxes and adds to the bulge is $M_1/M_2$. This differs with our models where 
the entire primary stellar disk is always destroyed in a major merger.
Furthermore, the fraction $F$ of the total gas mass, which loses angular momentum, 
falls to the nucleus, and is transformed into stellar mass in a nuclear 
starburst, is  $\sim(1-f_{\rm gas})\times(M_1/M_2)$. In particular,  $F$ is lower
for more gas-rich disks, causing a suppression of the burst efficiency in 
gas-rich systems, and a reduction in the stellar mass that ends up in the
bulge built during the  major merger. The predictions for the distribution of 
$B/T$ from the Hopkins \etal (2009) models are shown in the bottom of  
Table~\ref{tmodel1}, with major mergers considered as those with baryonic
galaxy-galaxy mass ratio $>$~/1/3.  Due to the reduced stellar mass that ends up in the 
bulge after a major merger, the  Hopkins \etal (2009) models tend to yield lower  $B/T$ 
after such a merger. Thus, the models produce overall fewer high $B/T > 0.4$ systems,  
and more intermediate  0.2~$< B/T <$~0.4 and low $B/T \leq 0.2$ systems (see column 
6 in Table~\ref{tmodel1}).  
Nonetheless, the  predictions are not very different from the models by  Khochfar
 \& Burkert (2005) and Khochfar \& Silk (2006), which we use in this paper.
In particular, the conclusion that the large frequency of high mass 
($M_\star \geq  1.0\times 10^{10} M_\odot$) spirals  with low $B/T \leq 0.2$ bulges 
can only be accounted for by spirals without a major merger since  $z\le 2$ also  
holds with the Hopkins \etal (2009) models. 

Thus, we  conclude that
{\it  the observed large frequency ($\sim$ 66\%) 
of high mass ($M_\star \geq  1.0\times 10^{10} M_\odot$) spirals with low 
present-day $B/T \leq 0.2$   can be accounted for in our and other 
hierarchical models by high mass spirals, which have not undergone a major merger 
since $z\le 2$, and most of which have not even experienced a major merger over 
the last 12 Gyr since $z\le 4$.
}
In the models, most of these present-day low $B/T \leq 0.2$ bulges are 
built by  minor mergers  since  $z\le4$.   
As noted earlier, our models explore bulge-building via minor and
major mergers, but do not explicitly incorporate secular processes
(see  $\S$~\ref{bulgeFormation}). In practice, secular processes 
may contribute to the building of  present-day low $B/T \leq 0.2$ 
bulges, and are particularly relevant at $z\le2$, where major 
mergers cannot build such bulges.

For completeness, we further explore how sensitive are our results to 
other  assumptions made in the data and models of  Khochfar \& Burkert (2005) 
and Khochfar \& Silk (2006):

\begin{itemize}
\vspace{-3mm}
\item 
How sensitive are the results to the mass ratio used to separate major and minor
mergers?  Fig.~\ref{pbulgeless2} is similar to  Fig.~\ref{pbulgeless1} 
except that the model now defines major mergers as those with  mass ratio  
$M_1/M_2 \ge 1/6$.  In this case, about 30\% of the model spirals 
undergo major mergers since $z\le4$ rather than $\sim20\%$. 
The overall model $F$ (black dashed line) now under-predicts the 
data $F$ by  about 10\%  for  $B/T >$~0.2. However, the main conclusion that 
bulges built by major mergers   since $z\le4$ 
cannot account for most of the low $B/T\le0.2$ 
bulges, present in a large percentage ($\sim66\%$) of spirals still holds.

\vspace{-3mm}
\item 
How sensitive are the results to the $B/T$  cut used to define spirals? 
Fig.~\ref{pbulgeless3} is similar to  Fig.~\ref{pbulgeless1} 
except that here spirals are considered to be systems with a 
$B/T \le0.55$ rather than 0.75 in the models, and a corresponding 
cut is applied to the data points. The results are similar to 
Fig.~\ref{pbulgeless1}

\vspace{-3mm}
\item 
How sensitive are the results to our assumed constant $H$-band  
mass-to-light ($M/L$)  for the bulge, disk, and bar? 
Fig.~\ref{pbulgeless4} is similar to  Fig.~\ref{pbulgeless1} 
except that the  $B/T$ of all the observed galaxies has been  multiplied 
by a factor of two, in order to test what would happen in the case 
where the $M/L$ ratio of the bulge in $H$-band is twice as high 
as that of the disk and bar.  This could happen in an extreme 
example where the dominant bulge stellar population was much older 
(e.g. 12 Gyr) than the age of the dominant disk stellar population  
(e.g., 3 Gyr). In such a case, the fraction of high mass spirals 
with $B/T\le$~0.2 would change from $\sim66\%$ in Fig.~\ref{pbulgeless1} 
to  $\sim55\%$, and deviate from the model overall model $F$ (black dashed line)  
by $\sim20\%$. However, the main conclusion that bulges built by major mergers  
since $z\le4$ cannot account for most of the low $B/T\le0.2$ bulges, 
present in a large percentage ($\sim55\%$) of spirals still holds.  
\end{itemize}

Finally, it is important to note that so far we have compared the data and model 
only in terms of  bulge $B/T$, but not in terms of bulge index $n$ or in 
terms of bar fraction.  In effect, we have only shown that the models 
reproduce a subset of the results outlined in  points (1)-(3) of 
$\S$~\ref{bulgeFormation}.  Since 
the semi-analytic models do not produce a distribution of bulge index $n$, 
we resort to  presenting only an indirect comparison in Table~\ref{tmodel2}.
We assume that bulges, which form in major mergers have a bulge $n>2.5$. 
This assumption is based the evidence presented in $\S$~\ref{bulgeFormation}.
Thus, in  Test 1 of Table~\ref{tmodel2}, we compare the fraction 
($\sim$~66\%) of galaxies in the semi-analytic models having 
$B/T \le0.2$ and  no  major merger since $z\le4$, to the observed fraction 
($\sim$~65\%) of  galaxies  with $B/T \le 0.2$ and bulge 
$n\le 2.5$.  There is close agreement between the 
two values. In  Test 2 of Table~\ref{tmodel2}, the fraction 
($\sim$~12.7\%) of model galaxies with  $B/T>0.4$ and no  major 
merger since $z\le4$ is a factor of $\sim$~3 higher than the fraction 
($\sim$~3.5\%) of high mass spirals with $B/T>0.4$  and  bulge $n>2.5$. 
Similarly, the fraction ($\sim$~1.6\%) of model galaxies with  
$B/T \le 0.2$  and  a past major merger since $z\le4$  is also a factor 
of  $\sim$~2 higher than the observed fraction 
($\sim$~0.9\%) of high  mass spirals with $B/T\leq0.2$ and bulge 
$n>2.5$ (Test 3, Table~\ref{tmodel2}). 
Thus, in terms of bulge $B/T$ and $n$, there is  good agreement 
between  data and model for Test 1 (involving model galaxies with no 
major mergers since $z\le4$). This suggests that  
the vast majority of bulges with  $B/T \le 0.2$ and $n \le 2.5$ 
likely formed in galaxies having had no major merger since $z\le4$.
However, the agreement is less good for Tests 2 and 3 (involving 
model galaxies with major mergers  since  $z\le4$) and this suggests
that the models may be building bulges a little bit too efficiently 
during a major merger, in agreement with the conclusion reached earlier.

What about the role of  bars in the formation of these bulges of low $B/T$ and 
low $n$? A detailed direct comparison with the semi-analytic models is 
not possible as the role of bars is not yet modeled, but related comparisons 
are possible. First, it is important to note that bar-driven gas inflow into 
the inner kpc and the subsequent building of disky stellar components or 
`pseudobulges' (see $\S$~\ref{sintro}) can happen in both isolated galaxies and 
in minor mergers ($\S$~\ref{bulgeFormation}), since bars can be spontaneously 
induced in an isolated disk or tidally induced during an interaction or minor 
merger. The triggering of a bar is favored in a prograde interaction or minor 
merger. Thus, bulge-building via induced bars is more likely to happen in prograde
rather than retrograde minor mergers. Statistically about half of minor mergers 
might be prograde or prograde-like, and half retrograde. Thus, one would expect 
bars to be induced in only half of the minor mergers. If this assumption is 
correct and if most  of the mass in bulges with  present-day $B/T \le 0.2$ is 
formed in minor mergers, then one would expect only about half of these bulges 
to host bars. 
This  is close to what is observed, as  shown by  Table~\ref{barFraction}.
We see that   $\sim63\%$ of high mass spirals with low $n\leq2$ bulges  
and $\sim68\%$ of spirals with low $B/T \le 0.2$ bulges host bars.
This suggests that  in high mass spirals, spontaneous and/or 
tidally induced bars \emph{may} play a part in forming up to two-thirds 
of $B/T \le 0.2$ or $n\leq2$ bulges. 
The remaining one-third of such bulges may have
been formed either by   mechanisms that  do  not involve bars (e.g., retrograde 
minor mergers) or by bars that are not long-lived.

\section{Summary}\label{ssumm}

The properties of galaxy components (bulges, disks, and bars) in the local Universe
provide key constraints for models of galaxy evolution.  Most previous $2D$
decompositions have focused on two-component bulge-disk decomposition, and 
ignored the contribution of the bar even in strongly barred galaxies. 
However, as shown by this work and other recent studies 
(e.g., Laurikainen \etal  2005; Laurikainen 
\etal 2007;  Reese \etal 2007),  it is important to include the bar component 
in the $2D$ decomposition, in order to correctly estimate  the  bulge-to-total 
ratio ($B/T$) and disk properties. In this paper we have developed an iterative 
$2D$, bulge-disk-bar decomposition technique using GALFIT and applied it to 
$H$-band images to  a complete sample (S1) of 143  bright ($M_B\leq-19.3$)  
low-to-moderately inclined ($i \le 70^\circ$)  spirals from the OSU Bright Spiral 
Galaxy Survey (OSUBSGS).  The sample has primarily spirals with Hubble type 
S0/a to Sc and stellar mass $M_\star \geq 1.0 \times 10^{10} M_\odot$. We 
performed two-component bulge-disk decomposition, as well as three-component 
bulge-disk-bar decomposition  on the $2D$ light distribution of all galaxies, 
taking into account the PSF. We use an exponential profile for the disk, 
and S\'ersic profiles for the bulge and bar. A number of quantitative indicators, 
including bar classification from ellipse fits, are used to pick either the 
bulge-disk-bar decomposition or bulge-disk decomposition, as the best final fit 
for a galaxy. Our main results are the following.

\begin{figure}[]
\centering
\plotone{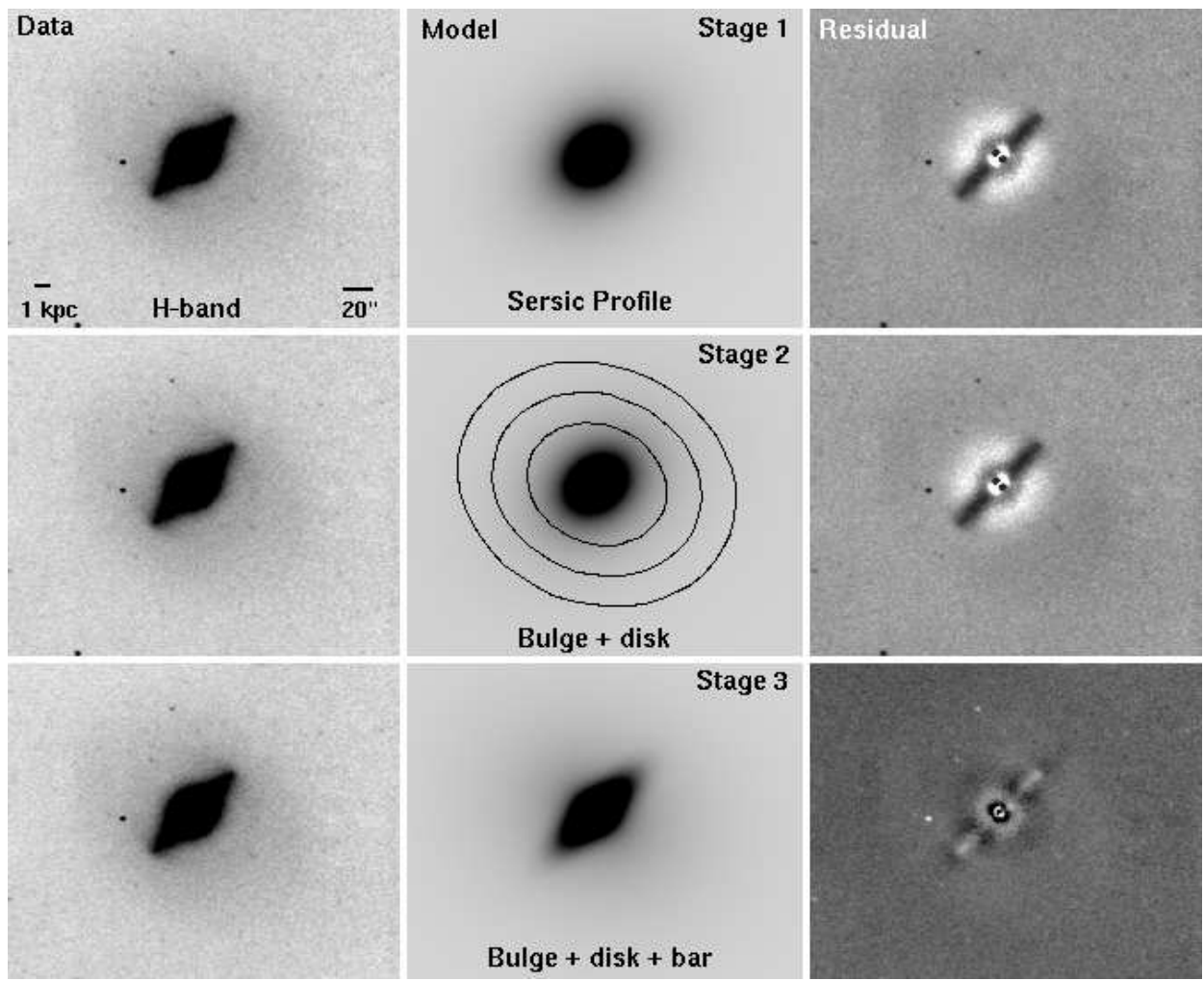}
\caption{Complete $2D$ decomposition for NGC 4643.
Note the prominent bar residuals in the residual for the Stage 1 and Stage 2 bulge-disk 
decomposition. This is a case where the prominent bar causes the Stage 2 bulge-disk fit to
artificially extend the bulge and inflate the $B/T$.  The disk fitted in Stage 2 has a low
surface brightness and is very extended, well beyond the real disk: the $b/a$ and $PA$ of the
fitted disk is shown as contours. Stage 3 bulge-disk-bar decomposition 
provides the best model. The $\chi^2$ for the Stage 1, Stage 2, and Stage 3
residual images are 7360.7, 7284.8, and 2111.59, respectively.
See Table~\ref{tngc4643} for the fit parameters.
\label{ngc4643}}
\end{figure}

\begin{figure}[]
\centering
\plotone{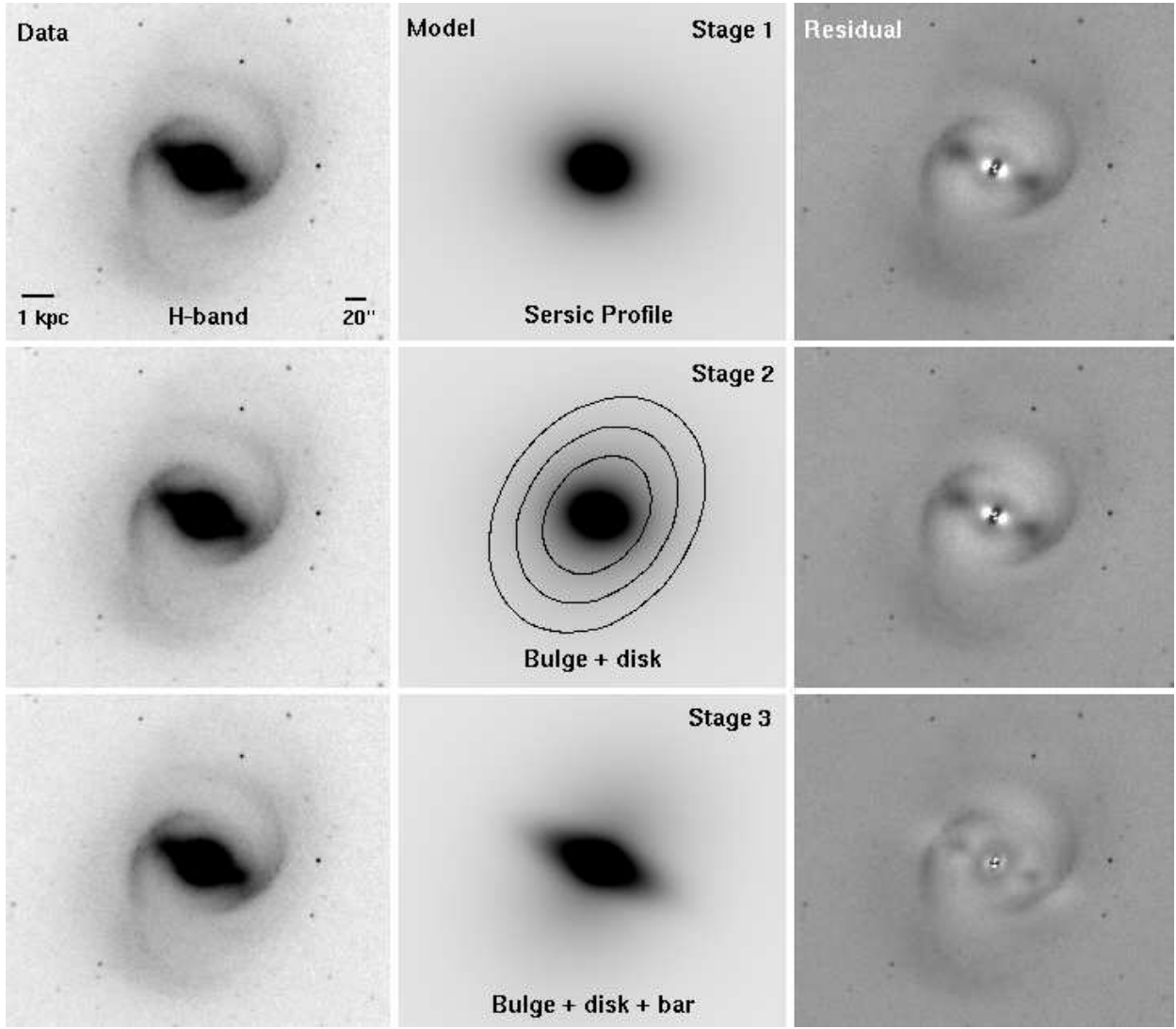}
\caption{The complete $2D$ decomposition for NGC 4548.
This is an extreme example where the prominent bar results in an extended
bulge and inflated $B/T$ in the Stage 2 bulge-disk fit.  Like NGC 4643 in 
Figure~\ref{ngc4643}, the disk fitted in Stage 2 has a low
surface brightness and is very extended: its $b/a$ and $PA$ are shown as 
contours.  Stage 3 bulge-disk-bar decomposition provides the best model. 
The $\chi^2_\nu$ for the Stage 1, Stage 2, and Stage 3
residual images are 7076.1, 6301.3, and 3260.4, respectively.
See Table~\ref{tngc4548} for the fit parameters.  \label{ngc4548}}
\end{figure}

\begin{figure}[] 
\centering
\scalebox{0.95}{\includegraphics*[0.28in,3.5in][9in,10in]{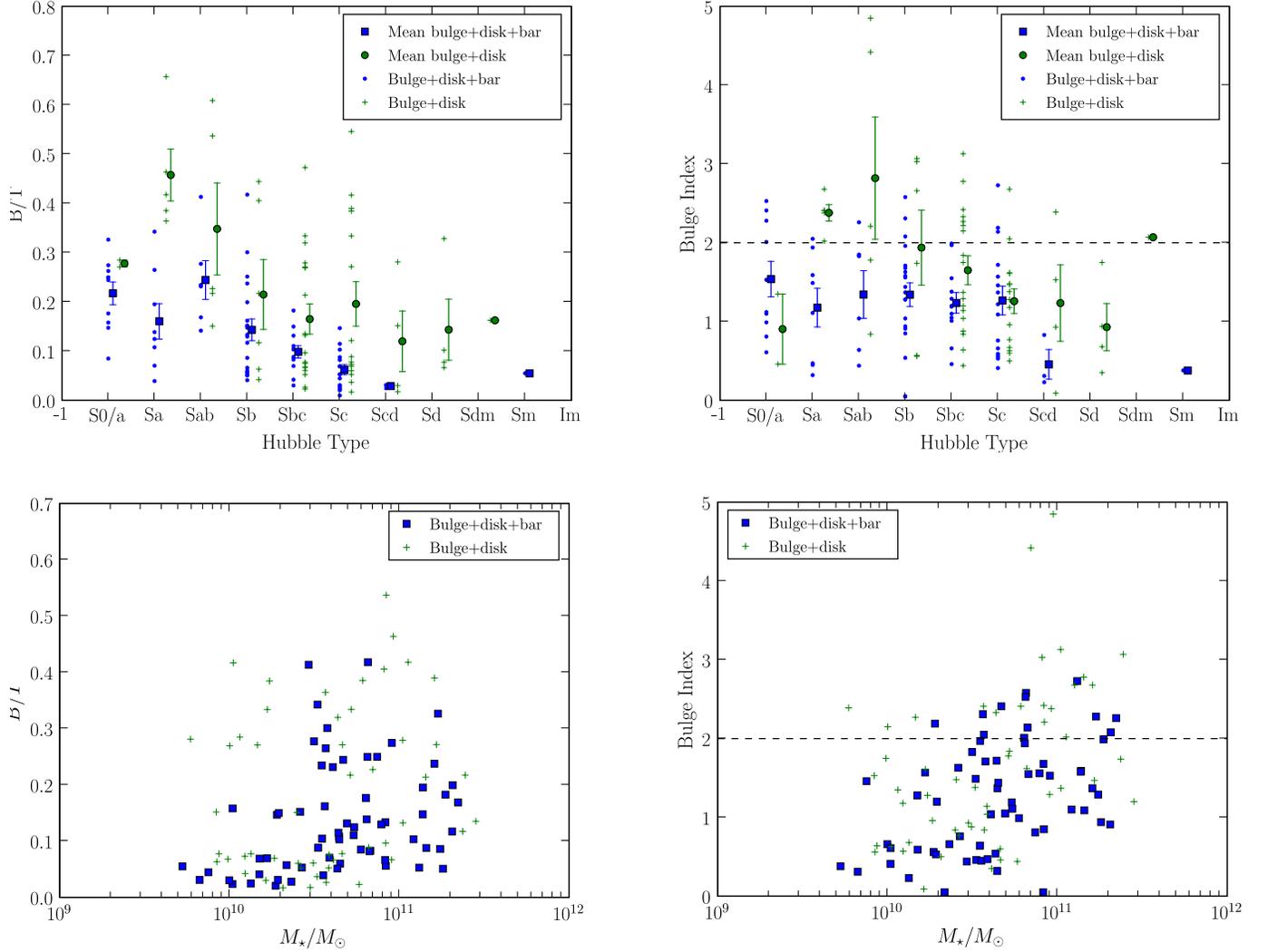}}
\caption{
The individual and mean  $B/T$ (left panels) and 
bulge S\'ersic index (right panels) are plotted  
as a function of Hubble type  for the sample S1 of bright galaxies, 
and as a function of  galaxy stellar mass for sample S2.
The error bars  indicate the standard deviation of the population around the mean in each bin.
The legend in each panel indicates the type of decomposition used for
each data point.
The mean $B/T$ and bulge index in barred galaxies differ
systematically from unbarred galaxies, but there is a large overlap
in the individual values.  
As many as  $\sim 69\%$ of bright spiral galaxies have $B/T \leq 0.2$; 
these bulges are pervasive and exist across the Hubble sequence.
Furthermore, as many as $\sim76\%$   of  bright  spirals have 
low $n \leq 2$ bulges. Such bulges exist  in barred and unbarred galaxies 
across a wide range of Hubble types.
}
\label{4p-bplot1}
\end{figure}

\begin{figure}[] 
\centering
\epsscale{0.80}
\plotone{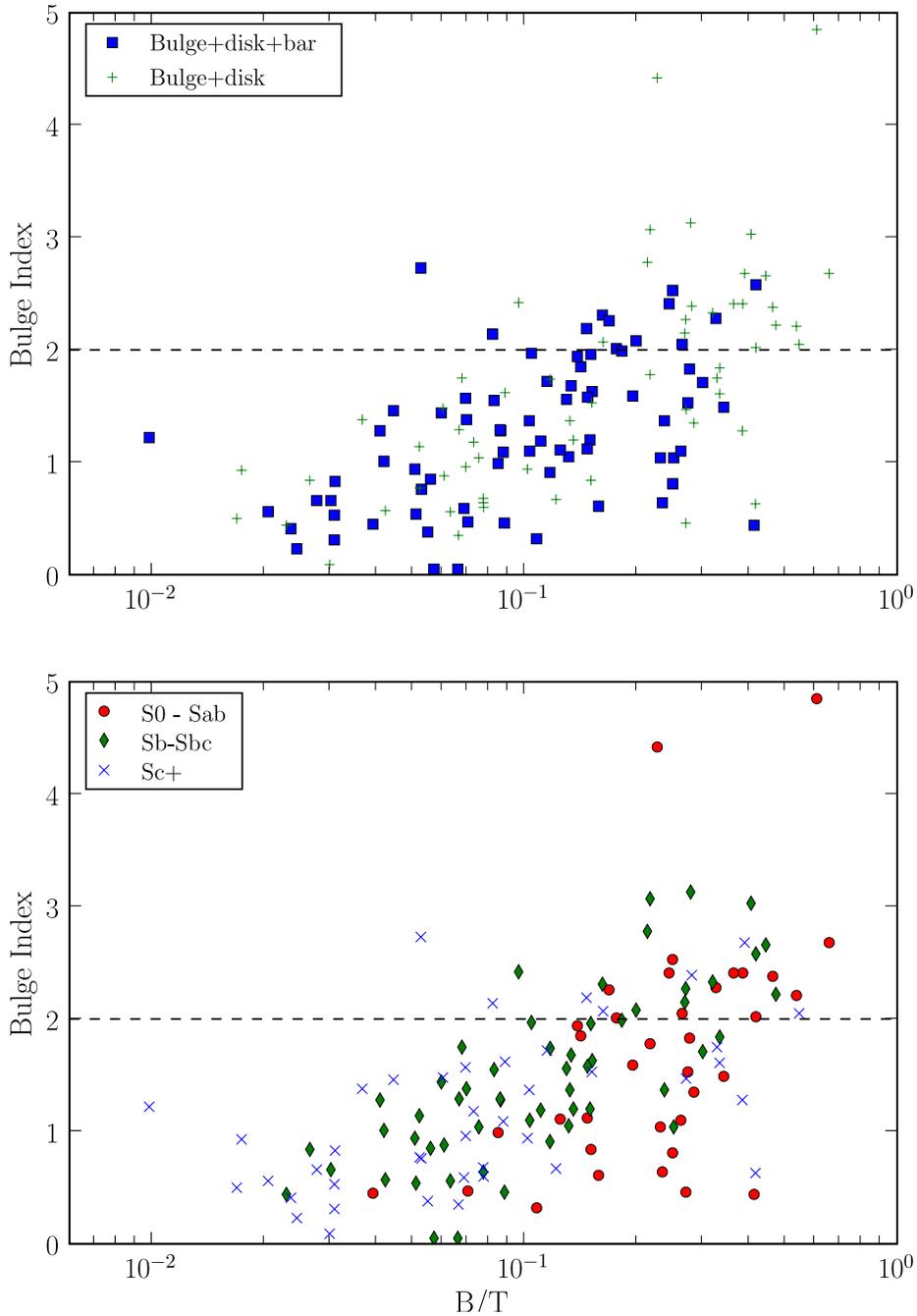}
\caption{ The relation between $B/T$ and bulge index is shown.
In the top panel, galaxies are coded according to bar class.  
The legend indicates the type of decomposition used for each data point.
In the lower panel, galaxies are coded according to Hubble type.  
A striking $\sim76\% $ of bright  spirals have       
low $n \leq 2$ bulges. Such bulges exist  in barred and unbarred galaxies 
across a wide range of Hubble  types, and  their $B/T$ range from 0.01 to 0.4, 
with most having $B/T \leq$ 0.2.
A  moderate fraction ($\sim22\%$) have intermediate $2<n<4$ bulges. 
These exist in barred and unbarred S0/a to Sd galaxies, and their $B/T$ spans a 
wide range  (0.05 to 0.5).
Only ($\sim1\%$) have $n\geq4$.
}
\label{2p-bplot2}
\end{figure}

\begin{figure}[]
\centering
\scalebox{0.95}{\includegraphics*[0.28in,3.5in][9in,10in]{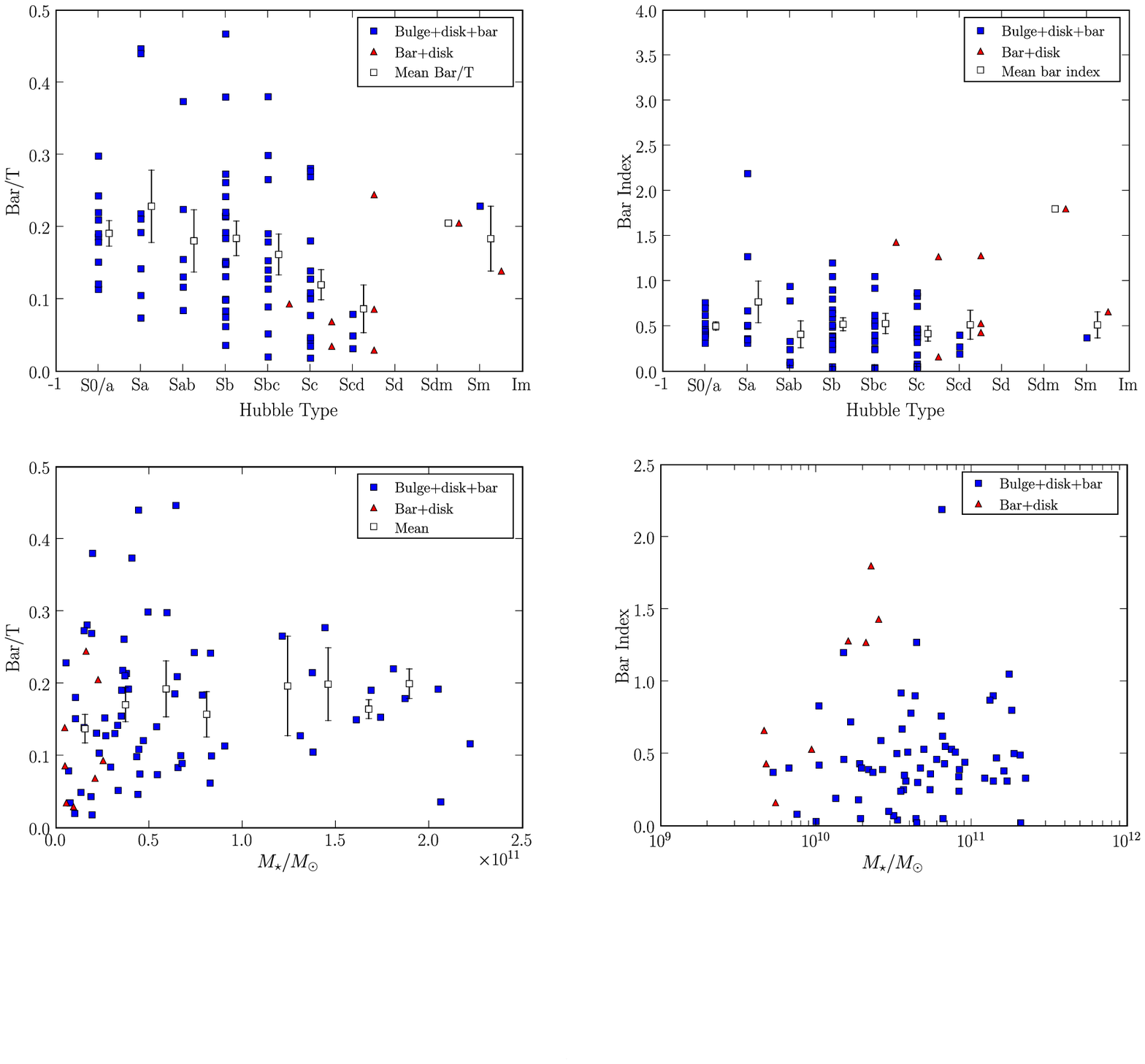}}
\caption{ 
The properties of bars are shown. 
The error bars  indicate the standard deviation of the population around the mean in each bin.
The legend in each panel indicates the type of decomposition used for
each data point.
Upper left:  Mean and individual Bar/$T$ plotted against Hubble type. 
Upper right: Mean and individual bar S\'ersic indices plotted against Hubble
type.
Lower left: Bar/$T$ plotted against total galaxy stellar mass.  The mean
Bar/$T$ in bins of stellar mass is indicated.
Lower right: Bar S\'ersic index plotted against total galaxy stellar mass. 
}
\label{fbar1}
\end{figure}

\begin{figure}[]
\centering
\scalebox{0.95}{\includegraphics*[0.28in,3.5in][9in,10in]{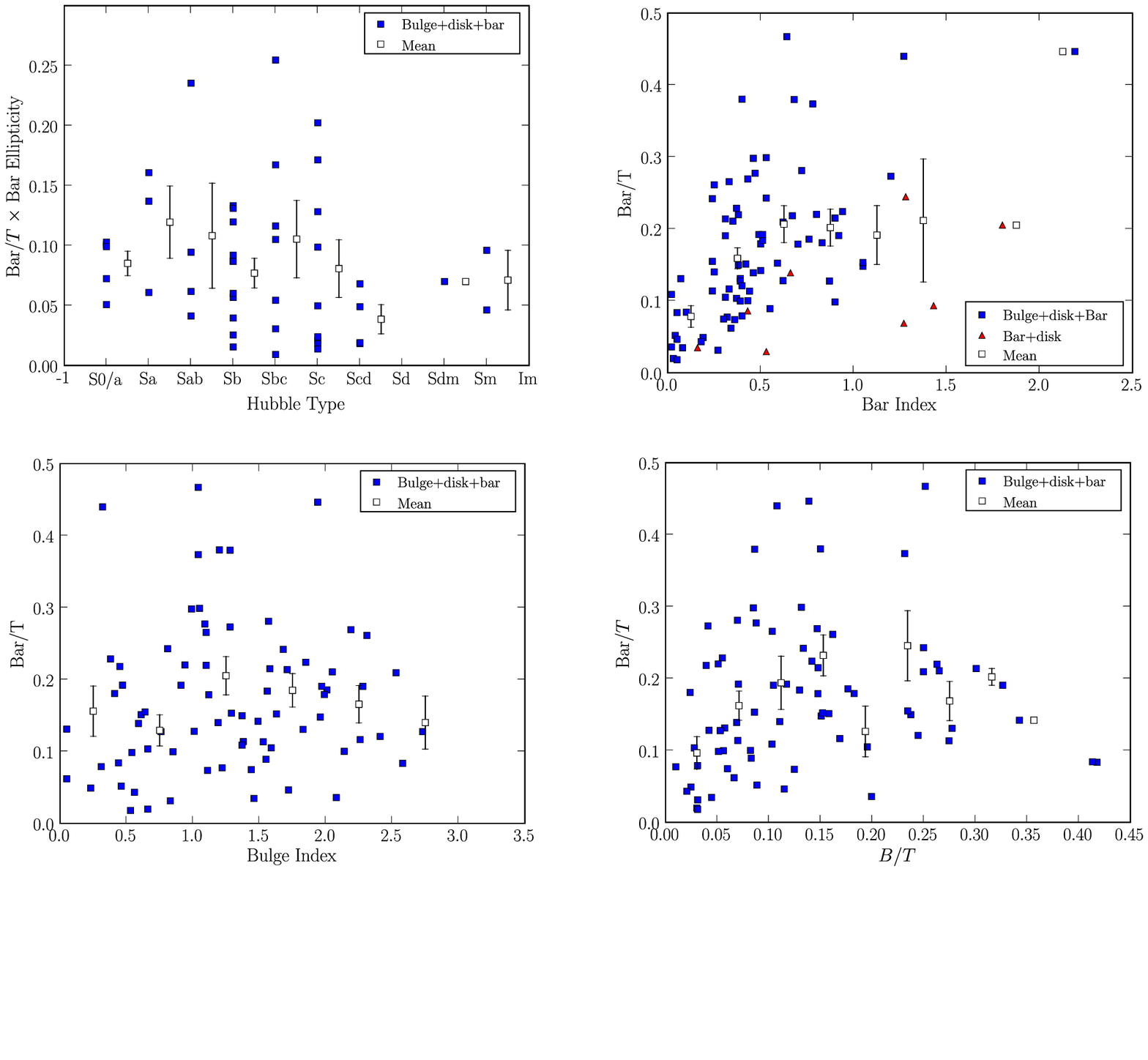}}
\caption{
Bar strength is plotted against Hubble type and the properties of bars 
are compared with bulges. The legend in each panel indicates the type of 
decomposition used for each data point.
Upper left: Bar strength, the product of Bar/$T$ and peak bar ellipticity
$e_{bar}$ from MJ07 is plotted against Hubble type.
Upper right:  Bar/$T$ is plotted against bar S\'ersic index.
Lower left: Bar/$T$ is plotted against bulge S\'ersic index.
Lower right: Bar/$T$ is plotted against  $B/T$.
In the first plot, mean bar strong is calculated for each Hubble type.
In the latter three plots, mean Bar/$T$ is calculated for bins along 
the ordinate axis.  The error bars  indicate the standard deviation of 
the population around the mean in each bin.
}
\label{fbar2}
\end{figure}

\begin{figure}[]
\centering
\plotone{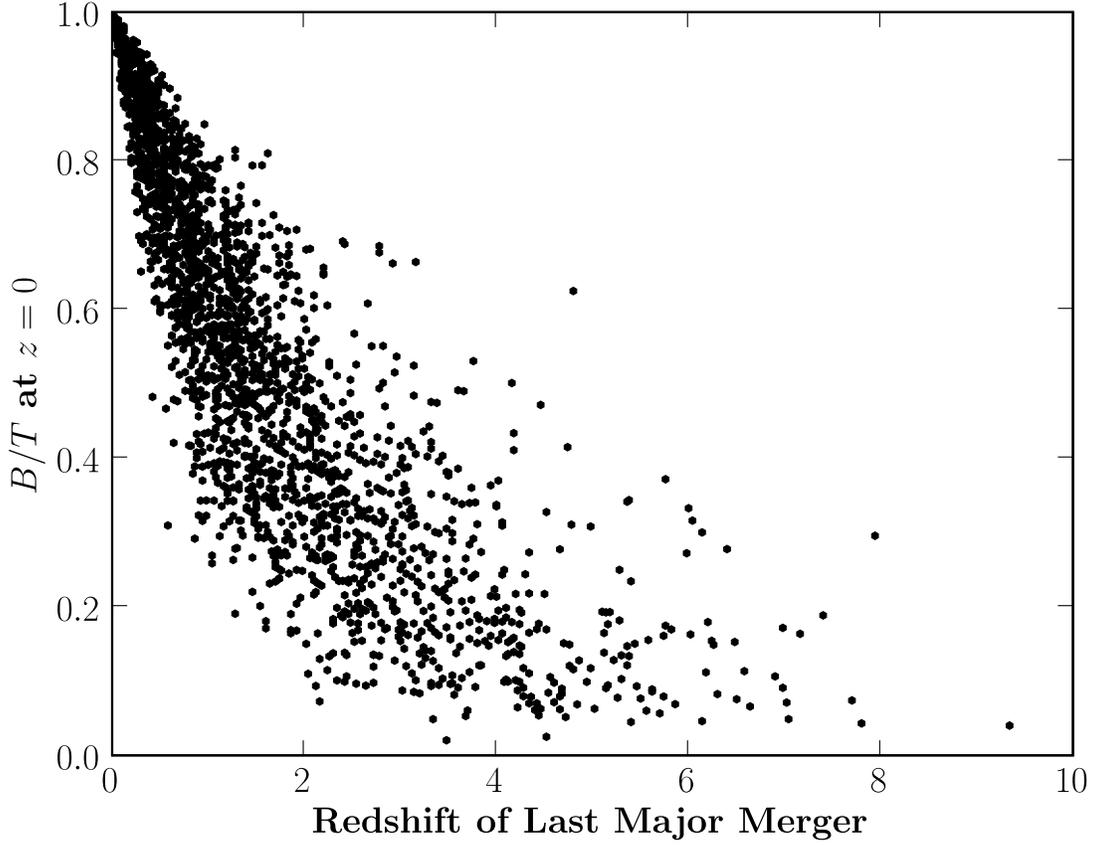}
\caption{
For those  high mass  ($M_\star \geq  1.0\times 10^{10} M_\odot$)
galaxies in the theoretical models that experienced a 
major merger (see $\S$~\ref{smodel1}), the $B/T$ of the remnant 
at $z\sim0$ is plotted against the redshift  
$z_{\rm last}$  of the last major merger. 
Systems  where the last major merger occurred at earlier times have 
had more time to grow a disk and thus have a lower $B/T$  at $z\sim$~0.
The dispersion in the present-day $B/T$ at a given $z_{\rm last}$ 
is  due to the different times spent by a galaxy in terms of being a satellite 
versus a central galaxy in a DM halo, since the cooling of gas and the 
growth of a  disk is stopped when a galaxy becomes a satellite. 
In the model, a  high mass galaxy that has undergone a major merger at 
$z \le 2$ has  a present-day  $B/T > 0.2$.  In effect, 
a  high mass spiral can have  a  present-day $B/T \le0.2$ only if its 
last major  merger occurred at $z > 2$  (lookback times $>10$ Gyr).
}  
\label{mergehis}
\end{figure}

\begin{figure}[]
\centering
\epsscale{0.65}
\plotone{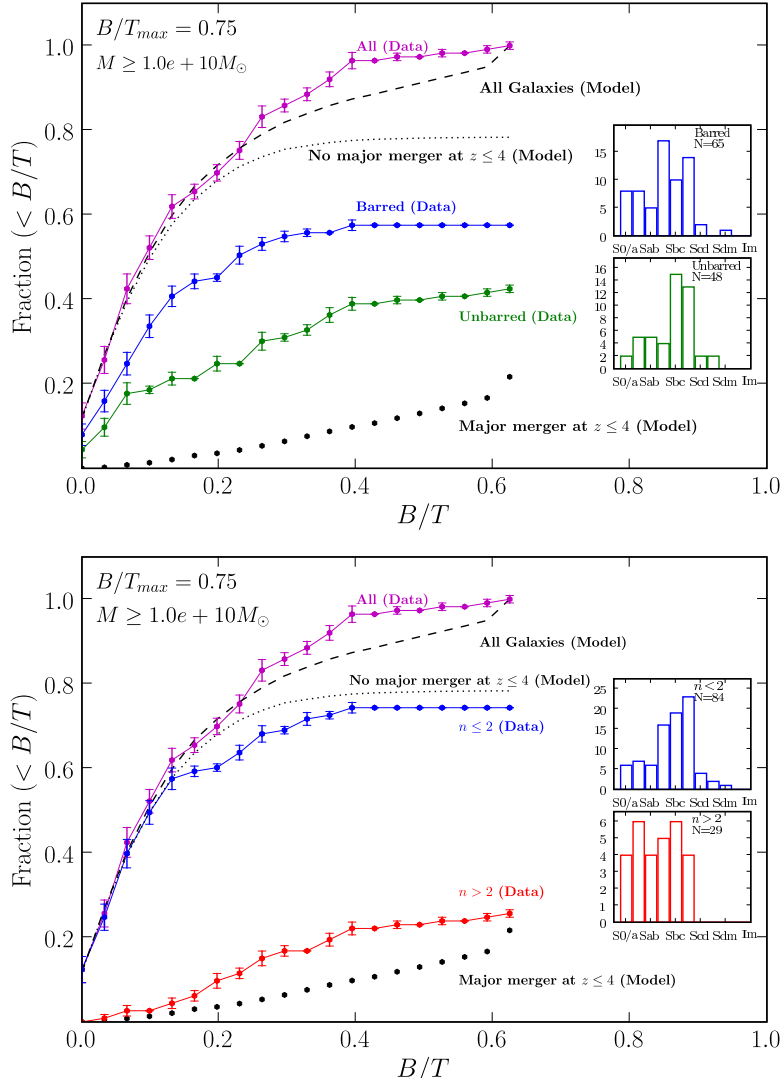}
\caption{
The {\it cumulative} fraction $F$ of high mass  ($M_\star \geq 1.0 
\times 10^{10} M_\odot$) spirals with present-day $B/T \le$ the x-axis value
is shown for the data (colored lines/ points) and
for the theoretical model  (black lines/points) described in  $\S$~\ref{smodel1}.
Model and data spirals are defined as systems with $B/T \le 0.75$. 
The magenta line shows $F$  from the data, while  the other two colored 
lines break this $F$ in terms of bar class (top panel) or  bulge $n$ (lower panel).
The black dashed line  shows $F$  from all  model
galaxies, while  the black dotted  line and black dots show the contribution
of model galaxies that  experienced, respectively, 
{\it no major merger} and {\it one or more major mergers} since $z \le 4$. 
Major mergers are defined here as those with  $M_1/M_2 \ge 1/4$. 
In the model, the fraction ($\sim$~1.6\%; see Table \ref{tmodel1}) 
of high mass spirals, which  have undergone a major merger since 
$z \le 4$  and host a bulge  with a  present-day $B/T \le 0.2$ is  
a factor of over 20 smaller than the observed fraction ($\sim$ 66\%) 
of high mass spirals  with present-day $B/T \le 0.2$.
Thus, bulges built  via major mergers  since $z\le 4$
seriously fail to account for most of the low  $B/T \le0.2$  bulges present 
in two-thirds of  high-mass spirals. 
}
\label{pbulgeless1}
\end{figure}

\begin{figure}[]
\centering
\epsscale{0.70}
\plotone{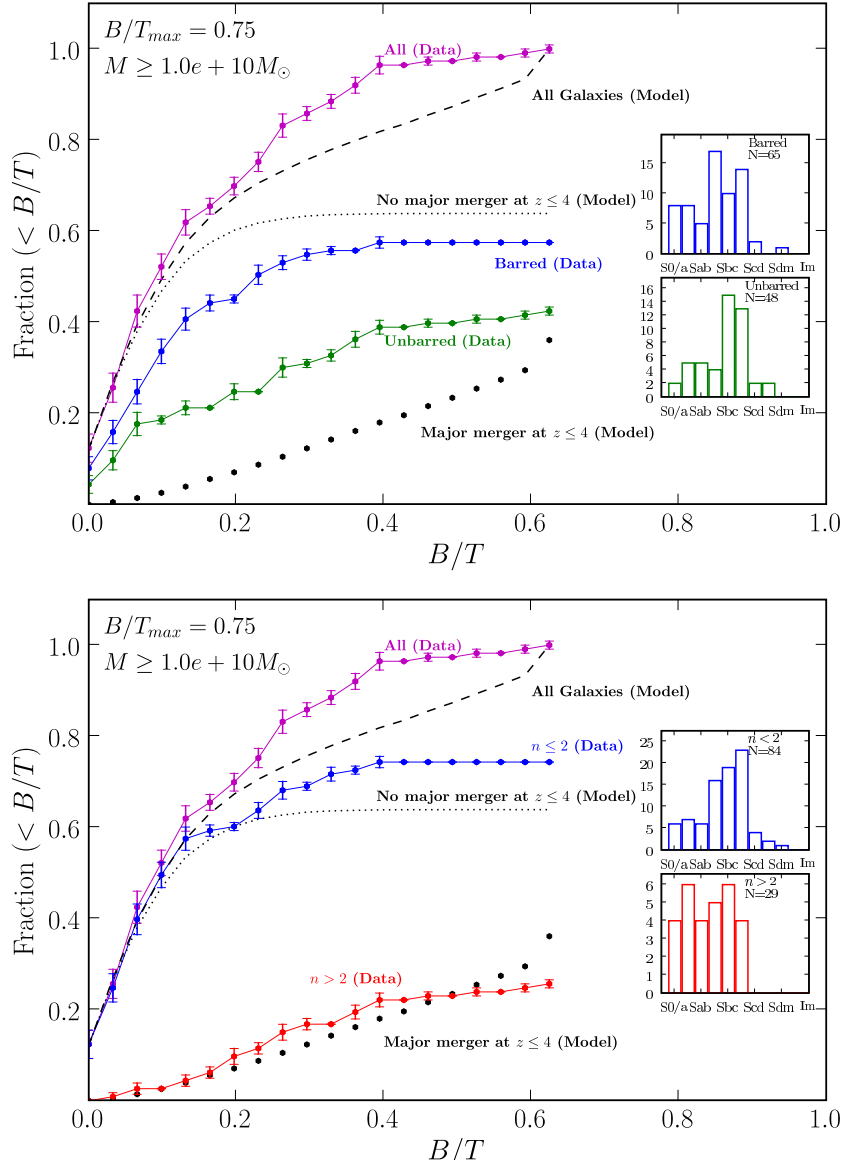}
\caption{This figure is similar to  Fig.~\ref{pbulgeless1}, 
except that the model now defines major mergers as those with  mass ratio  
$M_1/M_2 \ge 1/6$.  
}
\label{pbulgeless2}
\end{figure}

\begin{figure}[]
\centering
\epsscale{0.70}
\plotone{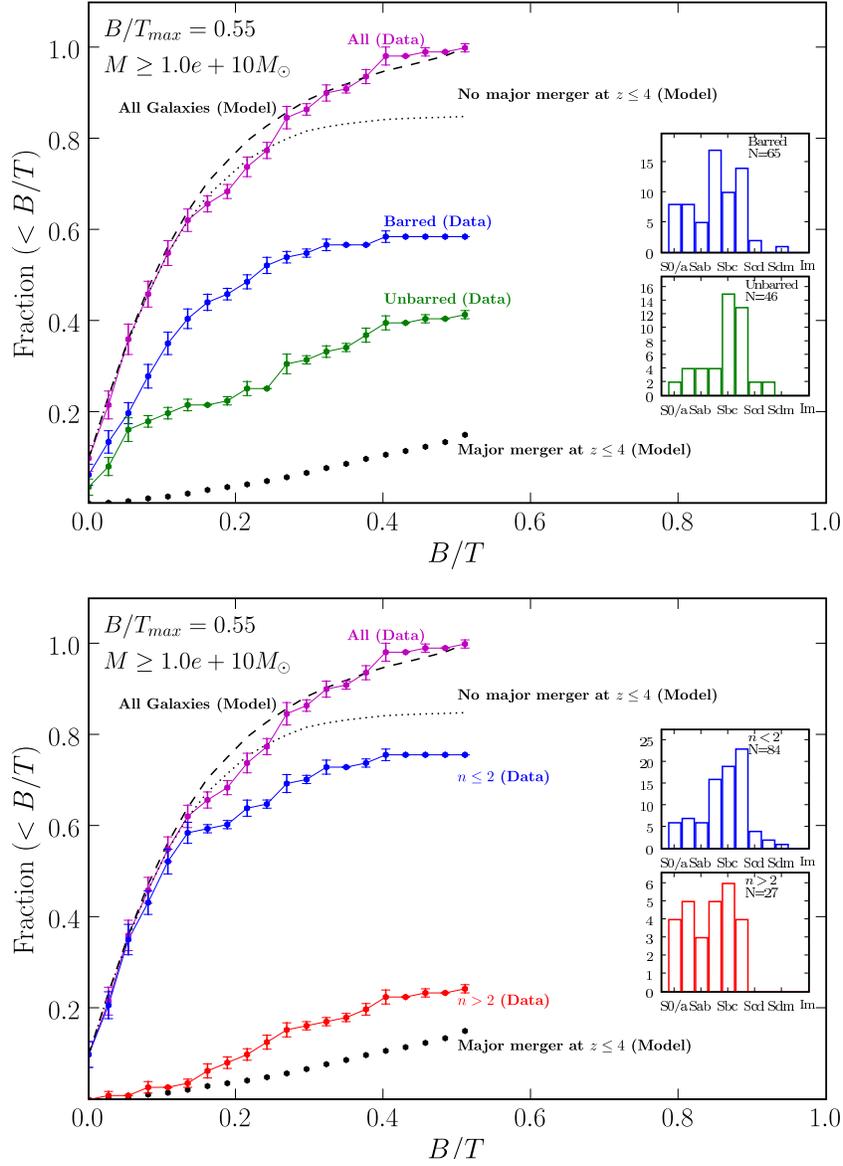}
\caption{This figure is similar to  Fig.~\ref{pbulgeless1}, 
except that here spirals are considered to be systems with a 
$B/T \le$~0.55 rather than 0.75 in the models, and a corresponding 
cut is applied to the data points. 
}
\label{pbulgeless3}
\end{figure}

\begin{figure}[]
\centering
\epsscale{0.70}
\plotone{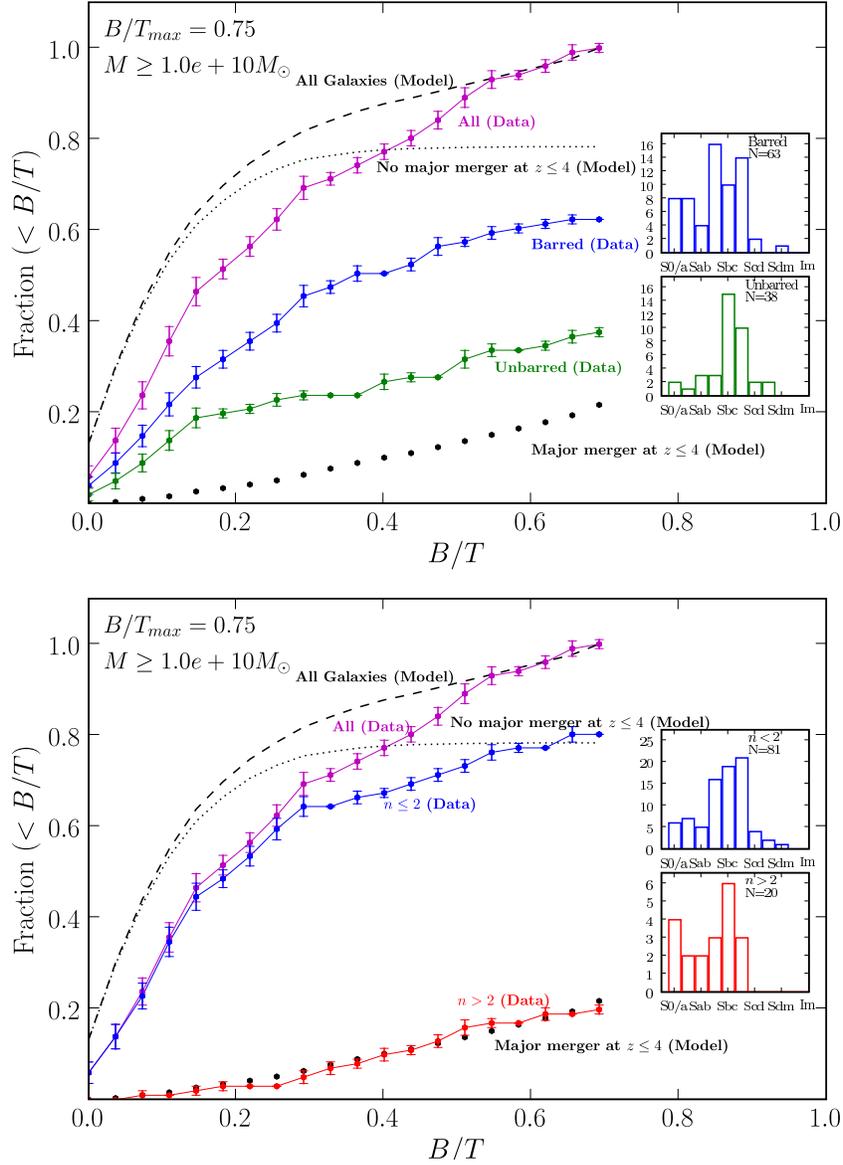}
\caption{This figure is similar to  Fig.~\ref{pbulgeless1}, 
except that  $B/T$ of all the observed galaxies has been  multiplied 
by a factor of two, in order to test what would happen in the case 
where the $M/L$ ratio of the bulge in $H$-band is twice as high 
as that of the disk and bar.  
This could happen in an extreme 
example where the dominant bulge stellar population was much older 
(e.g. 12 Gyr) than the age of the dominant disk stellar population  
(e.g., 3 Gyr). In such a case, the fraction of high mass spirals 
with $B/T\le 0.2$ would change from $\sim66\%$ in Fig.~\ref{pbulgeless1} 
to  $\sim50\%$. However, the main conclusion that 
bulges built by major mergers since $z\le 4$ 
cannot account for most of the low $B/T\le0.2$ bulges, 
present in a large percentage ($\sim55\%$) of spirals still holds.
}
\label{pbulgeless4}
\end{figure}

\begin{enumerate}
\item
We find that it is necessary to include  the bar component in $2D$ 
decomposition  of barred galaxies, otherwise, the  bulge-to-total ratio ($B/T$) 
will  be overestimated and the disk properties may be skewed. 
Examples of the effect of including the bar are shown 
for the prominently barred galaxies 
NGC 4643 (Figure~\ref{ngc4643}, Table~\ref{tngc4643}) 
and NGC 4548 (Figure~\ref{ngc4548}, Table~\ref{tngc4548}).
\item
We find that out of the 143 low-to-moderately inclined ($i\leq 70^\circ$) spirals
in our sample, 75 of 143  or $\sim52\%$ are better fit with a  Stage 3 bulge-disk-bar 
decomposition than a Stage 2 bulge-disk decomposition. There are also  
eight galaxies with pure bar-disk fits.  The resulting $H$-band bar 
fraction, defined as the fraction of disk galaxies that are barred, 
is $58.0 \pm 4.13\%$ (84 of 143). This fraction is in excellent agreement with the  
$H$-band bar  fraction of 60\% reported by MJ07, based on ellipse fits of the same 
OSUBSGS sample, with a more conservative inclination cut ($i\leq 60^\circ$).
\item
$H$-band images tend to  trace the overall mass fairly well and 
are not overly impacted by extinction and age gradients. We therefore 
assume a constant mass-to-light ($M/L$) in the  $H$-band for  the bulge, 
disk, and bar, and  assume their $H$-band light fraction is a measure of 
their mass fraction. For  our sample S1 of  143  bright ($M_B\leq-19.3$) 
low-to-moderately inclined  ($i \le 70^\circ$) spirals
with a mass-weighted mean Hubble type of Sab-Sb, we find that 
71.6\% of the stellar mass is in disks,  9.6\%  is in stellar bars and 
18.9\% is in bulges (with 10.4\% in $n>2$ bulges and 8.4\% in 
$n \leq 2$ bulges). 

If disks and bars are much younger (e.g., $\sim3$ Gyr old ) than bulges 
(e.g., $\sim12$ Gyr old), then  our prescription  would  underestimate  
the true $B/T$ by a factor of $\le$~2. On the other hand,
if the bulge is younger than the disk and happens to harbor a significant young 
population of  massive stars, then our prescription will  {\it overestimate} the 
true $B/T$ mass ratio, and  make our current results on the  high fraction 
of  low $B/T$ bulges  (see point 4 below) even stronger.  
\item
We explore the relationship between $B/T$, bulge S\'ersic index, and Hubble types  
(Fig.~\ref{4p-bplot1}  \&  Fig.~\ref{2p-bplot2}). 
Only a small fraction ($\sim1\%$) of bright spirals have high $n\ge4$
bulges; such bulges lie primarily in S0/a to Sab, and have a large $B/T > 0.2$. 
A  moderate fraction ($\sim22\%$)   have intermediate 
$2<n<4$ bulges; these exist in barred and unbarred S0/a to Sd,  and 
their $B/T$ spans a wide range  (0.03 to 0.5) with a mean of 0.29.
Finally, {\it a strikingly large fraction ($\sim76$\%) of bright  spirals have 
low $n \leq 2$ bulges; such bulges exist  in barred and unbarred galaxies 
across a wide range of Hubble types, and  their $B/T$  varies from 0.01 to 0.4, 
with most having $B/T \leq0.2$.} 
\item
Bulges with very high $B/T$ ($>0.4$)  exist primarily in galaxies 
with high mass ($M_\star > 6 \times 10^{10} M_\odot$) and early types (S0/a to Sab).
Bulges with very low  $B/T$ ($<0.1$) lie primarily in lower mass galaxies 
with later morphologies (Sb to Sc).
{\it 
As many as $\sim69\%$  of  bright 
spirals have bulges with $B/T \leq 0.2$: these bulges are pervasive   
and exist across  the whole spectrum of S0/a to Sd} (Figure~\ref{4p-bplot1}).
\item
Modeling bars with $2D$ decomposition allows us to measure 
bar properties and the bar-to-total ratio (Bar/$T$), which is a measure of bar 
strength. There is a wide range ($\sim0.03$ to  $\sim0.47$) in the  individual  
Bar/$T$ at a given Hubble type.  The mean Bar/$T$ remains fairly 
constant with Hubble type from Sa to Sb, but shows a possible weak decline by about 
0.1 from Sb to Sc (See Figure~\ref{fbar1} and  Figure~\ref{fbar2}).
The  bar fraction (Table~\ref{barFraction}) declines  with $B/T$; it is  
highest ($\sim68\%$)  for bright spirals with  $B/T \le 0.2$, and lower ($\sim36\%$)  
by nearly a factor of two in  spirals with  $B/T > 0.2$. 

It may be tempting to infer this to mean the formation 
of two-thirds of low-$B/T$ bulges is related  to
spontaneous or/and tidally induced bars. 
Such a cause-effect relationship is not the
only scenario consistent with this result.  It may also be possible that
bar instabilities are favored in galaxies with low $B/T$ and no inner Lindblad
resonances.  Under these conditions, the swing amplifier model with
a feedback loop may be responsible for bar formation and partly account 
for the high bar fraction in  galaxies of low $B/T$.

\item
We compare the observed  distribution of bulge $B/T$ and $n$ in 
high mass ($M_\star \geq  1.0\times 10^{10} M_\odot$) spirals with 
predictions from a set of $\Lambda$CDM  cosmological semi-analytical 
models  (Table~\ref{tmodel1}, Table ~\ref{tmodel2}, and 
Figs.~\ref{pbulgeless1} to \ref{pbulgeless4}).
Major mergers are  considered as those with stellar mass ratio 
$M_1/M_2 \ge 1/4$. 
In the models, a high mass spiral can have a bulge with a 
present-day low  $B/T \leq$~0.2  only if it did not undergo 
a major merger since $z\le 2$ (Fig.~\ref{mergehis}). 
The model merger history shows that only 
$\sim$~13.5\%  and $\sim$~20\% of the high mass spirals  experience 
major mergers since $z\le 2$ and  $z\le 4$, respectively.
The fraction ($\sim$~1.6\%) of high mass spirals which have 
undergone a major merger since $z\le 4$  and host a bulge with
a present-day low  $B/T \le 0.2$ is  {\it a factor of over thirty smaller}  
than the observed fraction  ($\sim66\%$) of high mass spirals  
with   $B/T \le 0.2$(Table~\ref{tmodel1}).
Thus, {\it  bulges built  via major mergers  since $z\le 4$, over the last
12 Gyr, seriously fail to account for most of the low  $B/T \le0.2$  bulges 
present in two-thirds of  high-mass spirals}. 
The overall picture that emerges is that  the observed large frequency 
($\sim$ 66\%) of high mass ($M_\star \geq  1.0\times 10^{10} M_\odot$) 
spirals with low present-day $B/T \leq 0.2$   can be accounted for in 
our hierarchical  models by 
{\it  
high mass spirals, which have not undergone a major merger since $z\le 2$, 
and most of which have not even experienced a major merger since $z\le 4$.
}
Most of these present-day low $B/T \leq 0.2$ bulges are likely to 
have been built by a combination of minor mergers and/or secular 
processes since  $z\le4$.

\end{enumerate}

\acknowledgments 
S.J. and T.W. acknowledge support from the National Aeronautics and Space
Administration (NASA) LTSA grant NAG5-13063, NSF grant AST-0607748,
and $HST$ grants GO-10395 and GO-10861 from STScI, which is operated
by AURA, Inc., for NASA, under NAS5-26555.
We thank Chien Peng for technical assistance in the operation of GALFIT. 
We acknowledge the usage of the Hyperleda database (http://leda.univ-lyon1.fr).
This research has made use of the NASA/IPAC Extragalactic Database (NED) which 
is operated by the Jet Propulsion Laboratory, California Institute of 
Technology, under contract with the National Aeronautics and Space 
Administration.


\clearpage

\clearpage

\clearpage

\clearpage

\clearpage

\clearpage

\clearpage

\clearpage

\clearpage

\clearpage

\clearpage

\clearpage

\clearpage

\clearpage

\clearpage

\clearpage

\clearpage

\clearpage

\clearpage

\clearpage

\clearpage

\clearpage

\clearpage

\clearpage

\clearpage

\clearpage

\clearpage


\clearpage
\begin{deluxetable}{lccccccc}
\tabletypesize{\scriptsize}
\tablewidth{0pt}
\tablecolumns{8}
\tablecaption{Sample S1 of Bright ($M_B\leq-19.3$) Low-to-Moderately Inclined ($i \le 70^\circ$) Spirals in  OSUBSGS  (N=143) \label{tosu}}
\tablehead{
  \colhead {Galaxy Name} &
  \colhead {Best Fit} &
  \colhead {Hubble Type} &
  \colhead {Bar Type} &
  \colhead { $D$} &
  \colhead { $M_B$  } &
  \colhead { $B-V$  } &
  \colhead { $M_\star$ } \\
  \colhead {} &
  \colhead {} &
  \colhead {(RC3)} &
  \colhead {(RC3)} &
  \colhead { (Mpc) } &
  \colhead { (mag) } &
  \colhead { (mag) } &
  \colhead { ($M_\odot$) } \\
  \colhead {(1)} & \colhead {(2)} & \colhead {(3)} & \colhead {(4)} &
  \colhead {(5)} & \colhead {(6)} & \colhead {(7)} & \colhead {(8)} 
}
\startdata
ESO138-10 & PSF+Bulge+Disk & SA(s)cd & A  & 15.75 & -20.2 & - & -\\ 
IC0239 & PSF+Bulge+Disk+Bar & SAB(rs)cd & AB  & 15.21 & -19.3 & 0.70 & 1.34e+10\\ 
IC4444 & Bulge+Disk & SAB(rs)bc & AB  & 28.82 & -20.9 & 0.64 & 4.35e+10\\ 
IC5325 & PSF+Bulge+Disk & SAB(rs)bc & AB  & 19.39 & -19.6 & 0.56 & 9.78e+09\\ 
NGC0150 & PSF+Bulge+Disk+Bar & SB(rs)bc & B  & 20.57 & -20.1 & 0.64 & 2.17e+10\\ 
NGC0157 & PSF+Bulge+Disk & SAB(rs)bc & AB  & 22.39 & -21.4 & 0.59 & 5.81e+10\\ 
NGC0210 & Bulge+Disk+Bar & SAB(s)b & AB  & 21.75 & -20.4 & 0.71 & 3.77e+10\\ 
NGC0278 & PSF+Bulge+Disk & SAB(rs)b & AB  & 12.64 & -19.5 & 0.64 & 1.24e+10\\
\enddata
\tablecomments{ Columns are:
(1) Galaxy name.
(2) The best fit chosen based on the criteria outlined in \S~\ref{schoice23}.
(3) Hubble type from  RC3 (de Vaucouleurs \etal 1991).
(4) RC3 bar type, which is based on visual inspection of optical images
and runs as `B'=`strongly barred', `AB'=`weakly barred', and `A'=`unbarred'.
(5) The distance in Mpc from the Nearby Galaxies Catalog (Tully 1988)
    scaled to a Hubble constant of 70 km s$^{-1}$ Mpc$^{-1}$.
    Distances are adjusted assuming the Galaxy is retarded by 300 km $s^{-1}$
    from the universal expansion by the mass of the Virgo Cluster.
    Exceptions are NGC 6753, NGC 6782, and NGC 6907 where    
    distances are from RC3,
    NGC 3504 where the distance is from Kenney \etal (1993), 
    and NGC 4314 where the distance is from Benedict \etal (1996). 
(6) Absolute $B$-band magnitude from Hyperleda. Corrections have been
    applied for galactic and internal extinction.  Galactic absorption
    is computed based on Schlegel \etal (1998).  Internal extinction is
    corrected for following Bottinelli \etal (1995).
    K-correction co-efficients are drawn from RC2 (de Vaucouleurs \etal 1976).
(7) $B-V$ color from Hyperleda.  Corrections for extinction and k-correction
    have been applied as in column 6.
(8) Stellar mass, calculated as outlined in \S~\ref{stellmass}.
}
\end{deluxetable}

\clearpage
\begin{deluxetable}{lccccccccccccc}
\tabletypesize{\scriptsize}
\rotate
\tablewidth{0pt}
\tablecolumns{14}
\tablecaption{Parameters From $2D$ Luminosity Decomposition for Sample S1 (N=143) \label{tmodparam}}
\tablehead{
  \colhead {Galaxy Name} &
  \colhead {Best Fit} &
  \colhead { $B/T$ } &
  \colhead { $D/T$ } &
  \colhead { Bar/$T$ } &
  \colhead { $B/D$ } &
  \colhead { Bulge $r_e$ } &
  \colhead { Bulge $r_e$ } &
  \colhead { Bulge $n$ } &
  \colhead { Disk h } &
  \colhead { Disk h} &
  \colhead { Bar $r_e$ } &
  \colhead { Bar $r_e$ } &
  \colhead { Bar $n$ } \\

  \colhead {           } &
  \colhead {        } &
  \colhead { \% } &
  \colhead { \% } &
  \colhead { \% } &
  \colhead {    } &
  \colhead { $\arcsec$ } &
  \colhead { kpc } &
  \colhead {         } &
  \colhead { $\arcsec$ } &
  \colhead {  kpc } &
  \colhead { $\arcsec$ } &
  \colhead {  kpc      } &
  \colhead {         } \\

  \colhead {(1)} & \colhead {(2)} & \colhead {(3)} & \colhead {(4)} &
  \colhead {(5)} & \colhead {(6)} & \colhead {(7)} & \colhead {(8)} &
  \colhead {(9)} & \colhead {(10)} & \colhead {(11)} & \colhead {(12)} 
  & \colhead {(13)} &\colhead {(14)}
}
\startdata
ESO138-10 & PSF+Bulge+Disk   & 16.25 & 83.75 & - & 0.19 &   10.94 & 0.86 & 2.07    & 27.10 & 2.14 &   - & - & - \\ 
IC0239 & PSF+Bulge+Disk+Bar   & 2.45 & 92.65 & 4.91 & 0.03 &   6.63 & 0.41 & 0.23    & 34.56 & 2.15 &   13.95 & 0.87 & 0.19 \\ 
IC4444 & Bulge+Disk   & 31.97 & 68.03 & - & 0.47 &   7.94 & 1.06 & 2.33    & 17.67 & 2.36 &   - & - & - \\ 
IC5325 & PSF+Bulge+Disk   & 6.79 & 93.21 & - & 0.07 &   15.17 & 1.56 & 1.75    & 23.18 & 2.39 &   - & - & - \\ 
NGC0150 & PSF+Bulge+Disk+Bar   & 5.72 & 81.18 & 13.11 & 0.07 &   4.22 & 0.44 & 0.05    & 33.57 & 3.52 &   17.82 & 1.87 & 0.39 \\ 
NGC0157 & PSF+Bulge+Disk   & 2.29 & 97.71 & - & 0.02 &   3.44 & 0.41 & 0.44    & 31.82 & 3.78 &   - & - & - \\ 
NGC0210 & Bulge+Disk+Bar   & 30.07 & 48.54 & 21.39 & 0.62 &   4.81 & 0.57 & 1.71    & 83.21 & 9.92 &   29.41 & 3.51 & 0.31 \\ 
NGC0278 & PSF+Bulge+Disk   & 4.23 & 95.77 & - & 0.04 &   3.46 & 0.15 & 0.57    & 13.48 & 0.58 &   - & - & - \\
\enddata
\tablecomments{ Columns are:
(1) Galaxy name.
(2) The best fit chosen based on the criteria outlined in \S~\ref{schoice23}.
(3) Bulge-to-total light ratio.  If a PSF is included in the model, the
    contribution from the PSF is also added in.
(4) Disk-to-total light ratio.
(5) Bar-to-total light ratio.
(6) Bulge-to-disk ratio
(7) Bulge effective radius in arcseconds.
(8) Bulge effective radius in kpc.
(9) Bulge S\'ersic index.
(10) Disk scalelength in arcseconds.
(11) Disk scalelength in kpc.
(12) Bar effective radius in arcseconds.
(13) Bar Bulge effective radius in kpc.
(14) Bar S\'ersic index.
}
\end{deluxetable}

\clearpage
\begin{deluxetable}{cccc cccc}
\tablewidth{0pt}
\tablecolumns{8}
\tablecaption{Decomposition For NGC 4643  \label{tngc4643}   }   
\tablehead
{
  \colhead {Fit} &
  \colhead {} &
  \colhead { $r_e$ or h } &
  \colhead { $r_e$ or h } &
  \colhead { $n$   } &   
  \colhead { $b/a$  } &
  \colhead { Position Angle  } &
  \colhead { Fractional light  } \\

  \colhead {} &
  \colhead {} &
  \colhead { ($\arcsec$) } &
  \colhead { (kpc) } &
  \colhead { } &
  \colhead { } &
  \colhead { } &
  \colhead { } \\

  \colhead {(1)} & \colhead {(2)} & \colhead {(3)} & \colhead {(4)} &
  \colhead {(5)} & \colhead {(6)} & \colhead {(7)} & \colhead {(8)}

}
\startdata
Stage 1 & S\'ersic & 27.90                  &  2.66              & 4.44  &  0.80 &  -51.03           & 100\%           \\
\hline
Stage 2 & Bulge    & 23.86                  &  2.30              & 4.16  &  0.80 &  -51.08           & 34.6\%          \\
        & Disk     & 335.88                 &  32.33             & 1.00  &  0.84 &   66.94           & 65.4\%          \\
\hline
Stage 3 & Bulge    & 5.43                   &  0.52              & 2.53  &  0.90 &   60.52           & 25.0\%          \\
        & Disk     & 48.22                  &  4.64              & 1.00  &  0.84 &   66.94           & 54.1\%          \\
        & Bar      & 21.30                  &  2.05              & 0.62  &  0.37 &  -45.84           & 20.9\%          \\
\enddata
\tablecomments{ Columns are:
(1) Indicates whether the model is for Stage 1 (pure S\'ersic profile), 
    Stage 2 (bulge+disk or bar+disk), or Stage 3 (bulge+disk+bar).
(2) One of pure S\'ersic profile, S\'ersic bulge, exponential disk,
    or S\'ersic bar.
(3) Effective radius of the bulge/bar or disk scalelength in arcseconds.
(4) Effective radius of the bulge/bar or disk scalelength in kpc.
(5) The S\'ersic index of the profile.  
(6) Axis ratio of the component.
(7) Position angle in degrees of the component.  Positive angles are East of North.
(8) The percentage of total galaxy light contributed by the component.
}
\end{deluxetable}

\clearpage
\begin{deluxetable}{cccc cccc}
\tablewidth{0pt}
\tablecolumns{8}
\tablecaption{Decomposition For NGC 4548  \label{tngc4548}} 
\tablehead
{
  \colhead {Fit} &
  \colhead {} &
  \colhead { $r_e$ or h } &
  \colhead { $r_e$ or h } &
  \colhead { $n$   } &
  \colhead { $b/a$  } &
  \colhead { Position Angle  } &
  \colhead { Fractional light  } \\

  \colhead {} &
  \colhead {} &
  \colhead { ($\arcsec$) } &
  \colhead { (kpc) } &
  \colhead { } &
  \colhead { } &
  \colhead { } &
  \colhead { } \\

  \colhead {(1)} & \colhead {(2)} & \colhead {(3)} & \colhead {(4)} &
  \colhead {(5)} & \colhead {(6)} & \colhead {(7)} & \colhead {(8)}
}
\startdata
Stage 1 & S\'ersic & 154.59                 &  5.19              & 5.19  &  0.80 &   78.31           & 100\%           \\
\hline
Stage 2 & Bulge    & 57.86                  &  1.94              & 4.32  &  0.76 &   75.77           & 61.5\%          \\
        & Disk     & 60.39                  &  2.03              & 1.00  &  0.75 &  -32.54           & 38.5\%          \\
\hline
Stage 3 & Bulge    & 6.98                   &  0.23              & 1.56  &  0.88 &  -66.50           & 13.0\%          \\
        & Disk     & 58.22                  &  1.96              & 1.00  &  0.75 &  -32.54           & 68.6\%          \\
        & Bar      & 44.91                  &  1.51              & 0.51  &  0.35 &   66.65           & 18.4\%          \\
\enddata
\tablecomments{ Columns are:
(1) Indicates whether the model is for Stage 1 (pure S\'ersic profile),   
    Stage 2 (bulge+disk or bar+disk), or Stage 3 (bulge+disk+bar).
(2) One of pure S\'ersic profile, S\'ersic bulge, exponential disk,
    or S\'ersic bar.
(3) Effective radius of the bulge/bar or disk scalelength in arcseconds.
(4) Effective radius of the bulge/bar or disk scalelength in kpc.
(5) The S\'ersic index of the profile.
(6) Axis ratio of the component.
(7) Position angle in degrees of the component.  Positive angles are East of North.
(8) The percentage of total galaxy light contributed by the component.
}
\end{deluxetable}

\clearpage
\begin{deluxetable}{cccc cccc}
\tablewidth{0pt}
\tablecaption{Decomposition For NGC 4902 \label{tngc4902} }
\tablecolumns{8}
\tablehead
{
  \colhead {Fit} &
  \colhead {} &
  \colhead { $r_e$ or h } &
  \colhead { $r_e$ or h } &
  \colhead { $n$   } &
  \colhead { $b/a$  } &
  \colhead { Position Angle  } &
  \colhead { Fractional light  } \\

  \colhead {} &
  \colhead {} &
  \colhead { ($\arcsec$) } &
  \colhead { (kpc) } &
  \colhead { } &
  \colhead { } &
  \colhead { } &
  \colhead { } \\

  \colhead {(1)} & \colhead {(2)} & \colhead {(3)} & \colhead {(4)} &
  \colhead {(5)} & \colhead {(6)} & \colhead {(7)} & \colhead {(8)}
}
\startdata
Stage 1 & S\'ersic & 154.4                  &   28.8             & 5.11  & 0.54    & 69.0              & 100\%           \\
\hline
Stage 2 & Bulge    & 11.5                   &   2.15             & 1.02  & 0.38    & 68.5              & 17.4\%          \\
        & Disk     & 32.8                   &   6.12             & 1.00  & 0.84    & 81.1              & 82.6\%          \\
\hline
Stage 3 & Bulge    & 4.31                   &   0.80             & 0.85  & 0.68    & -52.96            & 5.59\%          \\
        & Disk     & 29.2                   &   5.45             & 1.00  & 0.84    & 81.1              & 82.6\%          \\
        & Bar      & 14.1                   &   2.64             & 0.39  & 0.22    & 66.37             & 9.97\%          \\
\enddata
\tablecomments{ Columns are:
(1) Indicates whether the model is for Stage 1 (pure S\'ersic profile),   
    Stage 2 (bulge+disk or bar+disk), or Stage 3 (bulge+disk+bar).
(2) One of pure S\'ersic profile, S\'ersic bulge, exponential disk,
    or S\'ersic bar.
(3) Effective radius of the bulge/bar or disk scalelength in arcseconds.
(4) Effective radius of the bulge/bar or disk scalelength in kpc.
(5) The S\'ersic index of the profile.
(6) Axis ratio of the component.
(7) Position angle in degrees of the component.  Positive angles are East of North.
(8) The percentage of total galaxy light contributed by the component.
}
\end{deluxetable}

\clearpage
\begin{deluxetable}{cccccccccccccccc}
\tablewidth{0pt}
\tablecolumns{9}
\tablecaption{Checking GALFIT robustness with different input guesses for Stage 3\label{t1dfit}}
\tabletypesize{\scriptsize}
\tablehead
{
\colhead{} & \colhead{$B/T$} & \colhead{Bulge $r_e$} & \colhead{Bulge $n$} &
\colhead{$D/T$} & \colhead{Disk $h$} & \colhead{Bar/$T$} & \colhead{Bar $r_e$} &
\colhead{Bar $n$} \\

\colhead{} & \colhead{} & \colhead{('')} &  \colhead{} & 
\colhead{} & \colhead{('')} & \colhead{} & \colhead{('')} & \colhead{}  \\

\colhead {(1)} & \colhead {(2)} & \colhead {(3)} & \colhead {(4)} &
\colhead {(5)} & \colhead {(6)} & \colhead {(7)} & \colhead {(8)} & \colhead {(9)}
}
\startdata
\cutinhead{ \textbf{NGC 4548} }
Initial guesses from 1D decomposition & 17.5\% & 7.39 & 1.17 & 63.5\% & 28.4 & 19.0\% & 37.5 & 0.54 \\
Stage 3 Output           & 13.0\% & 6.98 & 1.56 & 68.6\% & 58.2 & 18.4\% & 44.9 & 0.51 \\
Initial guesses from from Stage 2           & 11.1\% & 7.50 & 1.70 & 69.9\% & 64.5 & 19.1\% & 37.5 & 0.54 \\
Stage 3 Output           & 13.0\% & 6.98 & 1.56 & 68.6\% & 58.2 & 18.4\% & 44.9 & 0.51 \\
\cutinhead{ \textbf{NGC 4643} }
Initial guesses from 1D decomposition & 33.6\% & 7.18 & 0.86 & 40.4\% & 37.5 & 26.0\% & 22.0 & 0.60 \\
Stage 3 Output           & 25.0\% & 5.43 & 2.53 & 54.1\% & 48.2 & 20.9\% & 21.3 & 0.62 \\
Initial guesses from Stage 2          & 24.1\% & 5.30 & 2.5  & 51.8\% & 46.4 & 24.1\% & 22.0 & 0.60 \\
Stage 3 Output           & 25.0\% & 5.43 & 2.53 & 54.1\% & 48.2 & 20.9\% & 21.3 & 0.62 \\
\enddata
\tablecomments{ For each galaxy, the rows are:
(1) The initial guesses to Stage 3 provided by 1D decomposition.
(2) The corresponding output for the input of row (1).
(3) The initial guesses to Stage 3 provided by Stage 2.
(4) The corresponding output for the input of row (3).\\
The columns are:
(2) Bulge-to-total luminosity fraction.
(3) Effective radius of the bulge in arcseconds.
(4) The S\'ersic index of the profile.
(5) Disk-to-total luminosity fraction.
(6) Disk scalelength in arcseconds.
(7) Bar-to-total luminosity fraction.
(8) Bar effective radius in arcseconds.
(9) Bar S\'ersic index.
} 
\end{deluxetable}

\clearpage
\begin{deluxetable}{lccccccccccccccc}
\tablewidth{0pt}
\tablecolumns{9}
\tablecaption{Parameter Coupling\label{tcoupling}}
\tabletypesize{\scriptsize}
\tablehead
{

\colhead{Type of Fit} & \colhead{$\chi^2$} & \colhead{Bulge $r_e$} & \colhead{Bulge $n$} &
\colhead{$B/T$} & \colhead{Disk $h$} & \colhead{$D/T$} & \colhead{Bar $r_e$} &
\colhead{Bar $n$} & \colhead{Bar/$T$} \\

\colhead{} & \colhead{} & \colhead{('')} &  \colhead{} & 
\colhead{} & \colhead{('')} & \colhead{} & \colhead{('')} & \colhead{} & \colhead{} \\

\colhead {(1)} & \colhead {(2)} & \colhead {(3)} & \colhead {(4)} &
\colhead {(5)} & \colhead {(6)} & \colhead {(7)} & \colhead {(8)} & 
\colhead {(9)} & \colhead {(10)}

}
\startdata
\cutinhead{ \textbf{NGC 3885} }
initial input & 11241.33 & 2.32 & 2.00 & 27.7\% & 10.44 & 72.3\% & - & - & -\\
n=free & 8210.12 & 3.13 & 0.46 & 27.1\% & 11.94 & 72.9\% & - & - & -\\
n=1 fixed & [8307.82] & [2.96] & [1.00] & [29.1\%] & [12.90] & [70.9\%] & - & - & -\\
n=4 fixed & [7373.55] & [6.28] & [4.00] & [53.7\%] & [20.45] & [46.3\%] & - & - & -\\
n=1 re-fit & 8210.12 & 3.13 & 0.46 & 27.1\% & 11.94 & 72.9\% & - & - & -\\
n=4 re-fit & 7258.46 & 7.11 & 4.49 & 57.3\% & 19.87 & 42.7\% & - & - & -\\
\cutinhead{ \textbf{NGC 7213} }
initial input & 605757.39 & 5.80 & 4.00 & 56.4\% & 9.25 & 43.6\% & - & - & -\\
n=free & 71170.10 & 19.55 & 2.68 & 65.7\% & 63.19 & 34.3\% & - & - & -\\
\enddata
\tablecomments{ For each galaxy, the rows are:
(1)  The model outputs (columns 2 to 10) from GALFIT
     based on the Stage 1 or Stage 2 fits. These outputs
     are used as initial guesses in the fits for rows 2 and 3.
(2)  The model outputs (columns 2 to 10) from GALFIT when
     the bulge $n$ is allowed to vary freely, and the input guesses are based on
     the parameters in columns 2 to 10 of row 1.
(3)  The model outputs (columns 2 to 10) from GALFIT when the bulge
     S\'ersic index is held fixed at $n=1$,  and the input guesses are based on
     the parameters in columns 2 to 10 of row 1. The model is not always physically
     meaningful because the bulge index is fixed.
(4)  As in row 3, except that the  bulge S\'ersic index is now held fixed at $n=4$.
      The model is not always physically meaningful because the bulge index is fixed.
(5) The model outputs (columns 2 to 10) from GALFIT, when
    the bulge $n$ is allowed to vary freely, but the input guesses are now based on
    the parameters in columns 2 to 10 of row 3.
(6) The model outputs (columns 2 to 10) from GALFIT, when
    the bulge $n$ is allowed to vary freely, but the input guesses are now based on
    the parameters in columns 2 to 10 of row 4.
The columns are:
(1) The type of fit whose model outputs are  shown in columns 2 to 10.
(2) The $\chi^2$ of the fit.
(3) Effective radius of the bulge in arcseconds.
(4) The S\'ersic index of the bulge (in the case of Stage 2 bulge-disk fits 
    in row 1 and in the case of all fits in rows 2 to 6), or the S\'ersic index 
    of a single component (in the case of Stage 1 fits in row 1).
(5) Bulge-to-total luminosity fraction.
(6) Disk scalelength in arcseconds.
(7) Disk-to-total luminosity fraction.
(8) Bar effective radius in arcseconds.
(9) Bar S\'ersic index.
(10) Bar-to-total luminosity fraction.
}
\end{deluxetable}

\clearpage
\begin{deluxetable}{cc}
\tablewidth{0pt}
\tablecolumns{2}
\tablecaption{Mass Breakdown of Galactic Structures \label{masstable} } 
\tablehead
{
     \colhead{Structure} & \colhead{Mass} 
}
\startdata
Bulges             & $18.9\% \pm 3.49\% $    \\
Disks              & $71.6\% \pm 4.02\% $     \\
Bars               & $9.58\% \pm 2.62\% $    \\
\hline
Bulges with $n>2$  & $10.4\% \pm 2.72\% $    \\
Bulges with $n\leq2$  & $8.42\% \pm 2.47\% $    \\
\enddata
\end{deluxetable}

\clearpage
\begin{deluxetable}{lc}
\tablewidth{0pt}
\tablecaption{$B/T$ and bulge $n$ in bright/high mass spirals \label{bulge1}}
\tablecolumns{2}
\tablehead
{
       \multicolumn{2}{c}{Sample ($M_B\leq-19.3$) }
}
\startdata
Fraction of spirals with bulge $n\geq4$   &  $ 1.40\% \pm 0.98\%$ \\
Fraction of spirals with bulge $2<n<4$    &  $ 22.4\% \pm 3.49\%$ \\
Fraction of spirals with bulge $n\leq2$   &  $ 76.2\% \pm 3.56\%$ \\
Fraction of spirals with bulge $n>2$   &  $ 23.8\% \pm 3.56\%$ \\
Fraction of spirals with $B/T \leq 0.2$  &   $ 68.5\% \pm 3.88\%$  \\
Fraction of spirals with $B/T > 0.2$  &   $ 31.1\% \pm 3.88\%$  \\
Fraction of spirals with $0.2 < B/T < 0.4$ & $ 23.1\% \pm 3.52\%$  \\
Fraction of spirals with $B/T \geq 0.4$  &   $ 8.39\% \pm 2.32\%$  \\
\cutinhead{Sample $M_\star \geq1\times10^{10}M_\odot$ }   
Fraction of spirals with bulge $n\geq4$  &   $ 1.77\% \pm 1.24\%$  \\
Fraction of spirals with bulge $2<n<4$   &   $ 23.9\% \pm 4.01\%$  \\
Fraction of spirals with bulge $n\leq2$  &   $ 74.3\% \pm 4.11\%$  \\
Fraction of spirals with bulge $n>2$   &  $ 25.7\% \pm 4.11\%$ \\
Fraction of spirals with $B/T \leq 0.2$  &   $ 66.4\% \pm 4.44\%$  \\
Fraction of spirals with $B/T > 0.2$  &   $ 33.6\% \pm 4.44\%$  \\
Fraction of spirals with $0.2 < B/T < 0.4$ & $ 25.7\% \pm 4.11\%$  \\
Fraction of spirals with $B/T \geq 0.4$  &   $ 7.96\% \pm 2.55\%$  \\
\enddata
\end{deluxetable}

\clearpage
\begin{deluxetable}{lc}
\tablecolumns{2}
\tablecaption{Bar Fraction as a Function of $B/T$ and Bulge Index  \label{barFraction}}
\tablewidth{0pt}
\tablehead
{
       \multicolumn{2}{c}{Sample ($M_B\leq-19.3$) }
}
\startdata
Bar fraction in spirals with bulge $n\geq4$   &  $ 0.00\% \pm 0.00\%$ \\
Bar fraction in spirals with bulge $2<n<4$    &  $ 37.5\% \pm 8.56\%$ \\
Bar fraction in spirals with bulge $n\leq2$   &  $ 65.1\% \pm 4.56\%$ \\
Bar fraction of spirals with bulge $n>2$   &  $ 35.3\% \pm 8.20\%$ \\
Bar fraction of spirals with $B/T \leq 0.2$  &   $ 68.4\% \pm 4.70\%$  \\
Bar fraction of spirals with $B/T > 0.2$  &   $ 35.6\% \pm 7.14\%$  \\
Bar fraction of spirals with $0.2 < B/T < 0.4$ & $ 42.4\% \pm 8.60\%$  \\
Bar fraction of spirals with $B/T \geq 0.4$  &   $ 16.7\% \pm 10.8\%$  \\
\cutinhead{ Sample $M_\star \geq1\times10^{10}M_\odot$ }        
Bar fraction of spirals with bulge $n\geq4$  &   $ 0.00\% \pm 0.00\%$  \\
Bar fraction of spirals with bulge $2<n<4$   &   $ 44.4\% \pm 9.56\%$  \\
Bar fraction of spirals with bulge $n\leq2$  &   $ 63.1\% \pm 5.27\%$  \\
Bar fraction of spirals with bulge $n>2$   &  $ 41.4\% \pm 9.15\%$ \\
Bar fraction of spirals with $B/T \leq 0.2$  &   $ 68.0\% \pm 5.39\%$  \\
Bar fraction of spirals with $B/T > 0.2$  &   $ 36.8\% \pm 7.83\%$  \\
Bar fraction of spirals with $0.2 < B/T < 0.4$ & $ 41.4\% \pm 9.15\%$  \\
Bar fraction of spirals with $B/T \geq 0.4$  &   $ 22.2\% \pm 13.9\%$  \\
\enddata
\end{deluxetable}

\clearpage
\begin{deluxetable}{cccccc}
\tablecolumns{6}
\tablecaption{$B/T$: Data versus Hierarchical Models of Galaxy Evolution \label{tmodel1}}
\tablewidth{0pt}
\tabletypesize{\scriptsize}
\tablehead
{
                                              &     Data                & Model  & Model spirals with     & Model spirals with   & All model \\
                                              &                         &        &  major mergers         & no major merger      & spirals \\
                                              &                         &        & since redshift $z_1$   & since redshfit $z_1$ & \\
               (1)                            &    (2)                  &   (3)  &  (4)   & (5) &  (6) \\
}
\startdata
                                               &                        &        &                        &                      &        \\
\multicolumn{3}{r}{}                           & \multicolumn{1}{c}{For $z_{\rm 1}\le2$}  & \multicolumn{1}{c}{For $z_{\rm 1}\le2$} & \multicolumn{1}{c}{} \\
Fraction of spirals with $B/T \leq 0.2$        &   66.4\% $\pm$ 4.44\%  &   SK   & 0.1\%                 & 67.2\%               & 67.3\% \\
Fraction of spirals with $0.2 < B/T \le 0.4$   &   25.7\% $\pm$ 4.11\%  &   SK   & 2.2\%                 & 16.6\%               & 18.8\% \\
Fraction of spirals with $0.4 < B/T \le 0.75$  &   8.0\% $\pm$ 2.5\%  &   SK   & 11.2\%                 & 2.6\%               & 13.8\% \\
Fraction of spirals with $B/T \le 0.75$        &   100\%                &   SK   & 13.5\%                 & 86.5\%               & 100.0\% \\
\hline                                               &                                 &                   &                             &        \\
\multicolumn{3}{r}{}                           & \multicolumn{1}{c}{For $z_{\rm 1}\le4$}  & \multicolumn{1}{c}{For $z_{\rm 1}\le4$}  & \multicolumn{1}{c}{} \\
\enddata
\tablecomments{
Column 2 shows the empirical fraction of  high mass spirals with a given $B/T$.
Columns 3 to 6 show the fraction of high mass spirals with a given $B/T$ in the models, which define spirals as 
systems with $B/T \le 0.75$.  Column 3 indicates  the model used: SK  refers to the models by Khochfar \& Burkert (2005) and 
Khochfar \& Silk (2006), while PH refers to those by Hopkins et al. (2009).  The models are  described in \S \ref{smodel1} and 
have different prescriptions for the amount of mass that gets converted into bulge stars.  Column 4 shows the fraction of 
spirals, which have a given $B/T$ and have had one or more major mergers since redshift $z_1$.  Column 5 shows the fraction of 
spirals, which have a given $B/T$ and have had no major merger since redshift $z_1$.  Column 6  is the sum of column 4 and 5,
and shows the fraction of spirals with a given $B/T$ in the models. It can be directly compared to the data in column 2. 
}
\end{deluxetable}

\clearpage
\begin{deluxetable}{ll}
\tablecolumns{2}
\tablecaption{Comparison of model merger fraction since $z\le4$ and bulge S\'ersic index\label{tmodel2}}
\tablewidth{0pt}
\tablehead
{
  \multicolumn{2}{c} {Stellar mass limit: $1.0\times10^{10}M_\odot$} \\
  \multicolumn{2}{c} { Model galaxies: 7674 } \\
  \multicolumn{2}{c} { Observed galaxies: 113 }
}
\startdata
\cutinhead{ \textbf{Test 1} }  \\
\% of model galaxies w/ $B/T \le 0.2$ with no major merger since $z\le4$: & 5040/7674 = $65.7\%  \pm 0.54\%$ \\
\% of observed galaxies w/ $B/T \le 0.2$ and bulge $n\le 3$:    & 75/113 = $66.4\%  \pm 4.44\%$ \\
\% of observed galaxies w/ $B/T \le 0.2$ and bulge $n \le 2.5$: & 74/113 =  $65.5\% \pm 4.47\%$  \\
\% of observed galaxies w/ $B/T \le 0.2$ and bulge $n \le 2$:   & 67/113 = $59.3\%  \pm 4.62\%$ \\
\cutinhead{ \textbf{Test 2} }  \\
\% of model galaxies w/ $B/T\geq0.4$ and a major merger since $z\le4$: & 971/7674 = $12.7\%  \pm 0.38\%$ \\
\% of observed spirals w/ $B/T\geq0.4$ and bulge $n>3$:& 2/113 = $1.77\%  \pm 1.24\%$ \\
\% of observed spirals w/ $B/T\geq0.4$ and bulge $n>2.5$:& 4/113 =  $3.54\% \pm 1.74\%$  \\
\% of observed spirals w/ $B/T\geq0.4$ and bulge $n>2$:& 7/113 =  $6.19\% \pm 2.27\%$  \\
\cutinhead{ \textbf{Test 3} }  \\
\% of model galaxies w/ $B/T \le 0.2$ and a major merger since $z\le4$: & 126/7674 = $1.64\%  \pm 0.15\%$ \\
\% of observed galaxies w/ $B/T\leq0.2$ and bulge $n>3$:& 0/113 =  $0.00\% \pm 0.00\%$ \\
\% of observed galaxies w/ $B/T\leq0.2$ and bulge $n>2.5$ & 1/113 = $0.88\% \pm 0.88\%$ \\
\% of observed galaxies w/ $B/T\leq0.2$ and bulge $n>2$ & 8/113 =    $7.08\% \pm 2.41\%$ \\
\enddata
\end{deluxetable}

\end{document}